\documentclass[12pt,preprint]{aastex}
\usepackage{morefloats}
\usepackage{rotating}
\shorttitle{Disk Masses}
\shortauthors{Mohanty}
\begin{document}
\def\zw{z_w }
\def\dop{\mathcal{D} }
\def\xangle{\vartheta }
\def\xanglew{\vartheta_w }
\def\Phid{{\Phi}_d }
\def\Phidx{{\Phi}_{\rm dx} }
\def\Phimx{{\Phi}_{\rm mx} }
\def\Phim{{\Phi}_{\rm m}}
\def\Phif{{\Phi}_{\rm f} }
\def\Phit{{\Phi}_t }
\def\Phiw{{\Phi}_w }
\def\Phir{{\Phi}_{\rm r} }
\def\rx{R_X }
\def\rk{R_k }
\def\del{{\bf{\nabla}} }
\def\delsq{{\nabla}^2 }
\def\vecr{{\bf{r}} }
\def\vecB{{\bf{B}} }
\def\vecBp{{\bf{B}_{\rm p}} }
\def\vecBa{{{B_{\varphi}}} }
\def\vecBr{{{B_{\varpi}}} }
\def\vecBz{{{B}_z} }
\def\malven{{\mathcal{M}}_A }
\def\rstar{R_{\ast} }
\def\rd{R_D }
\def\omstar{{\Omega}_{\ast} }
\def\omx{{\Omega}_X }
\def\ax{a_X }
\def\jbstar{\bar J_{\ast} }
\def\zmax{z_{\rm max} }
\def\deg{$^{\circ}$ }
\def\erhat{{\^{e}}$_{\varpi}$}
\def\eahat{{\^{e}}$_{\varphi}$}
\def\ezhat{{\^{e}}$_z$}
\def\mstar{M_{\ast} }
\def\rstar{R_{\ast} }
\def\tstar{T_{\ast} }
\def\lstar{L_{\ast} }
\def\msun{M_{\odot} }
\def\rsun{R_{\odot} }
\def\lsun{L_{\odot} }
\def\mdotd{\dot M_D }
\def\mdot{\dot M }
\def\rhos{{\rho}_0 }
\def\vs{v_0 }
\def\rhoa{{\rho}_{Am} }
\def\va{v_{Am} }
\def\bs{B_0 }
\def\ba{B_{Am} }
\def\rs{r_0 }
\def\ra{r_{Am} }
\def\rr{r_{A2} }
\def\vp{{\mathcal{V}}_0 }
\def\hatm{{\hat{m}}_d }
\def\hatmu{{\hat{m}}_u }
\def\sd{{{\sigma}}_d }
\def\hatml{{\hat{m}}_{lim,u} }
\def\sl{{{\sigma}}_{u} }

\title{Protoplanetary Disk Masses from Stars to Brown Dwarfs}
\author{Subhanjoy Mohanty\altaffilmark{1}, Jane Greaves\altaffilmark{2}, Daniel Mortlock\altaffilmark{1}, Ilaria Pascucci\altaffilmark{3}, Aleks Scholz\altaffilmark{4}, \\Mark Thompson\altaffilmark{5}, Daniel Apai\altaffilmark{3,6}, Giuseppe Lodato\altaffilmark{7}, Dagny Looper\altaffilmark{8}}
\altaffiltext{1}{Imperial College London, 1010 Blackett Lab, Prince Consort Rd, London SW7 2AZ, UK. s.mohanty@imperial.ac.uk}
\altaffiltext{2}{SUPA, Physics \& Astronomy, Univ. of St. Andrews, North Haugh, St. Andrews, Fife KY16 9SS, UK.}
\altaffiltext{3}{Dept. of Planetary Sciences and Lunar and Planetary Lab, Univ. of Arizona, Tucson AZ 85721, USA.}
\altaffiltext{4}{School of Cosmic Physics, Dublin Institute for Advanced Studies, 31 Fitzwilliam Place, Dublin 2, Ireland.}
\altaffiltext{5}{Ctr. for Astrophysics Research, Univ. of Hertfordshire, College Lane, Hatfield AL10 9AB, UK.}
\altaffiltext{6}{Dept. of Astronomy and Steward Observatory, Univ. of Arizona, Tucson, AZ 85721, USA.}
\altaffiltext{7}{Dipartimento di Fisica, Universit\`{a} Degli Studi di Milano, Via Celoria, 16 Milano, 20133, Italy.}
\altaffiltext{8}{Institute for Astronomy, Univ. of Hawai'i, 2680 Woodlawn Dr, Honolulu, HI 96822, USA.}

\begin{abstract}
We present SCUBA-2 850\,$\mu$m observations of 7 very low mass stars (VLMS) and brown dwarfs (BDs).  3 are in Taurus and 4 in the TW Hydrae Association (TWA), and all are classical T Tauri (cTT) analogs.  We detect 2 of the 3 Taurus disks (one only marginally), but none of the TWA ones.  For standard grains in cTT disks, our 3$\sigma$ limits correspond to a dust mass of $1.2 M_{\Earth}$ in Taurus and a mere $0.2 M_{\Earth}$ in the TWA (3--10$\times$ deeper than previous work).  We combine our data with other sub-mm/mm surveys of Taurus, $\rho$ Oph and the TWA to investigate the trends in disk mass and grain growth during the cTT phase.  Assuming a gas-to-dust mass ratio of 100:1 and fiducial surface density and temperature profiles guided by current data, we find the following.  {\it (1)} The minimum disk outer radius required to explain the upper envelope of sub-mm/mm fluxes is $\sim$100\,AU for intermediate-mass stars, solar-types and VLMS, and $\sim$20\,AU for BDs.  {\it (2)} While the upper envelope of apparent disk masses increases with $\mstar$ from BDs to VLMS to solar-type stars, no such increase is observed from solar-type to intermediate-mass stars.  We propose this is due to enhanced photoevaporation around intermediate stellar masses.  {\it (3)} Many of the disks around Taurus and $\rho$ Oph intermediate-mass and solar-type stars evince an opacity index of $\beta \sim$ 0--1, indicating significant grain growth.  Of the only four VLMS/BDs in these regions with multi-wavelength measurements, three are consistent with considerable grain growth, though optically thick disks are not ruled out.  {\it (4)} For the TWA VLMS (TWA 30A and B), combining our 850\,$\mu$m fluxes with the known accretion rates and ages suggests substantial grain growth by 10\,Myr, comparable to that in the previously studied TWA cTTs Hen 3-600A and TW Hya.  The degree of grain growth in the TWA BDs (2M1207A, SSPM1102) remains largely unknown.  {\it (5)} A Bayesian analysis shows that the apparent disk-to-stellar mass ratio has a roughly constant mean of log$_{10}$$[M_{disk}/\mstar] \approx -2.4$ all the way from intermediate-mass stars to VLMS/BDs, supporting previous qualitative suggestions that the ratio is $\sim$1\% throughout the stellar/BD domain. {\it (6)} Similar analysis shows that the disk mass in close solar-type Taurus binaries (sep $<$100\,AU) is significantly lower than in singles (by a factor of 10), while that in wide solar-type Taurus binaries ($\geq$100\,AU) is closer to that in singles (lower by a factor of 3). {\it (7)} We discuss the implications of these results for planet formation around VLMS/BDs, and for the observed dependence of accretion rate on stellar mass.                
\end{abstract}


\section{Introduction}
The masses of the primordial disks girdling newborn stars and brown dwarfs, and the degree of grain growth in these disks, are key to the processes of accretion, planet formation and migration within them.  Much work has gone into inferring disk masses and grain growth around solar-type and higher-mass stars \citep[e.g.,][(B90, AW07, AW07, R10a,b)]{1990AJ.....99..924B, 2005ApJ...631.1134A, 2007ApJ...671.1800A, 2010AA...521A..66R, 2010AA...512A..15R}, and such studies now extend into the substellar domain as well \citep{2003ApJ...593L..57K, 2006ApJ...645.1498S, 2009ApJ...701..698S}.  


Here we present the results of a JCMT/SCUBA-2 850 $\mu$m pilot survey of 7 very low mass stars (VLMS) and brown dwarfs (BDs), spanning 0.02--0.2 $\msun$ in mass and located in the $\sim$1 Myr-old Taurus star-forming region and the $\sim$10 Myr-old TW Hydrae Association (TWA).  All the sources are known to be accreting from surrounding primordial disks. The study, undertaken as part of SCUBA-2 first-light observations, is 3--10 times deeper than previous such surveys of young VLMS/BDs, and includes 4 out of the 5 confirmed VLMS/BD accretors in the TWA, none of which have been examined at these wavelengths before\footnote{The one other confirmed TWA VLMS accretor, Hen 3-600A, has been observed earlier and is included in our final analysis, along with another (higher mass) accretor in the region, TW Hya.  Two additional VLMS TWA members -- TWA 33 and 34 -- have been reported very recently by Schneider et al.\,(2012); they are not included in our analysis, but discussed briefly in \S2.1.  The only other (higher mass) accretor in the region, TWA 5A, has not been observed in the sub-mm/mm, and is also excluded from our study.  }.  

Our goals are twofold.  First, by combining our data with previous large sub-mm/mm disk surveys (specifically, those of AW05, AW07, Scholz et al.\,2006 and Schaefer et al.\,2009), we wish to investigate trends in disk mass and grain growth as a function of stellar mass, and the attendant implications for planet formation and accretion.  Our analysis generally follows that of AW05 and AW07, with a few significant differences: {\it (a)} we concentrate solely on Class II objects (or, roughly equivalently, classical T Tauri (cTT) sources), in order to avoid confusion by both envelope contamination in earlier evolutionary types and a possible lack of primordial disks altogether in more evolved sources; {\it (b)} we combine the Taurus and $\rho$ Oph samples of AW05 and AW07, and extend the study to much lower (sub)stellar masses by the addition of objects from our own and other surveys; {\it (c)} we take advantage of various more recent investigations of disk radii, surface densities and temperatures, in order to define a set of realistic fiducial disk parameters that allows us to focus on the primary unknowns of interest here -- disk mass and grain growth; and {\it (d)} we explicitly frame the analysis in terms of generalized equations that extend the Rayleigh Jeans (RJ) formalism of B90 to the non-RJ regime (which we show is especially vital for VLMS/BDs).                  

Second, and equally important, we wish to present a Bayesian framework for some of the above analysis, which makes maximal use of the non-detections/upper limits in the combined dataset.  The majority of disks around VLMS and BDs, as well as a significant fraction of those around solar-type and higher-mass stars, remain undetected in the sub-mm/mm.  While these non-detections clearly have something to say about the underlying disk mass distribution, the question is how to combine them with the detections in order to extract the maximum information encoded in all the data.  This is obviously not an issue restricted to studies of disk masses, but one that is central to all surveys that include upper limits.  Unfortunately, it has often not been addressed satisfactorily in the young stellar community.  Non-detections are frequently simply ignored, or upper limit values -- ranging arbitrarily from 2 to 5$\sigma$ -- used as true detections in order to estimate the sample distribution.  A more sophisticated approach has sometimes been to invoke survival analysis, based on the classic work by \citet{1985ApJ...293..192F}.  However, as the latter authors explicitly caution, this technique {\it (a)} is not appropriate when the upper-limits are correlated with the variable under discussion (e.g., if the upper limits on the observed flux are primarily due to the sensitivity of a flux-limited survey), which is clearly the case in many if not most astronomical studies; and {\it (b)} does not account for noise in the data.  A Bayesian approach, on the other hand, provides an elegant and intuitively simple way of dealing with both upper limits and noise.  We outline the general method, and apply it to our combined disk mass dataset; we hope that the community adopts the technique more widely for analysing analogous surveys.  

Our sample, observations and data reduction procedure are summarized in \S2, and our derivation of stellar parameters described in \S3.  \S4 provides a brief overview of the equations we use to model the disk spectral energy distribution, with more details in Appendix A.  We summarize our method for analysing grain growth and disk mass in \S5, and discuss our choice of fiducial disk parameters for this analysis in Appendix B.  Our results are described in \S6-8, which cover: the validity of the Rayleigh-Jeans approximation (\S6), trends in grain growth and disk mass (\S7), and a Bayesian analysis of the ratio of disk to stellar mass (\S8, with an outline of the general Bayesian technique in Appendix C).  The implications for planet formation and mass accretion rates are discussed in \S9, and our conclusions presented in \S10.           

\section{Data}
\subsection{SCUBA-2 Sample and Additional Sources}
We were awarded 10 hours of shared-risk (first light) time on JCMT/SCUBA-2 to investigate primordial disks around VLMS/BDs.  The observed sample consists of 7 young objects: CHFT-BD Tau 12 (M6.5), GM Tau (M6.5) and J044427+2512 (M7.25) in Taurus, and TWA 30A (M5), TWA 30B (M4), 2MASS 1207-3932 (M8) and SSSPM 1102-3431 (M8) in the TWA.  All are optically revealed sources (i.e., without surrounding remnant envelopes) known to host accretion disks (from optical/UV accretion signatures and infrared dust emission), i.e., they are Class II classical T Tauri (cTT) analogs.  This choice allows us to focus on disk properties, without confusion from either envelope emission (Class 0/I sources) or the absence of a disk altogether (Class III). J044427+2512 had been detected earlier over 450\,$\mu$m--3.7\,mm \citep{2006ApJ...645.1498S, 2008A&A...486..877B}, and was selected to test our sensitivity; GM Tau had been observed before at 1.3\,mm and 2.6\,mm but not detected \citep{2009ApJ...701..698S}; and the rest had never been observed earlier in the sub-mm/mm.  Our four TWA sources, combined with the previously observed Hen 3-600A and TW Hya (see below), comprised all known VLMS/BD accretors, and 6 out of the 7 known accretors of any stellar mass, in this Association at the time of observation.  The one confirmed TWA accretor excluded here is TWA 5A (Mohanty et al.\,2003), which is a roughly solar-type cTT multiple not yet observed in the sub-mm/mm\footnote{The TWA 5A system (see Torres et al.\,(2003) and references therein for component details) also has a BD companion TWA 5B; the latter, however, does not have measurable accretion (Mohanty et al.\,2003).}. Very recently, Schneider et al.\,(2012) have announced two new VLMS TWA members, TWA 33 and 34.  Their relatively weak H$\alpha$ emission suggests very little accretion (applying the H$\alpha$ equivalent width criterion devised by Barrado y Navascues \& Martin); nevertheless, ther mid-infrared excesses in $WISE$ bands indicate surviving primordial disks (Schneider et al.\,2012).  These two sources have not yet been observed in the sub-mm/mm, and were announced too late for inclusion in our survey. 

To our sample observed with SCUBA-2, we add most of the Taurus, $\rho$ Oph and TWA Class II and/or cTT analogs (i.e., optically-revealed accretors) observed in the sub-mm/mm (850\,$\mu$m and/or 1.3\,mm) so far.  The Taurus and $\rho$ Oph data are taken from the surveys by AW05, AW07, Scholz et al.\,(2006) and Schaefer et al.\,(2009).  From the first two, we select only those classified as Class II (based on power-law fits over 2--60 or 2--25\,$\mu$m; see AW05, AW07).  From Scholz et al.\,(2006), we only choose the nine objects classified as cTTs by \citet{2005ApJ...626..498M}, \citet{2005ApJ...625..906M} or \citet{2004ApJ...617.1216L} based on optical/infrared diagnostics: CFHT-BD Tau 4, J041411+2811, J043814+2611, J043903+2544, J044148+2534, J044427+2512, KPNO Tau 6, KPNO Tau 7 and KPNO Tau 12. These are all in Taurus, and many of them are also known to harbor disks from {\it Spitzer} mid-infrared data (Luhman et al.\,2010).  There is a potential danger that such a selection criterion might discard very weak accretors which nonetheless still harbor disks (which describes a number of Class II sources).  In practice, however, we find that this criterion includes {\it all} the sources Scholz et al.\,detect at 1.3\,mm, as well as a few which they do not.  In other words, the sources we have dropped present no evidence {\it at all} of disks, from optical to mm wavelengths, and there is no rationale for including them at this point: while a few may very well harbor disks that are currently completely undetected, the same may be said of objects classified as Class III by AW05 and AW07, which we have also ignored.  

From the Taurus survey by Schaefer et al.\,(2009), we include all sources with spectral type M4 or later (i.e., VLMS/BDs), that have have been classified as cTTs by \citet{2005ApJ...626..498M} or \citet{2003ApJ...592..266M}, and/or as Class II by Luhman et al.\,(2010; based on mid-infrared {\it Spitzer} data out to 24\,$\mu$m). This yields seven objects: CIDA 1, CIDA 14, FN Tau, FP Tau, GM Tau, MHO 5 and V410 Anon 13 (an eighth object fitting these criteria, CIDA 12, has been previously observed by AW05, and we use their value for consistency). We ignore the remaining thirteen stars (all earlier than M4) from Schaefer et al's survey, for the following reasons. Seven of these have also been observed by AW05, of which six are classified by them as Class II (GO Tau, DN Tau, IQ Tau, CIDA 7, CIDA 8 and CIDA 11); these are already included in our study (with AW05's values). Among the remaining six stars not observed by AW05, various issues arise in some (e.g., large spectral-type uncertainty, or not classified as cTTs or Class II), so we conservatively choose to discard all six. Given the large sample at these spectral types that our study already includes, from AW05 and AW07, this decision has no discernible impact on our results.           

Finally, we include the sub-mm/mm fluxes for the TWA cTTs Hen 3-600A and TW Hya from Zuckerman (2001) and Weintraub et al.\,(1989) respectively.

Our combined sample consists of 134 objects -- 48 in $\rho$\,Oph, 80 in Taurus and 6 in the TWA -- spanning masses of $\sim$15\,$M_J$ to 4\,$\msun$.  For uniformity, we recalculate the stellar and disk parameters for all our sources taken from the literature, based on the spectral types and disk fluxes cited by the authors and using the methods discussed in \S\S3 and 5.  The full final sample is listed in Table I, with our SCUBA-2 sub-sample marked with asterisks.  

\subsection{SCUBA-2 Observations and Fluxes}
Exhaustive descriptions of the SCUBA-2 instrument, its calibration and performance, and its pipeline reduction procedure are supplied by \citet{2013MNRAS.430.2513H}, \citet{2013MNRAS.430.2534D} and \citet{2013MNRAS.430.2545C} respectively; we refer the reader to these for details, and simply summarize our data acquisition and reduction procedures here.  Each of our sources was observed simultaneously at 450 and 850\,$\mu$m, with a FWHM beam-size at 850\,$\mu$m of 14$\arcsec$.  Integration times were $\sim$30--60 minutes per source, employing a constant speed `{\sc daisy}' scan pattern appropriate for point sources (Holland et al.\,2013).  The 450\,$\mu$m data were discarded due to weather-related noise issues, but the 850\,$\mu$m observations were suitable.  The data were reduced using the {\sc makemap} routine within the {\sc smurf} package in the SCUBA-2 pipeline {\sc orac-dr}, with a parameter file specifically designed for faint point sources; point-source detection and calibration was carried out using a Mexican hat-type `matched-filter' method, with Mars and Uranus as the primary absolute calibrators (Chapin et al.\,2013).  We note that the relative flux calibration uncertainties at 850\,$\mu$m with SCUBA-2 are better than 5\% (Dempsey et al.\,2013), significantly superior to the 10\% uncertainty at this wavelength in its predecessor SCUBA.  For our sources, which are all very faint, the errors are dominated by photon noise; the final 1$\sigma$ noise levels achieved range from $\sim$0.8 to 1.8\,mJy (Table I).  AW05 and AW07 mention the systematic calibration uncertainties for their sample ($\sim$10\% at 850\,$\mu$m and $\sim$20\% at 1.3\,mm), which dominate the errors for a number of their brighter sources, but do not include it in their calculations; for comparison to their results, we do not include these uncertainties either, but mention their effect at relevant junctures.  These do not affect our overall results and conclusions.  

Assuming optically thin isothermal dust at 20\,K with an opacity of $\kappa_{dust,[850]}$ $\approx$ 3.5\,cm$^2$\,g$^{-1}$ (standard cTT disk grain parameters used in the literature, e.g., \citet{2006ApJ...645.1498S}, AW05, AW07; see detailed discussion below in \S5), this translates to 3$\sigma$ detection limits on the dust mass of $\sim$0.15--0.20\,$M_{\Earth}$ in the TWA and $\sim$0.70--1.2\,$M_{\Earth}$ in Taurus (employing the distances supplied in Table I; for Taurus, we adopt a constant mean $d = 140$\,pc, while for the TWA sources, we use the individual values based on measured parallaxes).  This is 3--10$\times$ deeper than any previous survey of VLMS/BD disks.  For a standard gas-to-dust mass ratio of 100:1 (i.e., total opacity of $\kappa_{[850]} \approx$ 0.035\,cm$^2$\,g$^{-1}$), these numbers imply total disk mass detection limits of 0.05--0.06\,$M_J$ and 0.2--0.4\,$M_J$ in the TWA and Taurus respectively.  J044427+2512 was strongly detected at 9.85$\pm$0.76\,mJy, consistent with its previous 850\,$\mu$m detection at 10$\pm$1.5\,mJy \citep{2008A&A...486..877B}, while CFHT-BD Tau 12 was marginally detected at 4.06$\pm$1.32 (3.1$\sigma$).  The remaining 5 sources remained undetected at the 3$\sigma$ level.      

\section{Stellar Parameters}
We calculate stellar masses, radii and effective temperatures for all our sources using previously determined spectral types, a single assumed age for all sources in a given star-forming region or association, and the predictions of theoretical evolutionary tracks.  Spectral types are converted to effective temperatures using the conversion scheme in \citet{1995ApJS..101..117K} for types M0 and earlier, and the scheme devised by \citet{2003ApJ...593.1093L} for later types. Objects in Taurus and $\rho$ Oph are assumed to have an age of $\sim$1 Myr, and those in the TWA an age of $\sim$10 Myr.  The derived temperatures are then compared to solar-metallicity evolutionary model predictions for the assumed ages to infer stellar masses and radii.  We use {\it (a)} the models of \citet{2000A&A...358..593S} for stars with mass $>$\,1.4\,$\msun$; {\it (b)} the models of \citet{1998AA...337..403B} for stellar masses 0.08--1.4\,$\msun$: specifically, models with a mixing length to pressure scale height ratio of $\alpha_{mix} = 1.0$ for 0.08--$<$0.6\,$\msun$, and models with $\alpha_{mix} = 1.9$ (value required to fit the Sun) for 0.6--1.4\,$\msun$ \citep[see discussion in][]{2002A&A...382..563B}; and {\it (c)} the ``Dusty'' models of \citet{2000ApJ...542..464C} for masses $<$0.08\,$\msun$ (which are mid- to late-M types at ages of a few Myr, the spectral type regime where photospheric dust starts to become important).  

We divide our sample into 4 stellar mass bins: intermediate-mass stars ($>$1--4\,$\msun$), solar-type stars (0.3--1\,$\msun$), VLMS (0.075--$<$0.3\,$\msun$) and BDs ($\lesssim$0.075\,$\msun$).  Within each, we adopt the median stellar mass as a representative of its class: 2.5\,$\msun$ for intermediate-mass stars ($\Rightarrow$ $\rstar \approx 4\,\rsun$ and $\tstar \approx 5000$\,K at 1\,Myr); 0.75\,$\msun$ for solar-types (2\,$\rsun$, 4000\,K; the usual parameters adopted for solar-type cTTs in the literature); 0.2\,$\msun$ for VLMS (1.5\,$\rsun$, 3200\,K); and 0.05\,$\msun$ for BDs (0.55\,$\rsun$, 2850\,K).  These fiducial masses will serve to illustrate the trends in disk mass and grain growth implied by our disk models for each mass bin.  In Taurus, we have 11 intermediate-mass stars, 49 solar-types, 9 VLMS and 11 BDs; in $\rho$ Oph, we have 9, 32, 7 and 0 in the same bins; and in the TWA, we have 0, 1, 3 and 2 respectively.

\section{Disk Spectral Energy Distribution: Theory}
The theory of disk spectral energy distributions (SEDs) is outlined in Appendix A.  The main assumptions and resulting equations relevant to our analysis are as follow.

We assume that the disk surface density and temperature have power-law radial profiles\footnote{More precisely, truncated power-law profiles, since we impose a finite inner and outer radius for the disk, as noted further below.}:
$$ \Sigma(r) = \Sigma_0\left(\frac{r}{r_0}\right)^{-p}\,\,\,\,\,, \,\,\,\,\,T(r) = T_0\left(\frac{r}{r_0}\right)^{-q} \eqno(1) $$
where $\Sigma_0$ and $T_0$ are the values at the disk inner edge $r_0$.  We further assume that the opacity is a power law in frequency:
$$ \kappa_\nu = \kappa_f \,\left(\frac{\nu}{\nu_f}\right)^\beta \eqno (2) $$
where $\kappa_f$ is the opacity at some appropriate fiducial frequency $\nu_f$.  The spectral index of the emission is defined as 
$$ \alpha \equiv \frac{d({\rm ln}F_\nu)}{d({\rm ln}\,\nu)} \eqno(3) $$
where $F_{\nu}$ is the disk flux density at frequency $\nu$ measured by an observer.

In the Rayleigh-Jeans (RJ) limit, the flux density is given by a polynomial expression (B90), derived in Appendix A.  In this case, the spectral index reduces to
$$ \alpha \approx 2 + \frac{\beta}{1+\Delta} \eqno(4) $$
where $\Delta$ is the ratio of optically thick to optically thin emission (see Appendix A)\footnote{Note that some papers (e.g., B90) define $\alpha \equiv d({\rm ln}\,L_\nu)/d({\rm ln}\,\nu)$, where $L_\nu \propto \nu F_\nu$, which yields an extra additive factor of 1 in the expression for $\alpha$ compared to ours in equation (4).}.  For grains much larger than the wavelength observed, the opacity $\kappa_\nu$ is independent of the frequency (i.e., $\beta \sim 0$), yielding $\alpha \approx 2$ regardless of the disk optical thickness.  Conversely, optically thick emission ($\Delta \rightarrow \infty$) also implies $\alpha \approx 2$, independent of $\beta$ and hence grain size.  In the optically thin limit ($\Delta \rightarrow 0$), on the other hand, $\alpha \approx 2+\beta$.  

If the RJ approximation is {\it not} valid, then the general expression for the flux density is
$$F_\nu \approx \nu^3 \left(\frac{4\pi h}{c^2}\right) f_0 \left(\frac{{\rm cos}\,i}{D^2}\right) \left[ \frac{(2-p){\bar\tau}_\nu}{2} \,R_d^p \right] \left[\int_{r_1}^{R_d} \frac{r^{1-p}}{{\rm exp}[h\nu/kT(r)]-1}\,dr \right] (1+\Delta) \eqno(5)$$ 
for a disk with inner radius $r_0$, outer radius $R_d$, situated at a distance $D$ from the observer, and inclined at an angle $i$ relative to the line of sight (so that $i$=$90^{\circ}$ for an edge-on orientation).  $r_1$ is the radius at which the emission at $\nu$ changes from optically thick to thin, ${\bar\tau}_\nu$ is the average optical depth in the disk ($\equiv (\kappa_\nu M_d)/(\pi R_d^2\,{\rm cos}\,i)$, where $M_d$ is the disk mass), and $f_0 \sim 0.8$ is a correction factor (see Appendix A).  The product of all the terms outside the last parentheses is the optically thin contribution to the flux density; the ratio $\Delta$ of the optically thick to thin contributions is given by
$$ \Delta \equiv \left( \frac{2}{(2-p){\bar\tau}_\nu\, R_d^p} \right) \left[\frac{\int_{r_0}^{r_1} \frac{r}{{\rm exp}[h\nu/kT(r)]-1}\,dr}{\int_{r_1}^{R_d} \frac{r^{1-p}}{{\rm exp}[h\nu/kT(r)]-1}\,dr}\right] \eqno(6) $$
Equations (5) and (6) must be evaluated numerically, and the spectral index $\alpha$ computed directly from its definition, equation (3).  In particular, even if the opacity remains a power-law in frequency, $\alpha$ does not reduce to the simple form of equation (4), and will generally be smaller in the sub-mm/mm (because the spectrum is flatter) than the RJ asymptotic values of 2 and $2+\beta$ for optically thick and thin disks respectively.  We show in \S6 that the RJ limit is not ideal at 850\,$\mu$m and 1.3\,mm for our sample, and instead use the generalized forms of $F_\nu$ and $\alpha$ to estimate disk masses and grain growth in \S7, via the technique described below.

\section{Grain Growth and Disk Mass: Method of Analysis}
For sources observed at both 850\,$\mu$m and 1.3\,mm, we have an estimate of $F_\nu$ as well as $\alpha$.  Using the foregoing equations, we wish to investigate their disk masses and grain properties.  The latter, however, are specified by 3 unknown parameters -- $M_d$, $\kappa_f$ and $\beta$ -- while $F_\nu$ and $\alpha$ represent only two independent observables.  As such, we can only derive $\beta$ and the {\it product} $\kappa_f M_d$ from the observed SED \citep[e.g.,][R10a,b; note that only this product enters the expression for the observed flux, equation (5), through the quantity ${\bar\tau}_\nu$]{2004A&A...416..179N}.    


Even this, of course, still requires specifying the other 6 parameters that $F_\nu$ and $\alpha$ depend on: the disk inclination $i$, the disk inner and outer radii $r_0$ and $R_d$, the surface-density power-law index $p$, the temperature power-law index $q$, and the temperature at the disk inner edge $T_0$.  A rigorous determination of these entails spatially resolved photometry and spectroscopy from optical to mm wavelengths, which have not been obtained for the vast majority of sources.  Instead, we adopt fiducial parameters guided by theory and current observations, which suffices to estimate the broad trends in $\kappa_f M_d$ and $\beta$.  The detailed rationale behind our choices is presented in Appendix B; the final adopted values are
$$ i=60^\circ, \,\,r_0=5\rstar, \,\,R_d=100\,{\rm AU}, \,\,p=1, \,\,q=0.58, \,\,T_0=880(\tstar/4000\,{\rm K})\,{\rm K} \eqno(7) $$
Additionally, we stipulate that the disk temperature cannot fall below a minimum value $T_{min} \equiv 10$\,K: the temperature declines with radius as a power-law until this value is reached, and remains at a constant 10\,K at larger radii.  This is justified in Appendix A, on the basis of heating by the interstellar radiation field (ISRF).  Note that such a broken power-law does not do any violence to our generalized equations (5) and (6), where the precise form of $T(r)$ is unspecified.  Finally, we also test the consequences of varying $R_d$ over the range 10--300\,AU and $p$ over the range 0.5--1.5, for different stellar mass bins.  With these six parameters specified, we determine $\kappa_f M_d$ and $\beta$ by comparing the observed $\alpha$ and $F_\nu$ to the predictions of the generalized equations in \S4, as follows.  

With only two observed wavelengths, 850\,$\mu$m and 1.3\,mm, the spectral index defined by equation (3) reduces to 
$$\alpha \equiv \frac{{\rm ln}\,F_{[850]} - {\rm ln}\,F_{[1300]}}{{\rm ln}\,1300 - {\rm ln}\,850} \eqno(8) $$  
We use this expression to calculate $\alpha$ for both the data and the SED models they are compared to.  Above, and henceforth, the subscript [$\lambda$] denotes a quantity evaluated at a frequency $\nu = c/\lambda$.  For sources detected at 1.3\,mm but not at 850\,$\mu$m, we compute the 3$\sigma$ {\it upper} limits on $\alpha$ by replacing $F_{[850]}$ above by its 3$\sigma$ upper limit.  Similarly, for sources detected at 850\,$\mu$m but not at 1.3\,mm, we find 3$\sigma$ {\it lower} limits on $\alpha$ by replacing $F_{[1300]}$ by its 3$\sigma$ upper limit.  Sources with only upper limits at both 850\,$\mu$m and 1.3\,mm, or those not observed at all at one or the other wavelength, are excluded from this analysis\footnote{AW05 and AW07 -- the source of most of our flux data -- do not cite lower limits on $\alpha$ for sources detected at 850\,$\mu$m but not at 1.3\,mm, because they either compute $\alpha$ using shorter wavelength (350 and/or 450\,$\mu$m) data in these cases, or exclude these disks altogether (when shorter wavelength data are absent).  We however restrict ourselves to 850\,$\mu$m and 1.3\,mm, in order to remain in as optically thin a regime as possible.}.

Next, following common practice, we choose to normalize our opacity power-law, equation (2), at a fiducial frequency $\nu_f \equiv 2.3\times 10^{11}$\,Hz, corresponding to $\lambda_f = 1300\,\mu$m (1.3\,mm).  We denote $\kappa_f$, the (a priori unknown) opacity at this frequency, explicitly as $\kappa_{[1300]}$.  We further denote the {\it commonly adopted value} of this fiducial opacity by $\tilde{\kappa}_{[1300]} \equiv 0.023$\,cm$^2$g$^{-1}$ (e.g., B90).  Note that the latter value assumes an ISM-like gas-to-dust mass ratio of 100:1.    

Finally, we define the {\it apparent disk mass} $M_{d,\nu}$ as the mass derived from the observed flux at some frequency $\nu$, {\it assuming} that the emission {\it (a)} is optically thin, {\it (b)} arises from an isothermal region of the disk with known temperature $\tilde{T}$, and {\it (c)} is due to material with a {\it known} opacity $\tilde{\kappa}_\nu$.  Thus:
$$ M_{d,\nu} \equiv \frac{F_{\nu} D^2}{\tilde{\kappa}_{\nu} B_{\nu}(\tilde{T})} \eqno(9) $$
We adopt $\tilde{T} = 20$K and $\tilde\kappa_\nu = \tilde\kappa_{[1300]}$\,$(\nu/2.3\times10^{11}{\rm Hz})^{\beta}$, with $\tilde\kappa_{[1300]}$ as defined above and $\beta = 1$.  We will write $M_{d,\nu}$ explicitly as $M_{d,[850]}$ or $M_{d,[1300]}$ when $F_\nu$ is the flux density at 850\,$\mu$m or 1.3\,mm.  With $\tilde T$ and $\tilde\kappa_\nu$ fixed, {\it $M_{d,\nu}$ is simply a proxy for the specific disk luminosity $F_{\nu}D^2$}.  The advantage of using this formulation is that $M_{d,\nu}$ (evaluated at 850\,$\mu$m or 1.3\,mm, with $\tilde{T}$ and $\tilde\kappa_\nu$ similar or identical to our adopted values) is widely used as a simplistic estimate of the disk mass (e.g., AW05; AW07; Klein et al.\,2003; Scholz et al.\,2006).  This allows us to directly compare our results to literature values for disk masses.   

Turning now to the theoretical models, we define the {\it opacity-normalized disk mass} as 
$$M_d^\kappa \,\equiv\, \frac{\kappa_{[1300]} M_d}{\tilde\kappa_{[1300]}} \,=\, \left(\frac{\kappa_{[1300]}}{0.023\,{\rm cm}^2{\rm g}^{-1}}\right)M_d \eqno(10) $$  
$M_d^\kappa$ is merely a scaled proxy for the real variable $\kappa_{[1300]} M_d$, but is useful to employ in lieu of the latter as a more intuitive quantity (e.g., Natta et al.\,2004): specifically, $M_d^\kappa$ equals the real disk mass $M_d$ {\it if} the true opacity $\kappa_{[1300]}$ equals the fiducial value $\tilde{\kappa}_{[1300]}$.  {\it Without knowing the true value of the opacity, $M_d^\kappa$ is the closest we can get to the real disk mass}.

Our analysis now is straightforward.  We first show, in \S6, that the RJ approximation is not ideal at the disk temperatures expected in our sample.  We thus use the generalized equations (5), (6) and (3) (with the latter approximated by equation (8)).  From these, we calculate the $F_{\nu}$, $\Delta$ and $\alpha$ predicted by our six fixed disk parameters and a physically plausible range of $M_d^\kappa$  and $\beta$; we further convert the predicted $F_\nu$ to a predicted $M_{d,\nu}$ using equation (9).  Comparing these theoretical $M_{d,\nu}$ and $\alpha$ values to the values derived from the {\it observed} fluxes enables us to (i) gauge what fraction of the emission is optically thin, and (ii) estimate the true $M_d^\kappa$ and $\beta$ when the emission is predominantly optically thin, or put lower limits on $M_d^\kappa$ (and no constraints on $\beta$) when it is optically thick.    

Basically, all we are doing is calculating (within the context of our fiducial disk parameters) what the emitted flux and $\alpha$ {\it should be} for any specified $M_d^{\kappa}$ and $\beta$, and comparing these to the {\it observed} flux and $\alpha$ to infer what the $M_d^{\kappa}$ and $\beta$ for a source {\it really are}; i.e., a simple inversion.  The only potential confusion for the reader is that, instead of directly using flux in these comparisons, we use its scaled proxy $M_{d,\nu}$, a quantity which is often cited in the literature and thus useful to have at hand.  However, the various assumptions that go into calculating $M_{d,\nu}$ (fixed $T=20$\,K, fixed $\beta=1$) have no effect on our results: the predicted and observed fluxes are converted to predicted and observed $M_{d,\nu}$ using {\it exactly the same} multiplicative factors, so the end result is precisely the same as simply using flux instead.   


\section{Results I: Inaccuracy of the Rayleigh-Jeans Approximation}
Fig.\,1 shows the $\alpha$ predicted by the generalized equations for our fiducial disk model, equation (7), for the median stellar temperature and radius within each of our four stellar mass bins: intermediate-mass stars, solar-type stars, VLMS and BDs.  The abscissae of the plots show the corresponding minimum temperature in the disk (i.e., the temperature at the outer disk radius $R_d$ = 100\,AU); in VLMS and BDs, this value levels out at our fixed lower limit of 10\,K. Note that the plotted $\alpha$ refers to emission integrated over the {\it entire} disk, not just from $R_d$.   The opacity normalized disk mass, $M_d^\kappa$, is either fixed at a very large value, so that the entire disk is optically thick (i.e., the transition radius $r_1$ from equation (A4) formally satisfies $r_1 \gg R_d$; {\it top panel} of Fig.\,1), or fixed at a very small value, so that the entire disk is optically thin ($r_1 \ll r_0$; {\it bottom panel}).  $\beta$ is varied from 0 to 2.    

The top panel shows that, as expected, $\alpha$ in the optically thick limit is independent of the opacity index $\beta$.  However, the asymptotic RJ value in this case, $\alpha = 2$, is not achieved for any of our stars.  For the relatively hot disks around intermediate-mass stars, the deviation from RJ is quite small, $\sim$0.25. For the cooler disks around solar-type stars to BDs, on the other hand, the SED is markedly flatter, yielding an $\alpha$ significantly smaller -- by 0.5--0.8 -- than the RJ expectation.  Similarly, the bottom panel shows that while $\alpha$ does depend on $\beta$ in the optically thin limit, as expected, the asymptotic RJ value in this limit, $\alpha = 2 + \beta$, is not reached.  Again, the deviation is relatively small for intermediate-mass stars ($\sim$0.2), but appreciable in VLMS and BDs ($\sim$0.4--0.6).    

It is often assumed that the RJ limit applies at $\sim$mm wavelengths for disk temperatures $\gtrsim$15\,K.  Fig.\,1 shows that this is not very accurate, with a discernible $\alpha$ deviation of $\sim$0.2 even for intermediate-mass stars, where the minimum disk temperature is $\sim$20\,K (and most of the disk is considerably hotter still).  More importantly, disks significantly larger than 100\,AU around solar-type and intermediate mass stars, as are often observed, as well as even 100\,AU disks around VLMS/BDs (like those plotted in Fig.\,1), may have temperatures down to $\sim$10\,K in their outer regions, where most of the mass resides.  Fig.\,1 shows that the RJ approximation is severely strained under these circumstances\footnote{The blackbody function $B_{\nu}[T]$ peaks at $\nu \approx 6\times10^{11}$\,Hz for $T = 10$\,K, which is within a factor of 2--3 of the 3.5--2.3$\times10^{11}$\,Hz (i.e., 850--1300\,$\mu$m) range over which $\alpha$ is determined; $h\nu \ll kT$ is thus poorly satisfied, and it is unsurprising that the RJ approximation becomes inaccurate.}.  The danger in assuming RJ values is that a low observed $\alpha$ from an optically thin disk will lead one to infer a spuriously low $\beta$ (i.e., too much grain growth), and/or a spuriously high optically thick contribution.  Equally importantly, a number of disks appear to have $\alpha < 2$ (as we shall shortly see); these cannot be explained at all under the RJ assumption.    

Even in the general non-RJ case, however, we see that $\alpha$ remains very sensitive to $\beta$ for optically thin emission, with a roughly linear relationship between the two when all other disk parameters are fixed (i.e., analogous to the RJ case, except that now the precise value of $\alpha$ depends on the disk temperature as well, in both optically thin and thick limits).  Thus $\alpha$ from the general equations can still be used to probe grain growth, as we do next.  

\section{Results II: Grain Growth and Disk Mass}

\subsection{Spectral Slopes and Apparent Disk Masses from the Observed Fluxes}
We first discuss some general trends in the spectral slopes ($\alpha$) and apparent disk masses ($M_{d,\nu}$) that we derive from the observed fluxes, before comparing to model predictions.  

Fig.\,2 shows the $\alpha$ for our sample as a function of stellar mass, for sources observed at both 850\,$\mu$m and 1.3\,mm and detected at at least one of these two wavelengths.  Sources detected at both are shown as filled circles with 1$\sigma$ error bars\footnote{These errors refer to photon noise; systematic calibration uncertainties (\S2) yield in general an additional error of $\sim$0.5 in $\alpha$.  For sources plotted with errors $\lesssim$0.5, therefore, the total 1$\sigma$ error including systematics rises to $\sim$0.5--0.7; for sources plotted with larger errors, including the systematics has negligible effect.}, while 3$\sigma$ upper and lower limits are plotted as downward and upward triangles respectively.  Note that the plotted sources are mainly at $\mstar \gtrsim 0.5\,\msun$, since most lower mass stars and BDs have either not been observed at both wavelengths or are undetected at both (see Table I).  The lower limits are also concentrated at the lower end of this stellar mass range, where the disks are still bright enough to be detected at 850\,$\mu$m but too faint for 1.3\,mm; this reflects the empirical fact that disks become fainter with diminishing $\mstar$.

For the sub-sample detected at both wavelengths, we find a mean spectral index of $\langle\alpha\rangle = 1.98 \pm 0.06$ ($1\sigma$ error; mean and associated 3$\sigma$ error plotted in Fig.\,2), in agreement with the average $\alpha$ derived by AW05\footnote{Including the systematic uncertainties in calibration modestly increases our 1$\sigma$ error on $\langle\alpha\rangle$ from 0.06 to $\sim$0.08 (AW05's cited error would increase similarly), and does not change any of our conclusions.}.  While our mean is driven predominantly by stars $\gtrsim 0.5\,\msun$, the 4 VLMS/BDs with measured $\alpha$ are consistent with this value.  Also, while we have ignored upper and lower limits in this calculation, Fig.\,2 shows that all but two of the lower limits and two out of the four upper limits are compatible, within the errors, with this mean; including the four deviant points negligibly affects the outcome.  

In the {\it upper panels} of Fig.\,3, we plot $M_{d,[850]}$ and $M_{d,[1300]}$ for our sample (derived via equation (9), using the distances $D$ discussed in \S3), as a function of stellar mass.  Going from BDs to solar-type stars, there is a clear trend of $M_{d,\nu}$ increasing on average with higher $\mstar$; we return to this in \S8.  For now, note that this simply reflects an increase in the average emitted specific disk luminosity ($F_{\nu}D^2$) with stellar mass.  Conversely, there is a decline in the upper envelope of $M_{d,{\nu}}$ from the solar-type to intermediate mass stars, going approximately as $[M_{d,\nu}]_{max} \propto \mstar^{-1/2}$ (shown by a dotted line in Fig.\,3).  While this relationship is defined by only a handful of stars, we can at least state that there is no evidence of $[M_{d,\nu}]_{max}$ {\it increasing} with $\mstar$ for these stars; we discuss this in \S7.3.  

In the {\it lower panels} of Fig.\,3, we plot the corresponding $M_{d,[850]}/M_{\ast}$ and $M_{d,[1300]}/M_{\ast}$ as a function of stellar mass.  Excluding the four upper limits among the VLMS/BDs in TWA (green downward triangles), the mean $M_{d,\nu}/M_{\ast}$ appears roughly constant from BDs to solar-types, reflecting the increase in $M_{d,\nu}$ with $\mstar$ noted above.  Conversely, the decline in the upper envelope from solar-types to intermediate-mass stars is now more pronounced, as expected: $[M_{d,\nu}]_{max} \propto M_{\ast}^{-1/2}$ in the upper panel translates to a steeper slope $[M_{d,\nu}/M_{\ast}]_{max} \propto M_{\ast}^{-3/2}$ in the lower one.  The limiting mass ratio above which gravitational instabilities set in, $M_d/M_{\ast} \sim 0.1$ \citep[e.g.,][]{2005MNRAS.364L..91L}, is also marked for later reference.     

Finally, since the $M_{d,\nu}$ results at 1.3\,mm are nearly identical to those at 850\,$\mu$m, we focus henceforth on the latter for clarity.  To estimate the true grain growth and opacity-normalized disk masses, we must now compare these $M_{d,[850]}$ and $\alpha$ to the model predictions.         

\subsection{Model Predictions}
Our model calculations are carried out for four fiducial $\mstar$ (with corresponding $\rstar$ and $\tstar$ at 1 Myr), representing median values within the four stellar mass bins in our sample (see \S3): $2.5\,\msun$ (intermediate-mass stars), $0.75\,\msun$ (solar-types), $0.2\,\msun$ (VLMS) and $0.05\,\msun$ (BDs). For each of these $\mstar$, models are constructed for $\beta$ = 0--2 (lowest possible to standard ISM) in steps of 0.1, and $M_d^{\kappa}$ = $10^{-6}$--$1$\,$\msun$ in steps of 1 dex (sufficient to capture the observed range in $M_{d,[850]}$, as illustrated shortly).  Furthermore, {\it if} $M_d^{\kappa}$ represents the true disk mass $M_d$ (i.e., if the true opacity $\kappa_{[1300]}$ equals the fiducial value 0.023\,cm$^2$g$^{-1}$), then the disk would nominally become gravitationally unstable above $M_{d,GI}^{\kappa}\equiv 0.1\,\mstar$.  We explicitly include this value in our models for each stellar bin, for reasons explained further below.  We examine a range of $R_d$ over 10--300\,AU, and also vary the surface density exponent $p$ over 0.5--1.5 (instead of fixing it at 1) in some cases.  The rest of the parameters are as specified in equation (7).  Inserting these into equations (5), (6) and (8), we compute the theoretical $F_\nu$, $\Delta$ and $\alpha$, and further convert the predicted $F_{[850]}$ to a predicted $M_{d,[850]}$ with equation (9).  

The model predictions are plotted in Figs.\,4--7.  Each figure is for a fixed $\mstar$; the top and bottom sets of plots in each figure correspond to two illustrative values of $R_d$.  The four panels in each plot show (from left to right, top to bottom): 

\noindent {\it (1)} The predicted $F_{[850]}$ (arbitrarily scaled to the distance of Taurus for illustration) versus $\beta$, for the range of $\beta$ and $M_d^\kappa$ examined.  For optically thin emission, the flux density increases with both $M_d^\kappa$ and $\beta$.  As the optically thick contribution rises with $M_d^\kappa$, $F_{[850]}$ becomes less sensitive to both parameters, until finally, in the thick limit, the flux density is independent of $\beta$ and $M_d^\kappa$, and saturates at a level set by the disk temperature and radial extent.  

\noindent {\it (2)} The predicted $\alpha$ versus $M_{d,[850]}$ (calculated from the predicted $F_{[850]}$; independent of distance), for our grid of $\beta$ and $M_d^\kappa$.  As discussed in \S6, $\alpha$ increases roughly linearly with $\beta$ for optically thin disks, and saturates to a fixed value when the emission becomes optically thick.  Unlike in the RJ limit, however, the value of $\alpha$ in both cases depends on the disk temperature (and thus on $\tstar$, or equivalently $\mstar$, as well as on $R_d$).    

Overplotted in this panel for comparison are the $\alpha$ and $M_{d,[850]}$ derived from our {\it observed} fluxes, for sources in the relevant stellar mass bin (objects with upper/lower limits in $\alpha$ are excluded for clarity).  The horizontal dashed line marks the mean index for the full sample, $\langle\alpha\rangle \approx 2$.  The vertical dashed line denotes the maximum observed $M_{d,[850]}$ (equivalently, maximum observed 850\,$\mu$m flux) in our data for the relevant stellar mass bin.  This observed upper limit sets a lower boundary on the size $R_d$ of the brightest disks, as follows.  For a given surface density and temperature profile, a disk of a specified size has a maximum allowed flux, which is its optically thick limit, and this maximal value increases with disk size.  Thus a minimum disk size is required to explain the maximum observed flux (or equivalently, maximum observed $M_{d,[850]}$)\footnote{Disks that are fainter than the observed upper limit in flux may of course be smaller.}. This minimum $R_d$ changes as a function of the adopted density and temperature profiles.    

An additional constraint may be set by the disk mass.  The optically thick limit for a disk of a given size is reached above a threshold disk mass, and this threshold value increases with disk size.  Thus the minimum $R_d$ derived above also corresponds to a minimum disk mass.  However, one expects the gravitational instability limit, $M_d \sim 0.1\,\mstar$, to set a reasonable upper limit to the allowed disk mass.  Thus the minimum mass corresponding to the derived lower bound on $R_d$ must not greatly exceed the instability limit.  If it does, then some other parameter must be changed (e.g., $p$ or $q$) to resolve the discrepancy.  

This analysis is complicated by the fact that we do not know the real opacity $\kappa_{[1300]}$, so we cannot associate our disk models with a true mass $M_d$, but only with its opacity-normalized counterpart $M_d^{\kappa}$.  Nevertheless, the mass constraint described above can still be implemented by setting plausible limits on how far the true opacity may deviate from our fiducial value $\tilde{\kappa}_{[1300]} = 0.023$\,cm$^2$g$^{-1}$.  Using detailed self-consistent dust models, R10b show that the fiducial opacity may be an underestimation by up to a factor of 10 for $\beta$$\sim$2, and an overestimation by the same factor for $\beta$$\sim$0.  If the true opacity equalled the fiducial one, then $M_{d,GI}^{\kappa}$ ($\equiv 0.1\,\mstar$, as defined earlier) would represent a disk that was actually at the unstable limit; the R10b analysis suggests that in fact, the instability limit may lie anywhere in the range 0.1--10\,$M_{d,GI}^{\kappa}$.  As such, in determining the minimum disk radius, we must verify that the corresponding minimum $M_d^{\kappa}$ does not exceed at least the upper limit 10\,$M_{d,GI}^{\kappa}$.      

\noindent {\it (3)} Predicted $\Delta$ versus $M_{d,[850]}$, explicitly showing the relative contributions of optically thin and thick emission for each $M_d^\kappa$ and $\beta$.  The emission flux, and thus the corresponding $M_{d,[850]}$, saturates to a fixed value in the optically thick limit.  

\noindent {\it (4)}  Predicted ratio of $M_{d,[850]}$ to $M_d^\kappa$, versus $M_{d,[850]}$.  For hot stars with optically thin disks, the assumption of isothermal emission at $\tilde{T} = 20$\,K in deriving $M_{d,[850]}$ can cause the latter to exceed the true value $M_d^\kappa$, since a good fraction of the emission is from radii hotter than 20\,K for these sources.  For cooler stars/BDs in the optically thin limit, the same assumption can make $M_{d,[850]}$ an underestimation of $M_d^\kappa$, due to emission from radii cooler than 20\,K.  In optically thick disks, $M_{d,[850]}$ saturates and always represents a lower limit on the true $M_d^\kappa$.  

The dashed line in this panel plots the [$M_d^\kappa$, $\beta$] locus, derived from panel {\it (2)} above, corresponding to the mean spectral index of the sample $\langle\alpha\rangle \approx 2$.  This reveals the average trend in $M_{d,[850]}/M_d^\kappa$ among our sources in each stellar mass bin.       

\subsection{Implications of Model Comparisons to Data}

\subsubsection{Implications for Full Sample}
Using the techniques described above, and within the context of our fiducial disk model, we can in principle derive the opacity index $\beta$ and the opacity-normalized disk mass $M_{d}^{\kappa}$ (or only upper limits on the latter, if the disk is optically thick) for every source observed at both 850\,$\mu$m and 1.3\,mm.  However, given the uncertainties in each of the six parameters in our fiducial disk model (most of which have not been measured for the majority of our sources), this is not a very useful exercise.  The interested reader can of course simply read off the $\beta$ and $M_{d}^{\kappa}$ implied by our fiducial model for any source, by comparing the observed $M_{d,[850]}$ and $\alpha$ listed in Table I to the predicted values plotted in the top right panels of Figs. 4--7.  We concentrate here instead on determining the broad trends in disk and dust properties implied by our sample as a whole; these are likely to be more valid than the results for any particular source.  We find that the data--model comparisons in Figs.\,4--7 imply the following:

\noindent {\it (i) Minimum disk radius:}

\noindent In point {\it (2)} of \S7.2, we discussed how the maximum observed $M_{d,[850]}$ sets a lower bound on $R_d$ for any given surface density and temperature profile.  We assume below (unless noted otherwise) that the temperature profile is fixed (given by equation (7)), and find the minimum $R_d$ for the fiducial density exponent $p$=1 as well as for the limits $p$=0.5 and 1.5.  In all cases, we confirm that the minimum $M_d^{\kappa}$ associated with this $R_{d,min}$ is comfortably below the instability upper limit 10\,$M_{d,GI}^{\kappa}$, as required (see \S7.2); the one exception is noted explicitly and investigated.  We emphasize that these $R_{d,min}$ apply only to the brightest disks, and are moreover not necessarily the actual size of these disks, but only a lower limit assuming optically thick conditions.  Fainter disks may certainly be smaller; conversely, the brightest ones may be much larger than $R_{d,min}$ if they are in fact optically thin. 

\noindent $\bullet$ For intermediate-mass stars, assuming $p$=1, the maximum observed $M_{d,[850]}$ corresponds to the optically thick limit for disks with $R_d = 100$\,AU ({Fig.\,4, \it top plot}).  Thus the minimum disk size for $p$=1 is $R_{d,min} \sim 100$\,AU.  Changing $p$ to 0.5 or 1.5 alters this only by $\sim$$\pm$10\,AU.  

\noindent $\bullet$ For solar-type stars, adopting $p$=1, the optically thick flux from a 100\,AU model disk is only half the amount required to explain the four brightest objects, though the large number of next brightest sources are fully consistent with this radius (Fig.\,5, {\it top plot}).  To explain the four brightest, a disk size of at least 300\,AU is required for $p=1$ ({\it bottom plot}); worryingly in this case, the corresponding minimum disk mass is only marginally consistent with the instability upper limit of 10$M_{d,GI}^{\kappa}$.  It is thus worth investigating whether these sources are somehow anomalous. It turns out that at least three of them are: AS 205, EL 24, GG Tau A.  First, while AW05 and AW07 derive a median value of $T_1\sim150$\,K for the temperature normalization at 1\,AU based on a large sample of mainly solar-type stars -- a value we adopt (see Appendix B) -- AW07 derive a much higher $T_1 = 304$\,K for AS 205 and 229\,K for EL 24.  With these normalizations, these two stars become consistent with 100\,AU disks as well (not shown); indeed, AS 205 can even accomodate a 50\,AU disk (in line with the disk size expected given the presence of a companion $\sim$1.3$\arcsec$ away).  $R_d \sim 50$ and 100\,AU for AS 205 and EL 24 respectively is also in agreement with the spatially resolved data presented by Andrews et al. (2009, 2010).  Second, GG Tau A is a close binary, and the circumbinary material girdling it has a complicated ring+disk structure that is very poorly represented by our disk model here (Guilloteau et al.\,1999; Harris et al.\,2012).  

With these three sources removed / explained, only the fourth disk, around 04113+2758, remains enigmatic.  Without any further explicit information about its properties, we cannot comment on why it appears so (anomalously) bright.  However, we can now confidently state that, bar this one source, the brightest disks among solar-type stars are indeed consistent with $R_{d,min} \sim 100$\,AU, for $p$=1.  Varying $p$ between 0.5 and 1.5 changes this by $\sim$$\pm$10\,AU.  

\noindent $\bullet$ For VLMS, assuming $p$=1, the maximum observed $M_{d,[850]}$ corresponds to optically thick flux from disks with $R_d \approx 100$\,AU (Fig.\,6, {\it top plot}).  Thus $R_{d,min} \sim 100$\,AU for $p$=1; this changes by $\sim$$\pm$10\,AU for $p$ varying from 0.5 to 1.5.    

\noindent $\bullet$ For BDs, the same analysis implies $R_{d,min} \sim 20$\,AU for $p$=1 (Fig.\,7, {\it top plot}); this changes negligibly (by a few AU) for $p$ ranging from 0.5 to 1.5.  This is in good agreement with the size estimates for the two BD disks marginally resolved so far (both in Taurus): $\sim$20--40\,AU for 2MASS J0428+2611 \citep[from optical scattered-light imaging of the nearly edge-on disk]{2007ApJ...666.1219L}, and $\sim$15--30\,AU (and at least $>$10\,AU for all $p$ = 0--1.5) for J044427+2512 \citep[from high-angular resolution 1.3\,mm continuum imaging; this is also one of the two objects used by us to derive $R_{d,min}$ in Fig.\,7; see also {\it (ii)} below]{2013ApJ...764L..27R}.      

\noindent {\it (ii) Evidence for grain growth:}

\noindent $\bullet$ Among the solar-type and intermediate-mass stars, the individual measured $\alpha$ (as well as the mean value $\langle \alpha\rangle \approx 2$) correspond to $\beta \sim$ 0--1 for a large number of sources, {\it if} they are optically thin.  While a fraction of these sources may be optically thick instead, this cannot be true of all of them: mainly because some have already been spatially resolved into sizes too large to be optically thick (see detailed discussion in R10a,b). Thus $\beta$ is in fact likely to be low in a significant number of sources with $\alpha \lesssim 2$, which in turn implies substantial grain growth in these disks (as R10a,b show, all realistic grain models require a maximum grain size of $\gtrsim 1$\,mm to explain $\beta \lesssim 1$).  

\noindent $\bullet$ Among the Taurus and $\rho$ Oph VLMS/BDs, the evidence for grain growth is not so clear (TWA sources are discussed in \S7.3.2).  Only 4 of these objects have measured $\alpha$.  Of these, Fig.\,6 shows that the $\alpha \sim1.5$ observed in the two VLMS  -- SR 13 and WSB 60, both in $\rho$\,Oph -- corresponds to $\beta \sim 0$, i.e., large grains, if the disks are optically thin. Fig.\,6 also shows, though, that optically thin conditions are obtained only if these disks are significantly larger than 100\,AU; for $R_d \sim 100$\,AU, they become optically thick. With hardly any constraints so far on VLMS disk sizes via resolved observations, we cannot rule out optically thick disks mimicking grain growth in these two sources.  

Among the two BDs, one (CFHT-BD Tau 4, in Taurus) has a large $\alpha \sim 3.5$, consistent with $\beta \sim 2$, i.e., no substantial grain growth beyond ISM sizes (Fig.\,7). In the other (J044427+2512, also in Taurus), the small $\alpha \lesssim 1$ corresponds to $\beta \sim 0$, and thus considerable grain growth, if optically thin.  Fig.\,7 moreover implies optically thin conditions apply only if the disk is significantly larger than 20\,AU; for $R_d \sim 20$\,AU, the disk becomes optically thick. Very recently, Ricci et al.\,(2013) have resolved this disk in 1.3\,mm continuum emission, and estimate a (somewhat model-dependent) radius of $\sim$15--30\,AU, suggesting (in the context of our modeling) either optically thick conditions or large grains. Based on their own modelling, Ricci et al.\,(2013) argue in favour of grain growth: they find the disk would be optically thin for $R_d > 10$\,AU\footnote{Somewhat smaller than our 20\,AU limit, because they include the longer wavelength 3.7\,mm data from Bouy et al.\,(2008), and also because their integrated 1.3\,mm flux -- 5.2$\pm$0.3\,mJy -- is a bit lower than the 7.6$\pm$0.9\,mJy we use, from Scholz et al.\,(2006).}, and the size of their resolved disk is indeed $>$10\,AU for all plausible surface-density profiles ($p$ = 0--1.5). The balance of evidence therefore suggests large grains in this source; overall, the conclusion is that more spatially resolved data is sorely needed to verify disk sizes and hence degree of grain growth in the VLMS/BD regime.   


\noindent {\it (iii) Estimates of the Opacity Normalized Disk Mass $M_d^\kappa$:} 

\noindent $\bullet$ For intermediate-mass stars, the observed $\langle \alpha \rangle \approx 2$ corresponds to $M_{d,[850]}/M_d^\kappa \sim$ 2--0.8 for optically thin disks with $R_d$ = 100--300\,AU (Fig.\,4).  In other words, the apparent disk mass $M_{d,[850]}$ reasonably approximates the true opacity-normalized mass $M_d^\kappa$, and may be an overestimation by a factor of 2 for the lower end of disk sizes, $R_d \sim$ 100\,AU.  

\noindent $\bullet$ For solar-type stars with the same mean $\alpha$ (Fig.\,5), $M_{d,[850]}$ may underestimate the true $M_d^\kappa$ by a factor of $\sim$2--3 for disks extending out to 100--300\,AU. 

\noindent $\bullet$ For VLMS and BDs, the current paucity of sources with measured $\alpha$ precludes the derivation of any reliable average spectral slope for the full sample of these objects.  Nevertheless, the low value of $\alpha$ in three out of the four individual sources, combined with their derived $M_{d,[850]}$, implies that $M_{d,[850]}$ underestimates the true $M_d^{\kappa}$ of these three sources by a factor of $\sim$3--5, depending on disk size (i.e., whether optically thin or thick).

\noindent {\it (iv) Decline in the maximum apparent disk mass among intermediate-mass stars:}

As noted in \S7.1 (see Fig.\,3), the upper envelope of apparent disk masses (both $M_{d,[850]}$ and $M_{d,[1300]}$) inferred from the observed fluxes rises from BDs to solar-type stars, but, puzzingly, falls off (or at least plateaus) from solar-types to intermediate-mass stars.  The increase from BDs to solar-types is consistent with the theoretical expectation that more massive objects should form out of larger cores, and thus harbor larger and more massive disks.  Why should this trend appear to reverse (or level off) upon moving to intermediate-mass stars?  It cannot be due to changes in disk temperature: while we have assigned a constant $\tilde{T} = 20$\,K to all the disks in deriving our naive $M_{d,\nu}$ estimates, the disks around more massive stars should be hotter, and correcting for this exacerbates the decline in the upper envelope of disk masses from solar-types to intermediate-mass stars, instead of fixing the problem.  We offer one explanation, and examine two others which seem less likely.    

\noindent $\bullet$  Photoevaporation may actively decrease the disk mass around intermediate-mass stars.  Specifically, \citet{2009ApJ...705.1237G} model FUV/EUV/X-ray-driven photoevaporation, complemented by viscous spreading of the disk.  They find disk lifetimes of a few Myr, fairly independent of stellar mass, for $\mstar \lesssim 3\,\msun$; for $\mstar \gtrsim 3\,\msun$, however, the inferred lifetimes decrease strongly with rising $\mstar$ due to the increasing FUV/X-ray photoevaporative flux, falling to a few $\times$ 10$^5$\,Myr by 10\,$\msun$.  This is roughly consistent with the observations in Fig.\,3, which indicate an apparent depletion in disk mass by an age of $\sim$1 Myr for stars $>$1\,$\msun$, and significant depletion for stars $\gtrsim$3\,$\msun$.  If photoevaporation is indeed the culprit, then the data suggest that it becomes important at slightly lower masses than Gorti et al.\,find; given the large uncertainties in current predictions of photoevaporation rates (see discussion by Gorti et al.\,(2009) and \citet{2009ApJ...699.1639E}), this remains a possibility.

\noindent $\bullet$ There are two other hypotheses that we consider unlikely.  The first is that grain growth to sizes much larger than a few millimetres reduces the sub-mm/mm flux, yielding spuriously low $M_{d,\nu}$ estimates for the intermediate-mass stars.  In this case, one expects the entire grain size distribution in these disks to be skewed to larger values, compared to disks around solar-type stars (which do not show a comparable depression in $M_{d,\nu}$).  This would lead to a flatter sub-mm/mm spectral slope, i.e., smaller $\alpha$, for intermediate-mass stars relative to solar-types.  However, the data do not show any significant difference in $\alpha$ between the two stellar populations (see Figs.\,2, 4 and 5), making this scenario improbable. (This is not to say that grains have not grown large around both solar-type and intermediate-mass stars -- they almost certainly have in many cases, as discussed earlier; just that they have not grown {\it preferentially larger} around the latter stars).      

\noindent $\bullet$ The other possibility is that accretion onto the central star has depleted the disks around intermediate-mass stars more than around solar-types.  Assuming a standard $\alpha$ viscous accretion disk, with $T(r) \propto r^{-1/2}$ (approximately congruent with our adopted $q = 0.58$), the viscosity goes as $\nu \propto r$ (Hartmann et al.\,1998).  In this case, the disk mass at any time $t$ is given by (Hartmann et al.\,1998, 2006):
$$ M_d(t) = \frac{M_d(0)}{(1 + t/t_v)^{1/2}} \,\,\,\,\, \Rightarrow \,\,\,\,\, M_d(t) \approx M_d(0) \left(\frac{t_v}{t}\right)^{1/2} \,\,\, {\rm for}\,\, t \gg t_v \eqno(11) $$
where $M_d(0)$ is the initial disk mass (at $t = 0$), $t_v$ is the viscous timescale, and the second equality holds for evolution over a time $t$ much longer than $t_v$.  In the same limit, the instantaneous accretion rate is found by differentiating the above with respect to time:
$$ \mdot(t) \propto M_d(0) \left(\frac{t_v}{t^3}\right)^{1/2} \,\,\, {\rm for}\,\, t \gg t_v \eqno(12) $$
If we make the usual assumption that $M_d(0)$ either remains constant or increases with $\mstar$, then for roughly coeval sources (constant $t$), the observed falloff in the estimated disk mass, $M_{d,\nu}(t) \propto \mstar^{-1/2}$, implies that the viscous timescale should decrease at least as rapidly as $t_v \propto \mstar^{-1}$.  We cannot judge the plausibility of this per se, without detailed information about how the initial disk radius and viscosity change with stellar mass (see Hartmann et al.\,2006).  However, note that the instantaneous accretion rate has the same dependence on $t_v$ and $M_d(0)$ as the instantaneous disk mass.  Thus, $\mdot(t)$, the accretion rate observed at the {\it current} time, should decline in the same way with increasing stellar mass: $\mdot(t) \propto \mstar^{-1/2}$.  There is no observational evidence of this; if anything, the observed accretion rate {\it increases} (albeit with large scatter) going from solar-type to intermediate-mass stars (e.g., Muzerolle et al.\,2005).  Hence viscous accretion also seems unlikely to cause the falloff in $M_{d,\nu}$.  

\noindent {\it (v) Observed spread in the apparent disk mass:} 

Within each of our four stellar regimes -- intermediate-mass stars, solar-types, VLMS and BDs -- the $M_{d,\nu}$ (and $M_{d,\nu}/\mstar$) span $\gtrsim$\,2 orders of magnitude in the roughly coeval Taurus and $\rho$ Oph populations (Fig.\,3).  While the statistics in the older TWA are far too small for a meaningful general comparison, at least the same range is seen among the VLMS in this Association as well (detected Hen 3-600A versus upper limits for TWA30A and B; Fig.\,3).  We examine several mechanisms which might cause this spread.

\noindent $\bullet$ The most straightforward explanation is that similar mass stars are nevertheless born with a wide range of initial disk masses, due to differences in initial conditions.  The latter might be, e.g., a spread in the parent core properties, or dynamical interactions between several stellar embryos formed within a core \citep[e.g.,][]{2009MNRAS.392..590B}.  

\noindent $\bullet$ Another possibility is that disks with comparable initial masses, around coeval stars of a given mass, are depleted to varying degrees due to differences in their accretion and/or photoevaporation rates.  This requires variations in initial disk properties other than mass (e.g., outer radius and surface density profile), and/or stellar properties other than mass (e.g., photoionizing/photoevaporative X-ray/UV flux).  In this context, it is perhaps suggestive that, within a fixed (sub)stellar mass bin, young stars and BDs in a given star-forming region evince fractional X-ray luminosities ($L_X/L_{bol}$) spanning $\sim$1 dex \citep{2007A&A...468..391G}, and accretion rates spanning $\sim$2 dex (e.g., Mohanty et al.\,2005; Muzerolle et al.\,2005): ranges comparable to or not much smaller than that in $M_{d,\nu}$.  It appears possible that the spread in apparent disk masses is related to the range in photoevaporative or accretion efficiencies.

We note that AW05 and AW07 have tried to test this, by comparing their estimated $M_{d,\nu}$ in Taurus and $\rho$ Oph to the equivalent widths and luminosities of $H\alpha$ emission in the parent stars, where the latter is an indicator of ongoing accretion.  They find that stars without detectable accretion (i.e., weak-line T Tauris) are overwhelmingly likely to lack disks; however, within the population of accretors (classical T Tauris), they find no correlation between $M_{d,\nu}$ and the $H\alpha$ equivalent width or luminosity.  Prima facie, this suggests that the range in $M_{d,\nu}$ is independent of accretion.  However, while the presence of strong $H\alpha$ emission is an excellent {\it indicator} of accretion, the actual value of its equivalent width or luminosity is a poor {\it quantitative} measure of the accretion rate: while $\mdot$ is broadly correlated with the line width and luminosity, there is a $\sim$1--2 dex dispersion in the correlation \citep[e.g.,][]{2004A&A...416..179N, 2008ApJ...681..594H, 2009ApJ...696.1589H}.  Some of this is due to variations in $\mdot$ coupled with non-coeval measurements of $H\alpha$, and some due to variations in the line independent of $\mdot$.  Either way, $H\alpha$ widths and luminosities are of limited value in determining accurate $\mdot$, and it remains an open question whether the spread in $M_{d,\nu}$ is correlated with the accretion rate or not.   A more careful analysis, using better and more direct $\mdot$ indicators such as UV continuum excess emission, is required to resolve this issue.           

\noindent $\bullet$ Conversely, one may postulate that the true disk masses around coeval stars of a given mass are actually quite similar, but variations in the rate of grain growth cause a spread in the {\it apparent} disk masses (i.e., growth to sizes much larger than a few millimetres in some disks depresses their sub-mm/mm emission, yielding spuriously low $M_{d,\nu}$ estimates).  In this case, one expects a shift in the entire grain size distribution to larger values, and thus a smaller spectral slope $\alpha$, in stars with the lowest fractional apparent disk masses ($M_{d,\nu}/\mstar$).  

To test this, we plot $\alpha$ versus both $M_{d,[850]}/\mstar$ and $M_{d,[1300]}/\mstar$ in Fig.\,8.  We see that in the stars with measured $\alpha$, which predominantly account for the upper $\sim$1 dex in $M_{d,\nu}/\mstar$, the distribution of $\alpha$ is essentially flat: there is no sign of $\alpha$ decreasing in step with $M_{d,\nu}/\mstar$.  The situation for the lower $\sim$1 dex of $M_{d,\nu}/\mstar$ values is less clear.  The majority of these stars only have {\it lower} limits on $\alpha$ (detected at 850\,$\mu$m but not at 1.3\,mm).  While most of these limits fall well below the mean ($\alpha$$\sim$2) of the high fractional mass disks, this merely reflects the survey sensitivity thresholds; the true distribution of $\alpha$ here is unknown.  We note that for a small subset of these Taurus and $\rho$ Oph stars, R10a,b have obtained 3\,mm fluxes as well.  Their data point to a slightly smaller $\alpha_{1.3-3}$ among the fainter disks (i.e., those with lower $M_{d,\nu}$ in our formulation) compared to the brighter ones in Taurus, and no significant difference between the two populations in $\rho$ Oph.  Overall, their number statistics are too small to rule out the hypothesis that their Taurus and $\rho$ Oph samples are drawn from the same population, or to prove that fainter disks indeed have larger grains.  The bottom line is that more observations are needed to test the validity of this scenario.

\subsubsection{Implications for Individual Sources in the TWA}

\noindent {\it (i) Disk masses and grain growth for TWA VLMS (Hen 3-600A and TWA 30A,B):}

\noindent {$\bullet$} {\it Hen 3-600A}: This a VLMS accretor in the $\sim$10\,Myr-old TWA.  It is actually a spectroscopic binary, part of the hierarchical triplet Hen 3-600.  Only the primary system (A) appears to harbor a disk \citep{2010ApJ...710..462A}.  Not much is known about the individual components of the primary, but they seem to be of roughly equal mass \citep{2003AJ....125..825T}; the systemic spectral type of $\sim$M3 then implies individual masses of $\sim$0.2\,$\msun$ for an age of $\sim$10\,Myr, similar to TWA 30A and B.  The disk in this system shows significant grain growth, comparable to that in TW Hya, and has a large central hole extending out to $\sim$1.3\,AU \citep{2004ApJS..154..439U}.  Resolved sub-mm data show that the disk is observed nearly face-on, and is also quite small, with $R_d \sim$ 15--25\,AU, compatible with tidal truncation by the third component of the triplet (Andrews et al.\,2010).  The observed 850\,$\mu$m flux of 65\,mJy \citep{2001ARA&A..39..549Z} yields $M_{d,[850]} \sim 5\times 10^{-4}$\,$\msun$ (Fig.\,3).  Comparing this to our models for the aforementioned star/disk properties (not plotted), we find $M_d^\kappa \sim M_{d,[850]}$ for large grains ($\beta = 0$)\footnote{Fig.\,6 shows that, for a 100\,AU disk, the observed flux from Hen 3-600A corresponds to $M_d^\kappa \sim 2.5\,M_{d,[850]}$ for $\beta \sim 0$.  For a 15--25\,AU disk with the same flux, $M_d^\kappa$ is smaller, because the disk dust is overall hotter.}.  For TW Hya, \citet{2002ApJ...566..409W} deduced $\kappa_\nu \approx 0.008$\,cm$^2$g$^{-1}$ at $\sim$1\,mm, three times smaller than our fiducial value for $\tilde\kappa_{[1.3]}$; since the grains in Hen 3-600A appear similar, we use the Weinberger et al.\,value to arrive at a true disk mass estimate of $M_d \sim 1.5\times 10^{-3}$\,$\msun$.  

We test the validity of this estimate by examining the accretion rate.  In particular, note from equations (12) and (13) that the viscous accretion rate at any time $t$ is given by $\mdot \approx M_d(t)/t$, within a factor of order unity.  For Hen 3-600A, the $M_d$ inferred above, combined with $t \sim 10$\,Myr for the TWA, then implies $\mdot \sim 1.5\times 10^{-10}$\,$\msun$\,yr$^{-1}$.  This is in excellent agreement with the average $\mdot \sim 3\times 10^{-10}$, with a spread of $\sim \pm$0.5\,dex, found by \citet{2011A&A...526A.104C} and \citet{2009ApJ...696.1589H} for Hen 3-600A from a number of optical, X-ray and UV spectroscopic diagnostics.  This bolsters our confidence in the derived $M_d$; conversely, it suggests that the theoretical relationship between the accretion rate and disk mass may be profitably used to investigate disk properties.  We use this technique below.

\noindent {$\bullet$} {\it TWA 30A and B}: These two stars constitute a VLMS binary system in the TWA, with a projected separation of $\sim$3400\,AU and component masses of $\sim$0.1$\msun$ and 0.2\,$\msun$ respectively \citep{2010AJ....140.1486L, 2010ApJ...714...45L}.  Moreover, various photometric and spectroscopic features suggest that the disks around both components are seen close to edge-on (discussed further below), with the secondary (B) appearing significantly underluminous in the optical and NIR as a result \citep{2010AJ....140.1486L, 2010ApJ...714...45L}.  Our SCUBA-2 850\,$\mu$m observations yield a 3$\sigma$ upper limit of $M_{d,[850]} < 3\times10^{-5}$\,$\msun$ for both disks.  Comparing to our model predictions in Fig.\,6, we see that this corresponds to the optically thin regime for 100--300\,AU disks.  The true opacity-normalized disk masses in this case range from $M_d^\kappa \sim M_{d,[850]}$ (for $\beta = 2$, $R_d = 100$\,AU) to $\sim 5M_{d,[850]}$ (for $\beta = 0$, $R_d = 300$\,AU).  Very recently, our group has marginally resolved the disk around TWA 30B with {\it HST}, finding that it extends out to $\sim$30\,AU in scattered light (Bochanski et al.\,in prep.); the true extent (below our detection limit for scattered light) may be somewhat larger.  Our model predicts that the 850\,$\mu$m emission from a 30\,AU disk with the observed $M_{d,[850]}$ is still optically thin, but the true $M_d^\kappa$ in this case is smaller, ranging over $\sim$ 0.3--1\,$M_{d,[850]}$ (for $\beta$ = 2--0; not plotted).  For very nearly edge-on orientations, of course, equation (A1), which forms the basis of our model, is not strictly valid.  However, semi-analytic models by \citet{1999ApJ...519..279C} indicate that, while the observed optical and IR flux is severely depressed in the edge-on case compared to smaller inclinations, the flux at $\sim$mm wavelengths is reduced by only a factor of $\sim$2; detailed Monte Carlo radiative equilibrium calculations by \citet{2003ApJ...598.1079W} bear this out.  Consequently, we expect the true $M_d^\kappa$ in TWA 30A and B to be at most about twice as large as cited above.  In summary, we predict a 3$\sigma$ upper limit of $M_d^\kappa <$ (1.8--30)$\times 10^{-5}\,\msun$ $\sim$ 6--100\,$M_{\earth}$ (for $R_d \sim$ 30--300\,AU and $\beta \sim$ 2--0) for these two disks.  Are such puny disk masses likely for TWA 30A and B?   
 
To address this, note that our derived $M_d^\kappa$ upper limits, together with $t \sim 10$\,Myr, imply $\mdot < 1.8\times 10^{-12}$ -- $3\times10^{-11}$\,$\msun$\,yr$^{-1}$ for TWA 30A and B.  The question then becomes, are these rates plausible for the two stars?  The optical and NIR photometry and spectra obtained by \citet{2010AJ....140.1486L, 2010ApJ...714...45L} reveal very strong signatures of accretion and outflow.  The strength of emission lines that arise at some distance from the star, e.g. in the outflow or in the accretion funnels, may be partially attributed to the edge-on viewing angle, wherein the disk occults the star but not the line-emitting regions, artificially enhancing the line flux relative to the photospheric continuum.  The strength of accretion-related features that arise close to or on the star, however, cannot be ascribed to the geometry (since such features are suppressed by the edge-on disk just as much as the stellar continuum), but must be related to the actual accretion rate.  In particular, TWA30A evinces excess emission from accretion shocks on the stellar surface, in the form of high optical veiling (filling in of photospheric absorption lines) and line emission signatures; TWA 30B shows similar excess emission.  While Looper et al.\,have not calculated accretion rates, models by \citet{2003ApJ...592..266M} show that significant optical veiling is expected in VLMS only for $\mdot \gtrsim 10^{-10}$\,$\msun$\,yr$^{-1}$.  This is similar to the $\mdot$ for Hen 3-600A, but 3--50$\times$ greater than our {\it upper} limits on $\mdot$ for TWA 30A and B, calculated above assuming fiducial grain properties.       

The most straightforward way of resolving this discrepancy is to invoke considerable grain growth.  Specifically, note that the closest parity achieved above, between the $\mdot$ estimated from veiling versus that predicted from disk masses, is for very extended (300\,AU) disks with $\beta \sim 0$ (which already points to very large grains).  To make up the remaining factor of three difference, we require the absolute opacity at 1.3\,mm to be about three times smaller than our fiducial $\tilde\kappa_{[1300]}$.  These values of $\beta$ and $\kappa_{[1300]}$ are identical to those indicated above for Hen 3-600A.  Conversely, if the disks around TWA 30A and B extend only up to $\sim$30\,AU, comparable to the disk radius for Hen 3-600A and consistent with our scattered light image for 30B, then even with $\beta \sim 0$, $\kappa_{[1300]}$ would need to be 15$\times$ smaller than our fiducial value, or five times lower than estimated for Hen 3-600A (which is possible for grains a few centimetres in size or larger, depending on the grain geometry and composition; e.g., R10a,b; B90 and references therein).  To summarize, the similarity between the estimated $\mdot$ in Hen 3-600A and TWA 30A and B, and their approximate coevality, suggests similar disk masses; the relatively much fainter 850\,$\mu$m emission from TWA 30A and B then implies that their disks have undergone at least as much grain growth as that of Hen 3-600A (if the 30A,B disks are much larger than the latter), or significantly more growth (with possibly different grain geometry and composition as well) if all three disks are comparable in extent. 

\noindent {\it (ii) Disk masses and grain growth for TWA BDs (2MASS 1207A and SSSPM 1102):}

\noindent $\bullet$ {\it 2MASS 1207-3932A}: This object (henceforth 2M1207A) has been the subject of intense study over the last few years, as the nearest and oldest BD to exhibit prominent signatures of both accretion and outflow, and with a giant planetary-mass companion to boot.  Our SCUBA-2 850\,$\mu$m data for this source yield a 3$\sigma$ upper limit of approximately $M_{d,[850]} < 4\times10^{-5}$\,$\msun$ (Fig.\,3).  Since 2M1207A is less massive and older (and thus cooler and smaller) than the fiducial 0.05\,$\msun$ BD at $\sim$1\,Myr used in our model in Fig.\,7, we recalculate our model for its specific parameters: [SpT, age] $\approx$ [M8, 10\,Myr] $\Rightarrow$ [$\mstar$, $\rstar$, $\tstar$] $\approx$ [0.03\,$\msun$, 0.25\,$\rsun$, 2500\,K].  The disk size is an additional issue for this source.  Its companion lies at a projected separation of $\sim$40\,AU; if this were the true separation, tidal truncation would imply a maximum disk size of $\sim$13\,AU around the primary.  Conversely, a true separation $\gtrsim$100\,AU appears unlikely: dynamical analyses \citep[e.g.,][and references therein]{2007ApJ...660.1492C}  indicate that, given the very small total mass of this system, such a distended orbit would be very unstable to disruption by encounters with other cluster members over 10\,Myr.  Using 100\,AU as an upper limit for the separation yields a maximum disk size of $\sim$30\,AU.  

For these parameters, our model (not plotted) predicts an opacity-normalized disk mass ranging from $M_d^\kappa \sim$ 2--3\,$M_{d,[850]}$ ( for $\beta =$ 2--0, $R_d = 13$\,AU) to $M_d^\kappa \sim$ 3--10\,$M_{d,[850]}$ (for $\beta =$ 2--0, $R_d = 30$\,AU).  Spectroastrometry of the jet, as well as variations in the accretion funnel flow signatures, further imply that the disk is seen at a high inclination \citep{2007ApJ...659L..45W, 2006ApJ...638.1056S}\citep[though not so close to edge-on as to occlude the BD:][]{2007ApJ...657.1064M}.  Assuming a maximum correction factor of $\sim$2 to account for the viewing angle (see TWA 30AB above), we get $M_d^\kappa <$ (8--40)$\times10^{-5}$\,$\msun$ for $R_d = 13$--30\,AU and $\beta =$ 2--0. 

Very recently, \citet{2012ApJ...744L...1H} have observed 2M1207A at 70 and 160\,$\mu$m with {\it Herschel}.  Combining their data with earlier {\it Spitzer} fluxes, they estimate a most probable disk mass of $\sim$10$^{-5}$\,$\msun$, with a plausible range of a few$\times$10$^{-6}$--10$^{-4}$\,$\msun$.  These results are fully consistent with our estimate above\footnote{\citet{2012MNRAS.422L...6R} use their own {\it Herschel} data to infer a disk mass more than an order of magnitude higher than the upper limits Harvey et al.\,and we find; however, their {\it Herschel} fluxes are grievously inconsistent with those of Harvey et al.\,(2012), and appear to be vitiated by a misidentification of the source (as Riaz et al.\,(2012b) also suggest, in a later erratum to their original paper).}.  Finally, the accretion rate inferred for 2M1207A, from various optical and UV diagnostics, is $\mdot \sim 10^{-12}$--$10^{-11}$\,$\msun$\,yr$^{-1}$ (Mohanty et al.\,2005; Herczeg et al.\,2009 and references therein), suggesting a disk mass of $\sim$$10^{-5}$--$10^{-4}$\,$\msun$ for this 10\,Myr-old BD.  This is again consistent with both our and Harvey et al.'s results.  

Lastly, none of these data strongly constrain the degree of grain growth in this disk \citep[though the weakness/absence of the 10\,$\mu$m silicate feature indicates that grains have grown beyond at least a few microns:][]{2004A&A...427..245S, 2008ApJ...681.1584R, 2008ApJ...676L.143M}.  

\noindent $\bullet$ {\it SSSPM 1102-3431}: This BD (hereafter SSSPM 1102) is nearly identical to 2M1207A in its intrinsic substellar properties, and in the 3$\sigma$ flux upper limit we obtain at 850\,$\mu$m.  However, there are no equivalent observational constraints on the size of its disk.  Adopting our fiducial limits for BDs, $R_d \sim$ 20--100\,AU, we find $M_d^\kappa <$ 4--40$\times10^{-5}$\,$\msun$ (for the full range [$R_d$, $\beta$] = [20,2]--[100,0]).  While Harvey et al.\,(2012) cannot significantly restrict most of its disk parameters, their 160\,$\mu$m detection of SSSPM 1102 allows them to put a fairly firm {\it lower} limit on its disk mass at a few$\times10^{-6}$\,$\msun$, fully consistent with our upper limits.  Finally, using UV diagnostics, Herczeg et al.\,(2009) have determined an accretion rate of $\mdot \approx 1.6\times10^{-13}$\,$\msun$\,yr$^{-1}$, the least known so far for any object.  For an age of $\sim$10\,Myr, this indicates a disk mass of $\sim$few$\times10^{-6}$\,$\msun$, at the lower end of our and Harvey et al.'s estimates.  Taken together, these data suggest $10^{-6} < M_d < 10^{-5}$\,$\msun$.  Again, there are no firm constraints on grain sizes (except that, as in 2M1207A, the lack of 10\,$\mu$m silicate emission implies grain growth beyond at least a few microns; Morrow et al.\,2008).        

Finally, it is noteworthy that AW07 perform a similar comparison (albeit for a much larger number of stars) between the disk mass based on fiducial disk/dust parameters ($M_{d,\nu}$ in our nomenclature) and that implied by the accretion rate, to find that the accretion-based mass is on average an order of magnitude higher.  This is comparable to our results for TWA 30A and B; AW05, like us, propose that grain growth may be responsible.  Note that in our analysis of Hen 3-600A above, we do {\it not} find such an offset when we adopt the $\beta$ and $\kappa_{[1300]}$ appropriate for the very large grains known to exist in its disk; using the fiducial $\beta$ and $\kappa_{[1300]}$ instead would indeed yield a disk mass much lower than the accretion-based value.  This supports grain growth as the culprit underlying such offsets.  

\section{Results III.  Relationship between Disk Mass and Stellar Mass} 

In the bottom two panels of Fig.\,3, the mean values of both $M_{d,[850]}/\mstar$ and $M_{d,[1300]}/\mstar$ appear roughly constant with $\mstar$, among the approximately coeval Taurus and $\rho$ Oph populations.  This apparently flat distribution of $M_{d,\nu}/\mstar$ has been commented on in previous work as well; it suggests that on average, $M_{d,\nu}/\mstar \sim 10^{-2}$ (e.g., Scholz et al.\,2006).  However, the presence of a large number of upper limits, especially among the VLMS and BDs, makes the veracity of this claim hard to judge by eye alone.  Instead, we use a Bayesian analysis to test this.  The technique is described in Appendix C, and the results are discussed below.

We emphasize that we only analyze the distribution of the apparent disk mass $M_{d,\nu}$, and not of the true disk mass $M_d$, or even of the opacity normalized mass $M_d^\kappa$.  As we have discussed, translating the first into either of the latter two quantities requires knowledge of a number of disk parameters, which are unknown for most of our sample.  As such, our discussion above of various broad trends in $M_d^\kappa$ and $M_d$ suggested by the data is the best we can do; precise determination of these two quantities for all the individual stars, required for a statistical investigation of the underlying distribution, is not currently possible.  Nevertheless, the statistics of $M_{d,\nu}$ alone are still valuable as an initial indicator of the possible behaviour of the true disk mass.  Equally importantly, the analysis serves to illustrate the Bayesian techniques that can be applied to the true $M_d$ distribution when it is derived in the future, as well as to any other distribution that is both noisy and plagued by upper limits.        

To begin with, we combine our $M_{d,[850]}$ and $M_{d,[1300]}$ estimates to get the largest possible sample of apparent disk masses.  Specifically, we use $M_{d,[850]}$ if available, otherwise $M_{d,[1300]}$ (see Table I).  Furthermore, our sample includes a number of known binaries and higher-order multiples (Table I).  Within our Taurus and $\rho$ Oph sub-samples, most such systems have not been resolved in the sub-mm/mm data presented here\footnote{The exceptions are a handful of extremely wide systems (sep $\sim$1500--4000\,AU), marked with a $^{\ddag}$ in Table I: DH Tau AB / DI Tau AB; FV Tau AB / FV Tau c AB; FY Tau / FZ Tau; GG Tau Aab / GG Tau Bab; GH Tau AB / V807 Tau ABab; GK Tau / GI Tau; V710 Tau AB / V 710 Tau C; and V955 Tau AB / LkHa 332 G2 AB / LkHa 332 G1 AB; see Kraus et al.\, (2011) and Harris et al.\,(2012). The flux and binarity data in Table I pertain to the individual sub-systems DH Tau AB, FV Tau AB, FV Tau c AB, FY Tau, GG Tau Aab, GH Tau AB, GK Tau, and V710 Tau AB (note that for those sub-systems that are themselves binaries, the disk emission around the individual stars is {\it not} resolved in the flux measurements listed).  We treat these sub-systems as isolated systems in our binary analysis in the text (e.g., GH Tau AB, denoted as GH Tau in Table I, is treated as a close binary (as noted in Table I), instead of one part of an extremely wide multiple system).}.  As such, the $M_{d,\nu}$ we derive corresponds to the {\it total} apparent disk mass in these systems; we do not know how this is partitioned between the components\footnote{More recently, Harris et al.\,(2012) {\it have} resolved the sub-mm emission around the individual components of some of our Taurus binaries. We have not included their data here in the interests of homogeneity, since their study, focussed specifically on multiplicity effects, has different selection criteria compared to ours. However, we do refer to their results where appropriate.}.  In all these cases, we assume that the disk material is present solely around the primary, i.e., we adopt $M_{d,\nu}/\mstar \equiv M_{d,\nu}/M_{primary}$.  This is because binarity studies are very much incomplete for our sample (especially for the $\rho$ Oph sources and the VLMS/BDs): many of the objects assumed to be single here have not been subject to as wide a companion parameter search as the identified binaries, and many have not been examined for multiplicity at all.  For unidentified binaries in our sample, the $M_{d,\nu}/\mstar$ we calculate implicitly corresponds to assigning the total disk mass entirely to the primary\footnote{Since we infer stellar mass from the spectral type and age, the heightened luminosity of close binaries compared to isolated stars does not influence us, and the mass determined is essentially that of the primary.}; doing the same for the known binaries/multiples is thus necessary for uniformity.  


The final combined sample is plotted in Fig.\,9.  The full sample is shown in the top panel, and the ``single'' objects (i.e., sample with {\it known} binaries/multiples removed) in the bottom panel; some of the latter may have as yet unknown companions.  Note that most of the VLMS/BDs plotted as 3$\sigma$ upper limits have actual measured values at $<$$3\sigma$ significance (Table I).  Our plotting convention is simply to facilitate visual comparison to objects from the literature for which only 3$\sigma$ upper limits in flux have been published.  The actual measurements, where available, {\it are} used in our subsequent Bayesian analysis. 

For the analysis, we first divide our full sample into 4 populations: {\it (1)} Taurus solar-type stars (49 sources); {\it (2)} $\rho$ Oph solar-type stars (32 sources); {\it (3)} all (Taurus + $\rho$ Oph) VLMS/BDs (27 sources); and {\it (4)} all (Taurus + $\rho$ Oph) intermediate mass stars (20 sources).  Taurus and $\rho$ Oph objects, which are roughly coeval, are lumped together in the VLMS/BD and intermediate-mass bins to increase the sample sizes therein.  The small and much older set of six TWA objects is excluded from this analysis.  The population of Taurus solar-type stars, which is the best constrained (in that it includes the most data and the least number of upper limits), serves as a baseline against which the other populations are compared.  

The Taurus solar-types have also been subject to more thorough binarity surveys than the other populations. To evaluate the effects of multiplicity, therefore, we also compare: the full sample of Taurus solar-types to the single Taurus solar-types (18 sources); the Taurus solar-type close binaries (projected component separation $<$100\,AU; 13 sources) to the singles; and the Taurus solar-type wide binaries ($\geq$100\,AU; 12 sources) to the singles. 

For each population, we assume that $M_{d,\nu}/\mstar$ is described by an underlying lognormal distribution specific to that population. Our Bayesian analysis then reveals the probability distributions for the mean ($\mu$) and standard deviation ($s$) of this lognormal in each case. 

Finally, we include only the Gaussian photon noise in the observed fluxes in our error analysis, and not the systematic calibration uncertainties, nor the uncertainties in the assumed values for disk parameters used to calculate $M_{d,\nu}$ ($\tilde{T}$=20\,K, $\beta$=1, $\tilde\kappa_{[1.3]}$=0.023\,cm$^2$g$^{-1}$ and gas-to-dust ratio = 100:1; see \S5), nor the uncertainties in the $\mstar$ inferred in \S3.  Including the additional calibration uncertainty would mostly only strengthen our results, as pointed out at appropriate junctures below.  Moreover, our lack of knowledge about the true disk parameters for a large fraction of our sources prevents us from deriving $M_d$ or $M_d^\kappa$ for individual stars in the first place, which is why we concentrate here only on the apparent disk mass (we do point out the effect of using $\beta < 1$, as seems appropriate for many of our stars, on our results).  Similarly, while uncertainties and systematic errors are undoubtedly present in the evolutionary tracks used to calculate $\mstar$, these are difficult to quantify precisely with our present knowledge.  We therefore consider only the Gaussian noise in the flux here.  


Figs.\,10--15 show the outcomes of our analysis.  In Figs.\,10--12, we inter-compare the results for various sub-samples of the Taurus solar-type population (full sample, singles, close binaries, and wide binaries).  In Figs.\,13--15, we compare the result for the full sample of Taurus solar-types to that for each of the other populations ($\rho$ Oph solar-types, all intermediate masses, and all VLMS/BDs).  In each figure, the three panels show (clock-wise from top right): {\it (a)} the full 2-D probability distribution of the lognormal parameters $\mu$ (mean) and $s$ (standard deviation) for a given population of stars, with the contours enclosing 68.27\% (1$\sigma$), 95.45\% (2$\sigma$) and 99.73\% (3$\sigma$) of the distribution, {\it (b)} the 1-D probability distribution of the mean $\mu$, marginalised (integrated) over all standard deviations $s$ (i.e., the distribution of $\mu$ independent of the precise value of $s$), with contours again at 1--3$\sigma$, and similarly {\it (c)} the 1-D probability distribution of $s$ marginalised over $\mu$.  Note that the $\mu$ and $s$ of the actual lognormal distribution, equation (C9), are in natural log units (ln$[M_d/\mstar]$); in these figures, we plot them in base-10 (log$_{10}$$[M_d/\mstar]$) units instead (i.e., we plot $\mu$\,log$_{10}$e and $s$\,log$_{10}$e instead of $\mu$ and $s$), for greater intuition. We obtain 4 main results:

\noindent {\it (i) $M_{d,\nu}/\mstar$ for Taurus solar-type stars:}  

Fig.\,10 shows that the lognormal for the full sample of Taurus solar-type stars has a most likely mean of $\mu$\,log$_{10}$e = $-2.4^{+0.3}_{-0.4}$, and a most likely standard deviation of $s$\,log$_{10}$e = $0.7^{+0.4}_{-0.2}$ (where the ranges are the $\pm$3$\sigma$ spread in probable values around the peak of the 1-D $\mu$ and $s$ distributions).  Note that $\beta$ for many of the solar-type stars appears to lie in the range 0--1 (see \S7.3.2), lower than our adopted $\beta = 1$.  Hence the mean of the true disk to stellar mass ratio $M_d/\mstar$ (modulo the absolute value of the opacity $\kappa$) may be a factor of $\sim$3 higher for these stars (as noted in \S7.3.2), corresponding to $\mu$\,log$_{10}$e $\sim -2$.  

\noindent{\it (ii) Effects of binarity in Taurus solar-type stars:}  

Fig.\,10 also shows that the 1-D distribution of the mean for the Taurus single solar-types deviates by nearly 2$\sigma$ from that for the full Taurus sample of these stars, with the most probable mean for the singles occurring at $\mu$\,log$_{10}$e $\approx -2$ (2.5 times higher than in the full sample). Since the full sample comprises both singles and multiples, this result suggests that multiplicity has some effect on the disk mass distribution. To investigate this further, we follow AW05 in dividing the Taurus solar-type binaries into close and wide systems (projected separations of $<$100\,AU and $\geq$100\,AU respectively), and compare the two sub-samples separately to the Taurus solar-type singles. To keep the analysis clean, we exclude the handful of higher-order multiples (which can have both close and wide components) and spectroscopic binaries (whose circumbinary disks may be as bright as the disks in wide binaries and around single stars; Harris et al.\,2012).       

Fig.\,11 shows that the 1-D distribution of the mean for the Taurus solar-type close binaries deviates by $\sim$3$\sigma$ from that for the single stars, with the most probable mean for the close systems occurring at $\mu$\,log$_{10}$e $\approx -3$ (i.e., 10 times smaller than for the singles). The deviation between the wide binaries and singles is significantly smaller: Fig.\,12 shows that the 1-D distribution of the mean for the wides is within 2$\sigma$ of that for the singles, and peaks at $\mu$\,log$_{10}$e $= 2.5$ (i.e., 3 times smaller than in the singles). We note in passing that the 1-D distribution of the standard deviation shows much less variation between the singles and close and wide binaries: as Figs.\,11 and 12 reveal, the $s$ distributions in all three cases lie within $\sim$1$\sigma$ of each other.     

These results are consistent with those of AW05 and \citet{2012ApJ...751..115H}, who found (in samples consisting predominantly of solar-type stars, in line with the sample tested here) that the total disk mass in wide binaries is similar to (or modestly smaller than) around singles, while it is substantially smaller in close binaries. It is also worth recalling that the means cited above refer to log$_{10}$$[M_{d,\nu}/M_{primary}]$, i.e., are calculated under the assumption that all the disk material is associated with the primary. Since in reality the material is often distributed between both binary components (albeit possibly with disproportionately more around the primary; see Harris et al.\,2012), the disk mass per individual component stellar mass in binaries is likely to deviate even more from the disk to stellar mass ratio in singles, than the numbers above indicate. Overall, these effects probably arise from a combination of tidal truncation of disks in binaries and the binary formation mechanism itself; we refer the reader to Harris et al.\,(2012) for a more detailed discussion.       

Our goal here is to examine how the disk to stellar mass ratio changes as a function of stellar mass. Ideally, therefore, we should conduct the same fine-analysis -- distinguishing between singles, close binaries, and wides -- when comparing the Taurus solar-type population to the $\rho$ Oph sample, and to the other stellar mass bins. However, as noted earlier, binarity surveys at all stellar masses in $\rho$ Oph, and among VLMS/BDs in both Taurus and $\rho$ Oph, are far less complete than among the Taurus solar-types, precluding such a detailed investigation at present. Instead, we conservatively compare the full sample of Taurus solar-types, regardless of binarity, to the other populations (whose binarity fraction is largely unknown), with the stipulation (as before) that in binaries and higher-order multiples our $M_{d,\nu}/\mstar$ refers to the total disk mass per primary stellar mass.     

\noindent{\it (iii) $M_{d,\nu}/\mstar$ for different stellar mass bins:}  

\noindent $\bullet$ The $\rho$ Oph solar-type sample, and the [Taurus + $\rho$ Oph] VLMS/BD and intermediate-mass samples, are all somewhat less constrained than the Taurus solar-type population, due to a combination of fewer data points and more upper limits.  Nevertheless, the $\mu$ and $s$ of their lognormal models are all consistent with that of the Taurus solar-types to within 1$\sigma$ (both in the full 2-D and marginalised 1-D parameter spaces; Figs.\,13--15).  We thus conclude that the current data is consistent with a constant log$_{10}$$[M_{d,\nu}/\mstar]$ $\approx -2.4$ all the way from intermediate-mass stars to low-mass brown dwarfs.  Note that including the flux calibration uncertainties would only broaden each of the probability distributions, increasing the overlap with the Taurus solar-types. Whether the binary and single star populations in these different mass bins are also individually similar to the corresponding populations among the Taurus solar-types remains to be clarified in the future, with better binary statistics.


\noindent {\it (iv) Validity of a lognormal distribution:}

We have so far {\it assumed} that the underlying distribution of $M_d/\mstar$ for any stellar mass bin is a lognormal.  Since $M_d/\mstar$ starts out relatively large (with the disk making up most of the star+disk system in the earliest evolutionary phases) and evolves eventually to zero (as the material either accretes on the star, or forms planetary bodies, or is ejected from the system), it is reasonable to suppose that in roughly coeval objects known to have disks but be more evolved than the earliest (Class 0/I) phases, $M_d/\mstar$ will be clustered about some mean, with fewer systems remaining very far above or having evolved very far below this value.  The most natural such distribution is the lognormal, hence our choice\footnote{Conversely, for example, if our sample were drawn randomly from all ages without the existence of disks being a prerequisite for inclusion, then a power-law or exponential, rising with decreasing $M_d/\mstar$, would have been more appropriate instead, since older stars without disks vastly outnumber younger ones with.}.    

Nevertheless, our analysis thus far does not provide evidence that the distribution is a lognormal, but only supplies the most probable model parameters {\it if} it is one.  There are two ways to investigate the validity of the lognormal assumption itself.  In the Bayesian context, one would undertake a model comparison to derive the relative merits of various possibilities (e.g., lognormal versus power law versus exponential\footnote{Note that, in the presence of measurement errors, no analysis can prove or disprove the absolute merit of a particular model; one can only say whether one model is better or worse than another.}).  We defer such rigorous analysis to the future, when more data becomes available.  The other option is to compare the best-fit lognormals we have derived to the results of a survival analysis, where the latter yields an estimate of the underlying distribution of the data independent of any particular model, under the assumptions that the upper limits are randomly distributed and measurement errors are negligible.  The general invalidity of these assumptions is what led us to a Bayesian inference in the first place.  The comparison is still useful, however, both as a consistency check when upper limits do not dominate the data and measurement noise is relatively small, and as a means of underlining the limitations of survival analysis when this is not true.  

Fig.\,16 illustrates the comparison between our Bayesian results and the outcome of a survival analysis based on the KM estimator\footnote{Calculated using the ASURV code \citep{1990BAAS...22..917I}.}  (Feigelson \& Nelson 1985; AW05; AW07).  For each of the 4 stellar mass bins we have considered, we plot the cumulative distribution corresponding to the lognormal with the most probable mean and standard deviation inferred from the Bayesian 1-D marginalisations (see Figs.\,13--15). This is compared to the underlying cumulative distribution predicted by the KM estimator.  Note that the error bars on the latter only reflect the uncertainty in binning the (assumed random) upper limits in the survival analysis, and do not include any actual measurement noise in the data (which, as mentioned before, cannot be treated in survival analysis).    

We see that, for the Taurus solar-type stars and the Taurus + $\rho$ Oph intermediate-mass stars -- the two populations with the lowest fraction of upper limits and the least noise -- our lognormals are in excellent agreement with the KM estimator.  For the $\rho$ Oph solar-type stars -- comprising more upper limits, but still dominated by detections -- our lognormal deviates slightly more from the KM estimator, but is still in good overall agreement with it.  Finally, for the Taurus + $\rho$ Oph VLMS/BDs, in which upper limits outnumber detections and the noise is greatest, our most-probable lognormal departs significantly from the survival analysis prediction.  These results argue that {\it (a)} a lognormal distribution is indeed valid for the solar-type and intermediate-mass stars, and {\it (b)} by Occam's razor, it is also appropriate for VLMS/BDs, with the mismatch here between our lognormal and the KM estimator arising from the {\it breakdown of the assumptions underlying survival analysis itself}.  Rigorous evidence for the validity of a lognormal for the VLMS/BDs must await more detections.

\section{Implications for Planet Formation and Disk Accretion}
\citet{2007MNRAS.381.1597P} have carried out simulations of planet formation via core accretion around a range of stellar/sub-stellar masses.  They find that low-mass stars with $\mstar \sim 0.3\msun$ and a mean $M_d/\mstar$ $\sim$ 5\% can form terrestrial mass planets efficiently, as well as intermediate-mass and giant planets (though recovering the observed semi-major axis distribution of the latter two classes requires some fine-tuning of Type I migration).  0.05$\msun$ brown dwarfs with the same mean $M_d/\mstar$ (translating to disk masses of a few Jupiters), on the other hand, form terrestrial mass planets less frequently, with $M_{\rm planet}$ $\gtrsim 0.3M_{\earth}$ arising in $\sim$10\% of cases, while giant planets are completely absent in this domain (timescales for collisionally-driven planetesimal growth are longer around such low masses, and the disk gas dissipates before rocky cores become large enough to initiate runaway growth).      

Given that the upper envelope of our observed distribution of $M_{d,\nu}/\mstar$ for VLMS/BDs at $\sim$1\,Myr lies at about 1--2\%, and that our most probable mean $M_{d,\nu}/\mstar$ for these objects is only about $\sim$0.5\%, it would appear that terrestrial planets may be hard to form around VLMS, and be extremely rare around BDs (for 0.05$\msun$ BDs with mean $M_d$ a fraction of a Jupiter mass, Payne \& Lodato find $M_{\rm planet}$ $\gtrsim$ 0.3$M_{\earth}$ in only 0.035\% of cases).  However, for the few VLMS/BDs with measured continuum slopes $\alpha$, we showed that $M_{d,\nu}$ may underestimate the true opacity-normalized disk-mass $M_d^\kappa$ by a factor of $\gtrsim$\,3--5; in general, $M_{d,\nu}$ underestimates $M_d^\kappa$ by a factor of at least a few in VLMS/BDs if their average continuum slope is comparable to that in more massive stars, $\langle\alpha_{850-1.3}\rangle \sim 2$ (\S7.3.1).  Thus disk masses of a few percent of the (sub-)stellar mass may in fact be more appropriate than the naive estimate $M_{d,\nu}$.  Moreover, Payne \& Lodato's simulations refer to the initial $M_d$ (at $\sim$10$^4$ yr); the values we observe at $\sim$1 Myr must be somewhat lower due to subsequent accretion/outflow.  Putting the two effects together, an initial mean $M_d/\mstar \sim$ 5\% is plausible, implying that terrestrial mass planets may indeed form copiously around VLMS, and, albeit less frequently, around BDs as well.  This must be tested through future multi-wavelength observations that constrain the spectral slope and degree of grain growth in VLMS/BD disks. 

Our inferred disk masses have implications for accretion too.  Empirically, the accretion rate $\mdot$ onto the central star/BD seems to fall off with stellar mass roughly as $\mdot \propto \mstar^2$ (though with considerable scatter), all the way from intermediate mass stars to very low-mass BDs \citep{2005ApJ...626..498M, 2005ApJ...625..906M, 2006A&A...452..245N}.  \citet{2005ApJ...622L..61P} proposed that this is linked to Bondi-Hoyle accretion onto the disk from the surrounding molecular cloud as the star+disk system moves through the cloud.  However, there are severe problems with this hypothesis (see discussion by Mohanty et al.\,2005 and Hartmann et al.\,2006), and it is unlikely to be the main reason for the observed relationship. Dullemond et al.\,(2006) and \citet{2008ApJ...676L.139V, 2009ApJ...703..922V} have presented two alternate theories.  According to the former authors, the dependence of $\mdot$ on $\mstar$ ultimately results from the distribution of rotation rates of the parent molecular cloud cores; according to the latter, it derives from gravitational instability-driven accretion.  In both cases, $\mdot \propto M_d$; insofar as the $\mdot \propto \mstar^2$ trend is real, therefore, both require $M_d \propto \mstar^2$.  Our results, on the other hand, indicate a significantly shallower trend, $M_{d,\nu} \propto \mstar$, all the way from intermediate-mass stars to very low mass BDs, and therefore do not support either of these two theories.  Note that we have assumed a constant characteristic disk temperature $\tilde{T} = 20$\,K independent of stellar mass, for estimating $M_{d,\nu}$; this is likely to be systematically slightly in error, with the appropriate temperature being somewhat higher for the hot stars and lower for BDs, as discussed in \S\S7.3.2 and 9.  Correcting for this, however, leads to disks systematically {\it less} massive with increasing stellar mass than our naive $M_{d,\nu}$ estimate, making the true disk mass an even shallower function of stellar mass than we find, which is {\it opposite} to the required effect.  As such, theories requiring $M_d \propto \mstar^2$ to explain the observed $\mdot$ appear unsupported by data.         

Alternatively, \citet{2006ApJ...639L..83A} show that the $\mdot \propto \mstar^2$ relationship can be reproduced even if $M_d \propto \mstar$ (as our $M_{d,\nu}$ results suggest), if the initial disk {\it radius} increases with decreasing (sub)stellar mass.  They note that viscous spreading means that stellar disks should rapidly expand to BD disk sizes, and moreover that external factors such as binary truncation can strongly influence the disk size, so that the inverse scaling of the initial disk radius with $\mstar$ may not be easily observable.  However, their model does predict that few if any objects should have $\mdot < 10^{-12}$\,$\msun$\,yr$^{-1}$, and that the observed scatter in accretion rates (assumed to be age-related) should be smaller in BDs than in higher-mass stars.  Sensitive UV measurements by Herczeg et al. (2009) have revealed $\mdot \sim 10^{-12}$--$10^{-13}$\,$\msun$\,yr$^{-1}$ in three BDs so far, nominally at odds with this hypothesis, but more data is required to clarify whether such low rates are indeed common in BDs or a rarity as this scenario predicts.         
  
However, the trend $\mdot \propto \mstar^2$ itself remains open to question \citep[e.g.,][]{2006MNRAS.370L..10C}; for instance, it is unclear whether it applies to the average $\mdot$ at any given mass or only to the upper envelope, and also whether the exponent is indeed 2 everywhere or smaller (e.g., Vorobyov \& Basu (2009) argue that $\mdot \propto \mstar^{1.3}$ among solar-type stars).  A more careful consideration of the selection effects and upper limits in the data (as suggested by Clarke \& Pringle 2006), as well as more sensitive $\mdot$ observations, are required to resolve this question.      

\section{Conclusions}
We have obtained SCUBA-2 850$\mu$m data for 7 accreting VLMS and BDs in Taurus and the TWA, and combined our observations with other recent sub-mm/mm surveys of Taurus, $\rho$ Ophiuchus and the TWA to investigate the trends in disk mass and grain growth during the cTT phase.  Assuming a disk gas-to-dust ratio of 100:1 and fiducial surface density and temperature profiles guided by current observations ($\Sigma \propto r^{-1}$, $T \propto r^{-0.6}$), and using generalized equations for the disk emission, we find that:  

\noindent $\bullet$ The Rayleigh-Jeans approximation is less than ideal for disks around solar-type and intermediate-mass stars, and even worse VLMS and BD disks.  

\noindent $\bullet$ The minimum disk outer radius required to explain the upper envelope of sub-mm/mm fluxes is $\sim$100\,AU for intermediate-mass stars, solar-types and VLMS, and $\sim$20\,AU for BDs.  

\noindent $\bullet$ There is a marked flattening/decrease in the upper envelope of observed fluxes going from solar-type to intermediate-mass stars, given roughly by $F_{\nu} \propto \mstar^{-1/2}$; this may be due to greater photoevaporation with increasing stellar mass.  Grain growth and accretion, while present, appear unlikely to cause this trend.  

\noindent $\bullet$ Many of the (likely) optically thin disks around Taurus and $\rho$ Oph intermediate-mass and solar-type stars evince an opacity power-law index of $\beta \sim$ 0--1 (as found in previous studies as well), suggesting substantial grain growth.  The situation for the VLMS/BDs in these regions is not clear: most have been observed at only one wavelength, and the few with a small measured $\alpha$ are consistent with either grain growth or optically thick disks.  

\noindent $\bullet$ We have observed the TWA VLMS TWA30A and B only at 850\,$\mu$m, but a comparison of their derived apparent disk masses and their observed accretion rates and age suggests substantial grain growth, similar to that in the TWA accretors Hen 3-600A and TW Hya.  

\noindent $\bullet$ For the TWA BD 2M1207A, an analogous analysis suggests $M_d/\mstar \sim$ 10$^{-3}$, comparable to the lower end of measured values among Taurus and $\rho$ Oph VLMS/BDs; the disk mass for the similar TWA BD SSSPM1102 appears to be a factor of 10 smaller still (of order 1\,$M_{\Earth}$). The available data put no strong constraints on grain growth in the two disks.  

\noindent $\bullet$ The observed spread of $\gtrsim$\,2 dex in apparent disk masses, in all stellar mass bins, may reflect a spread in initial disk masses, or in accretion and/or photoevaporation rates.  There is no strong evidence at present of this spread being caused by a reduction in disk flux due to grain growth, but more data on faint disks is required to test this proposition.  

We have also examined the relationship between apparent disk mass and stellar mass through a Bayesian analysis (with the caveat that the trend in the real disk mass may be different, depending on grain properties and the absolute value of the opacity, which are unknown for a large fraction/most of the sources).  We find the following: 

\noindent $\bullet$ Among Taurus solar-type stars, the disk mass in close binary systems (projected separation $<$100\,AU) is $\sim$10$\times$ smaller than around single stars, while the disk mass in wide binaries ($\geq$100\,AU) is closer to that around singles ($\sim$3$\times$ smaller), in line with previous studies.    

\noindent $\bullet$ The apparent disk-to-stellar mass ratio is consistent with a lognormal distribution with a mean of log$_{10}$$[M_{disk}/\mstar] \approx -2.4$, all the way from intermediate-mass stars to VLMS/BDs, in agreement with previous qualitative suggestions that the ratio is roughly 1\% throughout the stellar/substellar domain.

\noindent $\bullet$ We caution against the application of survival analysis techniques to astrophysical datasets in which upper limits are not random, but dependent on survey sensitivities.  

Finally, we have examined the implications of our results for planet formation around VLMS/BDs, and accretion rates as a function of stellar mass.  We find that:

\noindent $\bullet$ While the apparent disk masses, $M_{d,\nu}$, suggest that there may not be enough material around VLMS/BDs to efficiently form terrestrial mass planets, our detailed analysis suggests that these estimates might be systematically too low by factors of a few to 10.  If so, terrestrial mass planets may form copiously around VLMS, and in the most massive BD disks as well.  

\noindent $\bullet$ The observations do not support theories which require $M_d \propto \mstar^2$ in order to explain the apparent empirical correlation $\mdot \propto \mstar^2$ from intermediate-mass stars to BDs.  Either the mechanism behind this relationship is different (e.g., related to a mass-dependent variation in disk size), or the relationship itself is more nuanced (either not as universal as thought, or driven by observational selection effects).                    

\bigskip
\bigskip
{\it Acknowledgements:} 
We thank the anonymous referee for a very close reading of the paper and extremely useful comments.  S.M.\,is very grateful to the JCMT for awarding shared-risk SCUBA-2 time to this project, and to the {\it International Summer Institute for Modeling in Astrophysics} (ISIMA) for affording him the time and research environment necessary to take this work forward; he also acknowledges the funding support of the STFC grant ST/H00307X/1. Part of the work by A.S. was funded by the Science Foundation Ireland through grant 10/RFP/AST2780. 

\bibliographystyle{apj}
\bibliography{smohanty}

\begin{thebibliography}{65}
\expandafter\ifx\csname natexlab\endcsname\relax\def\natexlab#1{#1}\fi

\bibitem[{{Alexander} \& {Armitage}(2006)}]{2006ApJ...639L..83A}
{Alexander}, R.~D., \& {Armitage}, P.~J. 2006, \apjl, 639, L83

\bibitem[{{Andrews} \& {Williams}(2005)}]{2005ApJ...631.1134A}
{Andrews}, S., \& {Williams}, J. 2005, ApJ, 631, 1134 (AW05)

\bibitem[{{Andrews} \& {Williams}(2007)}]{2007ApJ...671.1800A}
---. 2007, ApJ, 671, 1800 (AW07)

\bibitem[{{Andrews} {et~al.}(2010){Andrews}, {Czekala}, {Wilner}, {Espaillat},
  {Dullemond}, \& {Hughes}}]{2010ApJ...710..462A}
{Andrews}, S.~M., {Czekala}, I., {Wilner}, D.~J., {Espaillat}, C., {Dullemond},
  C.~P., \& {Hughes}, A.~M. 2010, \apj, 710, 462

\bibitem[{{Baraffe} {et~al.}(1998){Baraffe}, {Chabrier}, {Allard}, \&
  {Hauschildt}}]{1998AA...337..403B}
{Baraffe}, I., {Chabrier}, G., {Allard}, F., \& {Hauschildt}, P.~H. 1998, A\&A,
  337, 403

\bibitem[{{Baraffe} {et~al.}(2002){Baraffe}, {Chabrier}, {Allard}, \&
  {Hauschildt}}]{2002A&A...382..563B}
---. 2002, \aap, 382, 563

\bibitem[{{Bate}(2009)}]{2009MNRAS.392..590B}
{Bate}, M.~R. 2009, \mnras, 392, 590

\bibitem[{{Beckwith} {et~al.}(1990){Beckwith}, {Sargent}, {Chini}, \&
  {Guesten}}]{1990AJ.....99..924B}
{Beckwith}, S.~V.~W., {Sargent}, A.~I., {Chini}, R.~S., \& {Guesten}, R. 1990,
  \aj, 99, 924

\bibitem[{{Bouy} {et~al.}(2008){Bouy}, {Hu{\'e}lamo}, {Pinte}, {Olofsson},
  {Barrado Y Navascu{\'e}s}, {Mart{\'{\i}}n}, {Pantin}, {Monin}, {Basri},
  {Augereau}, {M{\'e}nard}, {Duvert}, {Duch{\^e}ne}, {Marchis}, {Bayo},
  {Bottinelli}, {Lefort}, \& {Guieu}}]{2008A&A...486..877B}
{Bouy}, H., {et~al.} 2008, \aap, 486, 877

\bibitem[{{Chabrier} {et~al.}(2000){Chabrier}, {Baraffe}, {Allard}, \&
  {Hauschildt}}]{2000ApJ...542..464C}
{Chabrier}, G., {Baraffe}, I., {Allard}, F., \& {Hauschildt}, P. 2000, \apj,
  542, 464

\bibitem[{{Chapin} {et~al.}(2013){Chapin}, {Berry}, {Gibb}, {Jenness}, {Scott},
  {Tilanus}, {Economou}, \& {Holland}}]{2013MNRAS.430.2545C}
{Chapin}, E.~L., {Berry}, D.~S., {Gibb}, A.~G., {Jenness}, T., {Scott}, D.,
  {Tilanus}, R.~P.~J., {Economou}, F., \& {Holland}, W.~S. 2013, \mnras, 430,
  2545

\bibitem[{{Chauvin} {et~al.}(2005){Chauvin}, {Lagrange}, {Dumas}, {Zuckerman},
  {Mouillet}, {Song}, {Beuzit}, \& {Lowrance}}]{2005A&A...438L..25C}
{Chauvin}, G., {Lagrange}, A.-M., {Dumas}, C., {Zuckerman}, B., {Mouillet}, D.,
  {Song}, I., {Beuzit}, J.-L., \& {Lowrance}, P. 2005, \aap, 438, L25

\bibitem[{{Chiang} \& {Goldreich}(1999)}]{1999ApJ...519..279C}
{Chiang}, E.~I., \& {Goldreich}, P. 1999, \apj, 519, 279

\bibitem[{{Clarke} \& {Pringle}(2006)}]{2006MNRAS.370L..10C}
{Clarke}, C.~J., \& {Pringle}, J.~E. 2006, \mnras, 370, L10

\bibitem[{{Close} {et~al.}(2007){Close}, {Zuckerman}, {Song}, {Barman},
  {Marois}, {Rice}, {Siegler}, {Macintosh}, {Becklin}, {Campbell}, {Lyke},
  {Conrad}, \& {Le Mignant}}]{2007ApJ...660.1492C}
{Close}, L.~M., {et~al.} 2007, \apj, 660, 1492

\bibitem[{{Curran} {et~al.}(2011){Curran}, {Argiroffi}, {Sacco}, {Orlando},
  {Peres}, {Reale}, \& {Maggio}}]{2011A&A...526A.104C}
{Curran}, R.~L., {Argiroffi}, C., {Sacco}, G.~G., {Orlando}, S., {Peres}, G.,
  {Reale}, F., \& {Maggio}, A. 2011, \aap, 526, A104

\bibitem[{{Dempsey} {et~al.}(2013){Dempsey}, {Friberg}, {Jenness}, {Tilanus},
  {Thomas}, {Holland}, {Bintley}, {Berry}, {Chapin}, {Chrysostomou}, {Davis},
  {Gibb}, {Parsons}, \& {Robson}}]{2013MNRAS.430.2534D}
{Dempsey}, J.~T., {et~al.} 2013, \mnras, 430, 2534

\bibitem[{{Ercolano} {et~al.}(2009){Ercolano}, {Clarke}, \&
  {Drake}}]{2009ApJ...699.1639E}
{Ercolano}, B., {Clarke}, C.~J., \& {Drake}, J.~J. 2009, \apj, 699, 1639

\bibitem[{{Feigelson} \& {Nelson}(1985)}]{1985ApJ...293..192F}
{Feigelson}, E.~D., \& {Nelson}, P.~I. 1985, \apj, 293, 192

\bibitem[{{Gorti} {et~al.}(2009){Gorti}, {Dullemond}, \&
  {Hollenbach}}]{2009ApJ...705.1237G}
{Gorti}, U., {Dullemond}, C.~P., \& {Hollenbach}, D. 2009, \apj, 705, 1237

\bibitem[{{Grosso} {et~al.}(2007){Grosso}, {Briggs}, {G{\"u}del}, {Guieu},
  {Franciosini}, {Palla}, {Dougados}, {Monin}, {M{\'e}nard}, {Bouvier},
  {Audard}, \& {Telleschi}}]{2007A&A...468..391G}
{Grosso}, N., {et~al.} 2007, \aap, 468, 391

\bibitem[{{Harris} {et~al.}(2012){Harris}, {Andrews}, {Wilner}, \&
  {Kraus}}]{2012ApJ...751..115H}
{Harris}, R.~J., {Andrews}, S.~M., {Wilner}, D.~J., \& {Kraus}, A.~L. 2012,
  \apj, 751, 115

\bibitem[{{Harvey} {et~al.}(2012){Harvey}, {Henning}, {M{\'e}nard}, {Wolf},
  {Liu}, {Cieza}, {Evans}, {Pascucci}, {Mer{\'{\i}}n}, \&
  {Pinte}}]{2012ApJ...744L...1H}
{Harvey}, P.~M., {et~al.} 2012, \apjl, 744, L1

\bibitem[{{Herczeg} {et~al.}(2009){Herczeg}, {Cruz}, \&
  {Hillenbrand}}]{2009ApJ...696.1589H}
{Herczeg}, G.~J., {Cruz}, K.~L., \& {Hillenbrand}, L.~A. 2009, \apj, 696, 1589

\bibitem[{{Herczeg} \& {Hillenbrand}(2008)}]{2008ApJ...681..594H}
{Herczeg}, G.~J., \& {Hillenbrand}, L.~A. 2008, \apj, 681, 594

\bibitem[{{Holland} {et~al.}(2013){Holland}, {Bintley}, {Chapin},
  {Chrysostomou}, {Davis}, {Dempsey}, {Duncan}, {Fich}, {Friberg}, {Halpern},
  {Irwin}, {Jenness}, {Kelly}, {MacIntosh}, {Robson}, {Scott}, {Ade},
  {Atad-Ettedgui}, {Berry}, {Craig}, {Gao}, {Gibb}, {Hilton}, {Hollister},
  {Kycia}, {Lunney}, {McGregor}, {Montgomery}, {Parkes}, {Tilanus}, {Ullom},
  {Walther}, {Walton}, {Woodcraft}, {Amiri}, {Atkinson}, {Burger}, {Chuter},
  {Coulson}, {Doriese}, {Dunare}, {Economou}, {Niemack}, {Parsons},
  {Reintsema}, {Sibthorpe}, {Smail}, {Sudiwala}, \&
  {Thomas}}]{2013MNRAS.430.2513H}
{Holland}, W.~S., {et~al.} 2013, \mnras, 430, 2513

\bibitem[{{Isobe} \& {Feigelson}(1990)}]{1990BAAS...22..917I}
{Isobe}, T., \& {Feigelson}, E.~D. 1990, in Bulletin of the American
  Astronomical Society, Vol.~22, Bulletin of the American Astronomical Society,
  917--918

\bibitem[{{Kenyon} \& {Hartmann}(1995)}]{1995ApJS..101..117K}
{Kenyon}, S.~J., \& {Hartmann}, L. 1995, \apjs, 101, 117

\bibitem[{{Klein} {et~al.}(2003){Klein}, {Apai}, {Pascucci}, {Henning}, \&
  {Waters}}]{2003ApJ...593L..57K}
{Klein}, R., {Apai}, D., {Pascucci}, I., {Henning}, T., \& {Waters},
  L.~B.~F.~M. 2003, \apjl, 593, L57

\bibitem[{{Kraus} {et~al.}(2011){Kraus}, {Ireland}, {Martinache}, \&
  {Hillenbrand}}]{2011ApJ...731....8K}
{Kraus}, A.~L., {Ireland}, M.~J., {Martinache}, F., \& {Hillenbrand}, L.~A.
  2011, \apj, 731, 8

\bibitem[{{Lodato} {et~al.}(2005){Lodato}, {Delgado-Donate}, \&
  {Clarke}}]{2005MNRAS.364L..91L}
{Lodato}, G., {Delgado-Donate}, E., \& {Clarke}, C.~J. 2005, \mnras, 364, L91

\bibitem[{{Looper} {et~al.}(2010{\natexlab{a}}){Looper}, {Bochanski},
  {Burgasser}, {Mohanty}, {Mamajek}, {Faherty}, {West}, \&
  {Pitts}}]{2010AJ....140.1486L}
{Looper}, D.~L., {Bochanski}, J.~J., {Burgasser}, A.~J., {Mohanty}, S.,
  {Mamajek}, E.~E., {Faherty}, J.~K., {West}, A.~A., \& {Pitts}, M.~A.
  2010{\natexlab{a}}, \aj, 140, 1486

\bibitem[{{Looper} {et~al.}(2010{\natexlab{b}}){Looper}, {Mohanty},
  {Bochanski}, {Burgasser}, {Mamajek}, {Herczeg}, {West}, {Faherty}, {Rayner},
  {Pitts}, \& {Kirkpatrick}}]{2010ApJ...714...45L}
{Looper}, D.~L., {et~al.} 2010{\natexlab{b}}, \apj, 714, 45

\bibitem[{{Luhman}(2004)}]{2004ApJ...617.1216L}
{Luhman}, K.~L. 2004, \apj, 617, 1216

\bibitem[{{Luhman} {et~al.}(2003){Luhman}, {Stauffer}, {Muench}, {Rieke},
  {Lada}, {Bouvier}, \& {Lada}}]{2003ApJ...593.1093L}
{Luhman}, K.~L., {Stauffer}, J.~R., {Muench}, A.~A., {Rieke}, G.~H., {Lada},
  E.~A., {Bouvier}, J., \& {Lada}, C.~J. 2003, \apj, 593, 1093

\bibitem[{{Luhman} {et~al.}(2007){Luhman}, {Adame}, {D'Alessio}, {Calvet},
  {McLeod}, {Bohac}, {Forrest}, {Hartmann}, {Sargent}, \&
  {Watson}}]{2007ApJ...666.1219L}
{Luhman}, K.~L., {et~al.} 2007, \apj, 666, 1219

\bibitem[{{Mohanty} {et~al.}(2005){Mohanty}, {Jayawardhana}, \&
  {Basri}}]{2005ApJ...626..498M}
{Mohanty}, S., {Jayawardhana}, R., \& {Basri}, G. 2005, ApJ, 626, 498

\bibitem[{{Mohanty} {et~al.}(2007){Mohanty}, {Jayawardhana}, {Hu{\'e}lamo}, \&
  {Mamajek}}]{2007ApJ...657.1064M}
{Mohanty}, S., {Jayawardhana}, R., {Hu{\'e}lamo}, N., \& {Mamajek}, E. 2007,
  \apj, 657, 1064

\bibitem[{{Morrow} {et~al.}(2008){Morrow}, {Luhman}, {Espaillat}, {D'Alessio},
  {Adame}, {Calvet}, {Forrest}, {Sargent}, {Hartmann}, {Watson}, \&
  {Bohac}}]{2008ApJ...676L.143M}
{Morrow}, A.~L., {et~al.} 2008, \apjl, 676, L143

\bibitem[{{Muzerolle} {et~al.}(2003){Muzerolle}, {Hillenbrand}, {Calvet},
  {Brice{\~n}o}, \& {Hartmann}}]{2003ApJ...592..266M}
{Muzerolle}, J., {Hillenbrand}, L., {Calvet}, N., {Brice{\~n}o}, C., \&
  {Hartmann}, L. 2003, \apj, 592, 266

\bibitem[{{Muzerolle} {et~al.}(2005){Muzerolle}, {Luhman}, {Brice\~{n}o},
  {Hartmann}, \& {Calvet}}]{2005ApJ...625..906M}
{Muzerolle}, J., {Luhman}, K., {Brice\~{n}o}, C., {Hartmann}, L., \& {Calvet},
  N. 2005, ApJ, 625, 906

\bibitem[{{Natta} {et~al.}(2004){Natta}, {Testi}, {Neri}, {Shepherd}, \&
  {Wilner}}]{2004A&A...416..179N}
{Natta}, A., {Testi}, L., {Neri}, R., {Shepherd}, D.~S., \& {Wilner}, D.~J.
  2004, \aap, 416, 179

\bibitem[{{Natta} {et~al.}(2006){Natta}, {Testi}, \&
  {Randich}}]{2006A&A...452..245N}
{Natta}, A., {Testi}, L., \& {Randich}, S. 2006, \aap, 452, 245

\bibitem[{{Padoan} {et~al.}(2005){Padoan}, {Kritsuk}, {Norman}, \&
  {Nordlund}}]{2005ApJ...622L..61P}
{Padoan}, P., {Kritsuk}, A., {Norman}, M.~L., \& {Nordlund}, {\AA}. 2005,
  \apjl, 622, L61

\bibitem[{{Payne} \& {Lodato}(2007)}]{2007MNRAS.381.1597P}
{Payne}, M.~J., \& {Lodato}, G. 2007, \mnras, 381, 1597

\bibitem[{{Reipurth} \& {Zinnecker}(1993)}]{1993A&A...278...81R}
{Reipurth}, B., \& {Zinnecker}, H. 1993, \aap, 278, 81

\bibitem[{{Riaz} \& {Gizis}(2008)}]{2008ApJ...681.1584R}
{Riaz}, B., \& {Gizis}, J.~E. 2008, \apj, 681, 1584

\bibitem[{{Riaz} {et~al.}(2012){Riaz}, {Lodato}, {Stamatellos}, \&
  {Gizis}}]{2012MNRAS.422L...6R}
{Riaz}, B., {Lodato}, G., {Stamatellos}, D., \& {Gizis}, J.~E. 2012, \mnras,
  422, L6

\bibitem[{{Ricci} {et~al.}(2013){Ricci}, {Isella}, {Carpenter}, \&
  {Testi}}]{2013ApJ...764L..27R}
{Ricci}, L., {Isella}, A., {Carpenter}, J.~M., \& {Testi}, L. 2013, \apjl, 764,
  L27

\bibitem[{{Ricci} {et~al.}(2010{\natexlab{a}}){Ricci}, {Testi}, {Natta}, \&
  {Brooks}}]{2010AA...521A..66R}
{Ricci}, L., {Testi}, L., {Natta}, A., \& {Brooks}, K.~J. 2010{\natexlab{a}},
  A\&A, 521, 66 (R10a)

\bibitem[{{Ricci} {et~al.}(2010{\natexlab{b}}){Ricci}, {Testi}, {Natta},
  {Neri}, {Cabrit}, \& {Herczeg}}]{2010AA...512A..15R}
{Ricci}, L., {Testi}, L., {Natta}, A., {Neri}, R., {Cabrit}, S., \& {Herczeg},
  G.~J. 2010{\natexlab{b}}, A\&A, 512, 15 (R10b)

\bibitem[{{Schaefer} {et~al.}(2009){Schaefer}, {Dutrey}, {Guilloteau}, {Simon},
  \& {White}}]{2009ApJ...701..698S}
{Schaefer}, G.~H., {Dutrey}, A., {Guilloteau}, S., {Simon}, M., \& {White},
  R.~J. 2009, \apj, 701, 698

\bibitem[{{Scholz} \& {Jayawardhana}(2006)}]{2006ApJ...638.1056S}
{Scholz}, A., \& {Jayawardhana}, R. 2006, \apj, 638, 1056

\bibitem[{{Scholz} {et~al.}(2006){Scholz}, {Jayawardhana}, \&
  {Wood}}]{2006ApJ...645.1498S}
{Scholz}, A., {Jayawardhana}, R., \& {Wood}, K. 2006, \apj, 645, 1498

\bibitem[{{Siess} {et~al.}(2000){Siess}, {Dufour}, \&
  {Forestini}}]{2000A&A...358..593S}
{Siess}, L., {Dufour}, E., \& {Forestini}, M. 2000, \aap, 358, 593

\bibitem[{{Sterzik} {et~al.}(2004){Sterzik}, {Pascucci}, {Apai}, {van der
  Bliek}, \& {Dullemond}}]{2004A&A...427..245S}
{Sterzik}, M.~F., {Pascucci}, I., {Apai}, D., {van der Bliek}, N., \&
  {Dullemond}, C.~P. 2004, \aap, 427, 245

\bibitem[{{Torres} {et~al.}(2003){Torres}, {Guenther}, {Marschall},
  {Neuh{\"a}user}, {Latham}, \& {Stefanik}}]{2003AJ....125..825T}
{Torres}, G., {Guenther}, E.~W., {Marschall}, L.~A., {Neuh{\"a}user}, R.,
  {Latham}, D.~W., \& {Stefanik}, R.~P. 2003, \aj, 125, 825

\bibitem[{{Uchida} {et~al.}(2004){Uchida}, {Calvet}, {Hartmann}, {Kemper},
  {Forrest}, {Watson}, {D'Alessio}, {Chen}, {Furlan}, {Sargent}, {Brandl},
  {Herter}, {Morris}, {Myers}, {Najita}, {Sloan}, {Barry}, {Green}, {Keller},
  \& {Hall}}]{2004ApJS..154..439U}
{Uchida}, K.~I., {et~al.} 2004, \apjs, 154, 439

\bibitem[{{Vorobyov} \& {Basu}(2008)}]{2008ApJ...676L.139V}
{Vorobyov}, E.~I., \& {Basu}, S. 2008, \apjl, 676, L139

\bibitem[{{Vorobyov} \& {Basu}(2009)}]{2009ApJ...703..922V}
---. 2009, \apj, 703, 922

\bibitem[{{Weinberger} {et~al.}(2002){Weinberger}, {Becklin}, {Schneider},
  {Chiang}, {Lowrance}, {Silverstone}, {Zuckerman}, {Hines}, \&
  {Smith}}]{2002ApJ...566..409W}
{Weinberger}, A.~J., {et~al.} 2002, \apj, 566, 409

\bibitem[{{Weintraub} {et~al.}(1989){Weintraub}, {Sandell}, \&
  {Duncan}}]{1989ApJ...340L..69W}
{Weintraub}, D.~A., {Sandell}, G., \& {Duncan}, W.~D. 1989, \apjl, 340, L69

\bibitem[{{Whelan} {et~al.}(2007){Whelan}, {Ray}, {Randich}, {Bacciotti},
  {Jayawardhana}, {Testi}, {Natta}, \& {Mohanty}}]{2007ApJ...659L..45W}
{Whelan}, E.~T., {Ray}, T.~P., {Randich}, S., {Bacciotti}, F., {Jayawardhana},
  R., {Testi}, L., {Natta}, A., \& {Mohanty}, S. 2007, \apjl, 659, L45

\bibitem[{{Whitney} {et~al.}(2003){Whitney}, {Wood}, {Bjorkman}, \&
  {Cohen}}]{2003ApJ...598.1079W}
{Whitney}, B.~A., {Wood}, K., {Bjorkman}, J.~E., \& {Cohen}, M. 2003, \apj,
  598, 1079

\bibitem[{{Zuckerman}(2001)}]{2001ARA&A..39..549Z}
{Zuckerman}, B. 2001, \araa, 39, 549

\end{thebibliography}
\clearpage

\begin{sidewaystable}
\centering
\caption{Sub-mm/mm Fluxes, Binarity and Derived Properties for Class II / cTTs in $\rho$ Oph, Taurus and the TWA}
{\scriptsize
\begin{tabular}{|l|l|l|l|l|l|l|l|l|l|l|l|l|l|l|l|l|l|l|l|l|l|l|l|l|}
\hline
ID\tablenotemark{(a)} & SpT & $T_{eff}$ & $M_{\ast}$ & $F_{[850]}$\tablenotemark{(b)} & 1$\sigma$ error\tablenotemark{(b)} & $F_{[1300]}$\tablenotemark{(b)} & 1$\sigma$ error\tablenotemark{(b)} & $\alpha_{850-1.3}$\tablenotemark{(c)} & $M_{d,\nu}$\tablenotemark{(d)} & 1$\sigma$ error & region & dist & Ref\tablenotemark{(e)} & mult\tablenotemark{(f)} & Ref\tablenotemark{(g)} \\
 &  & (K) & ($\msun$) & (mJy) & (mJy) & (mJy) & (mJy) & & ($M_{Jup}$) & ($M_{Jup}$) &  & (pc) &  & & \\
\hline
AS 205 & K5 & 4350 & 1.000 & 891 & 11 & 450 & 10 & 1.61 $\pm$ 0.06 & 59.0070 & 0.7285 & $\rho$ Oph & 150. & 1, 1  & w,sb & 1 \\
\hline
AS 209 & K5 & 4350 & 1.000 & 551 & 10 & 300 & 10 & 1.43 $\pm$ 0.09 & 36.4903 & 0.6623 & $\rho$ Oph & 150. & 1, 1  & - & - \\
\hline
DoAr 16 & K6 & 4205 & 0.850 & 47 & 8 & $<$50 & - & $> -0.15$ & 3.1126 & 0.5298 & $\rho$ Oph & 150. & 1, 1  & - & - \\
\hline
DoAr 24 & K5 & 4350 & 1.000 & - & - & $<$30 & - & - & $<$6.0340 & - & $\rho$ Oph & 150. & -, 1  & - & -\\
\hline
DoAr 24E & K1 & 5080 & 2.481 & 158 & 6 & 70 & 20 & 1.92 $\pm$ 0.68 & 10.4636 & 0.3974 & $\rho$ Oph & 150. & 1, 1  & w & 1 \\
\hline
DoAr 25 & K5 & 4350 & 1.000 & 461 & 11 & 280 & 10 & 1.17 $\pm$ 0.10 & 30.5300 & 0.7285 & $\rho$ Oph & 150. & 1, 1  & - & -\\
\hline
DoAr 28 & K5 & 4350 & 1.000 & - & - & $<$75 & - & - & $<$15.0851 & - & $\rho$ Oph & 150. & -, 1  & - & -\\
\hline
DoAr 32 & K6 & 4205 & 0.850 & - & - & $<$45 & - & - & $<$9.0511 & - & $\rho$ Oph & 150. & -, 1  & - & -\\
\hline
DoAr 33 & K4 & 4590 & 1.295 & 79 & 7 & 40 & 10 & 1.60 $\pm$ 0.62 & 5.2318 & 0.4636 & $\rho$ Oph & 150. & 1, 1  & - & -\\
\hline
DoAr 44 & K3 & 4730 & 1.538 & 181 & 6 & 105 & 11 & 1.28 $\pm$ 0.26 & 11.9868 & 0.3974 & $\rho$ Oph & 150. & 1, 1  & - & -\\
\hline
DoAr 52 & M2 & 3560 & 0.579 & - & - & $<$55 & - & - & $<$11.0624 & - & $\rho$ Oph & 150. & -, 1  & - & -\\
\hline
EL 18 & K6 & 4205 & 0.850 & - & - & $<$10 & - & - & $<$2.0113 & - & $\rho$ Oph & 150. & -, 1  & - & -\\
\hline
EL 24 & K6 & 4205 & 0.850 & 838 & 8 & 390 & 10 & 1.80 $\pm$ 0.06 & 55.4970 & 0.5298 & $\rho$ Oph & 150. & 1, 1  & - & -\\
\hline
EL 26 & M0 & 3850 & 0.691 & $<$51 & - & 15 & 5 & $< 2.88$ & 3.0170 & 1.0057 & $\rho$ Oph & 150. & 1, 1  & - & -\\
\hline
EL 27 & K8 & 3955 & 0.740 & 678 & 10 & 300 & 10 & 1.92 $\pm$ 0.09 & 44.9009 & 0.6623 & $\rho$ Oph & 150. & 1, 1  & - & -\\
\hline
EL 31 & M0 & 3850 & 0.691 & - & - & $<$10 & - & - & $<$2.0113 & - & $\rho$ Oph & 150. & -, 1  & - & -\\
\hline
EL 32 & K7 & 4060 & 0.785 & - & - & $<$50 & - & - & $<$10.0567 & - & $\rho$ Oph & 150. & -, 1  & - & -\\
\hline
EL 36 & A7 & 7850 & 3.639 & - & - & $<$10 & - & - & $<$2.0113 & - & $\rho$ Oph & 150. & -, 1  & - & -\\
\hline
GSS 26 & K8 & 3955 & 0.740 & 298 & 7 & 125 & 20 & 2.04 $\pm$ 0.38 & 19.7352 & 0.4636 & $\rho$ Oph & 150. & 1, 1  & - & -\\
\hline
GY 284 & M3 & 3415 & 0.478 & - & - & 130 & 10 & - & 26.1475 & 2.0113 & $\rho$ Oph & 150. & -, 1  & - & -\\
\hline
IRS 2 & K3 & 4730 & 1.538 & - & - & $<$25 & - & - & $<$5.0284 & - & $\rho$ Oph & 150. & -, 1  & - & -\\
\hline
IRS 37 & M4 & 3270 & 0.286 & 93 & 8 & $<$10 & - & $> 5.25$ & 6.1590 & 0.5298 & $\rho$ Oph & 150. & 1, 1  & - & -\\
\hline
IRS 39 & M2 & 3560 & 0.579 & 63 & 5 & $<$15 & - & $> 3.38$ & 4.1722 & 0.3311 & $\rho$ Oph & 150. & 1, 1  & - & -\\
\hline
IRS 49 & K8 & 3955 & 0.740 & 52 & 5 & 25 & 5 & 1.72 $\pm$ 0.52 & 3.4437 & 0.3311 & $\rho$ Oph & 150. & 1, 1  & - & -\\
\hline
RNO 90 & G5 & 5770 & 3.220 & 162 & 4 & 25 & 5 & 4.40 $\pm$ 0.47 & 10.7285 & 0.2649 & $\rho$ Oph & 150. & 1, 1  & - & -\\
\hline
ROXs 25 & M2 & 3560 & 0.579 & - & - & 30 & 5 & - & 6.0340 & 1.0057 & $\rho$ Oph & 150. & -, 1  & - & -\\
\hline
ROXs 42C & K6 & 4205 & 0.850 & - & - & $<$30 & - & - & $<$6.0340 & - & $\rho$ Oph & 150. & -, 1  & - & -\\
\hline
ROXs 43A & G0 & 6030 & 3.354 & - & - & $<$35 & - & - & $<$7.0397 & - & $\rho$ Oph & 150. & -, 1  & - & -\\
\hline
SR 4 & K5 & 4350 & 1.000 & 142 & 7 & 31 & 6 & 3.58 $\pm$ 0.47 & 9.4040 & 0.4636 & $\rho$ Oph & 150. & 1, 1  & - & -\\
\hline
SR 9 & K5 & 4350 & 1.000 & $<$25 & - & 15 & 5 & $< 1.20$ & 3.0170 & 1.0057 & $\rho$ Oph & 150. & 1, 1  & - & -\\
\hline
SR 10 & M2 & 3560 & 0.579 & - & - & $<$25 & - & - & $<$5.0284 & - & $\rho$ Oph & 150. & -, 1  & - & -\\
\hline
SR 13 & M4 & 3270 & 0.286 & 118 & 6 & 60 & 10 & 1.59 $\pm$ 0.41 & 7.8146 & 0.3974 & $\rho$ Oph & 150. & 1, 1  & c & 1 \\
\hline
SR 21 & G3 & 5830 & 3.251 & 397 & 6 & 95 & 15 & 3.37 $\pm$ 0.37 & 26.2916 & 0.3974 & $\rho$ Oph & 150. & 1, 1  & w & 2 \\
\hline
SR 22 & M4 & 3270 & 0.286 & 31 & 3 & $<$20 & - & $1.03$ & 2.0530 & 0.1987 & $\rho$ Oph & 150. & 1, 1  & - & -\\
\hline
Wa-Oph 4 & K4 & 4590 & 1.295 & - & - & $<$13 & - & - & $<$2.6148 & - & $\rho$ Oph & 150. & -, 1  & - & -\\
\hline
Wa-Oph 5 & M2 & 3560 & 0.579 & - & - & $<$25 & - & - & $<$5.0284 & - & $\rho$ Oph & 150. & -, 1  & - & -\\
\hline
Wa-Oph 6 & K6 & 4205 & 0.850 & 379 & 7 & 130 & 10 & 2.52 $\pm$ 0.19 & 25.0995 & 0.4636 & $\rho$ Oph & 150. & 1, 1  & - & -\\
\hline
WL 10 & K8 & 3955 & 0.740 & - & - & $<$60 & - & - & $<$12.0681 & - & $\rho$ Oph & 150. & -, 1  & - & -\\
\hline
WL 14 & M4 & 3270 & 0.286 & - & - & 30 & 10 & - & 6.0340 & 2.0113 & $\rho$ Oph & 150. & -, 1  & - & -\\
\hline
WL 18 & K7 & 4060 & 0.785 & - & - & 85 & 10 & - & 17.0965 & 2.0113 & $\rho$ Oph & 150. & -, 1  & w & 1 \\
\hline
WSB 19 & M3 & 3415 & 0.478 & $<$78 & - & - & - & - & $<$5.1656 & - & $\rho$ Oph & 150. & 1, 1  & - & -\\
\hline
WSB 37 & M5 & 3125 & 0.150 & - & - & $<$25 & - & - & $<$5.0284 & - & $\rho$ Oph & 150. & -, 1  & - & -\\
\hline
WSB 46 & M2 & 3560 & 0.579 & - & - & $<$20 & - & - & $<$4.0227 & - & $\rho$ Oph & 150. & -, 1  & - & -\\
\hline
WSB 49 & M4 & 3270 & 0.286 & - & - & $<$25 & - & - & $<$5.0284 & - & $\rho$ Oph & 150. & -, 1  & - & -\\
\hline
WSB 52 & K5 & 4350 & 1.000 & - & - & 51 & 10 & - & 10.2579 & 2.0113 & $\rho$ Oph & 150. & -, 1  & - & -\\
\hline
WSB 60 & M4 & 3270 & 0.286 & 149 & 7 & 89 & 2 & 1.21 $\pm$ 0.12 & 9.8676 & 0.4636 & $\rho$ Oph & 150. & 1, 1  & - & -\\
\hline
WSB 63 & M2 & 3560 & 0.579 & - & - & $<$25 & - & - & $<$5.0284 & - & $\rho$ Oph & 150. & -, 1  & - & -\\
\hline
YLW 16C & M1 & 3705 & 0.596 & - & - & 60 & 5 & - & 12.0681 & 1.0057 & $\rho$ Oph & 150. & -, 1  & - & -\\
\hline
\end{tabular}}
\end{sidewaystable}

\begin{sidewaystable}
\centering
{\scriptsize
\begin{tabular}{|l|l|l|l|l|l|l|l|l|l|l|l|l|l|l|l|l|l|l|l|l|l|l|l|l|l|}
\hline
ID\tablenotemark{(a)} & SpT & $T_{eff}$ & $M_{\ast}$ & $F_{[850]}$\tablenotemark{(b)} & 1$\sigma$ error\tablenotemark{(b)} & $F_{[1300]}$\tablenotemark{(b)} & 1$\sigma$ error\tablenotemark{(b)} & $\alpha_{850-1.3}$\tablenotemark{(c)} & $M_{d,\nu}$\tablenotemark{(d)} & 1$\sigma$ error & region & dist & Ref\tablenotemark{(e)} & mult\tablenotemark{(f)} & Ref\tablenotemark{(g)} \\
 &  & (K) & ($\msun$) & (mJy) & (mJy) & (mJy) & (mJy) & & ($M_{Jup}$) & ($M_{Jup}$) &  & (pc) & & &  \\
\hline
04113+2758 & M2 & 3560 & 0.579 & - & - & 410 & 40. & - & 71.8364 & 7.0084 & Taurus & 140. & -, 2 & w & 3 \\
\hline
04278+2253 & F1 & 7050 & 3.567 & 36. & 7. & - & - & - & 2.0768 & 0.4038 & Taurus & 140. & 2, - & - & -\\
\hline
AA Tau & K7 & 4060 & 0.785 & 144. & 5. & 88 & 9. & 1.16 $\pm$ 0.25 & 8.3073 & 0.2884 & Taurus & 140. & 2, 2 & - & -\\
\hline
AB Aur & A0 & 9520 & 3.791 & 359. & 67. & 103 & 18. & 2.94 $\pm$ 0.60 & 20.7107 & 3.8652 & Taurus & 140. & 2, 2 & - & -\\
\hline
BP Tau & K7 & 4060 & 0.785 & 130. & 7. & 47 & 0.7 & 2.39 $\pm$ 0.13 & 7.4997 & 0.4038 & Taurus & 140. & 2, 2 & - & -\\
\hline
CFHT-BDTau 4 & M7 & 2880 & 0.057 & 10.8 & 1.8 & 2.38 & 0.75 & $3.56 \pm 0.84$ & 0.6230 & 0.1038 & Taurus & 140. & 3, 4 & - & -\\
\hline
$^\ast$CFHT-BDTau 12 & M6.5 & 2935 & 0.074 & 4.06 & 1.32 & - & - & - & 0.2342 & 0.0762 & Taurus & 140. & 5, - & - & -\\
\hline
CIDA 1 & M5.5 & 3058 & 0.125 & - & - & 13.5 & 2.8 & - & 2.3653 & 0.4906 & Taurus & 140. & -, 6 & - & -\\
\hline
CIDA 3 & M2 & 3560 & 0.579 & $<$9. & - & - & - & - & $<$0.5192 & - & Taurus & 140. & 2, - & - & -\\
\hline
CIDA 8 & M4 & 3270 & 0.286 & 27. & 3. & - & - & - & 1.5576 & 0.1731 & Taurus & 140. & 2, - & - & -\\
\hline
CIDA 9 & M0 & 3850 & 0.691 & 71. & 7. & - & - & - & 4.0960 & 0.4038 & Taurus & 140. & 2, - & w & 4 \\
\hline
CIDA 11 & M3 & 3415 & 0.478 & $<$8. & - & - & - & - & $<$0.4615 & - & Taurus & 140. & 2, - & c & 4 \\
\hline
CIDA 12 & M4 & 3270 & 0.286 & $<$7. & - & - & - & - & $<$0.4038 & - & Taurus & 140. & 2, - & - & -\\
\hline
CIDA 14 & M5 & 3125 & 0.150 & - & - & $<$4.5 & - & - & $<$0.7884 & - & Taurus & 140. & -, 6 & - & -\\
\hline
CI Tau & K7 & 4060 & 0.785 & 324. & 6. & 190 & 17. & 1.26 $\pm$ 0.22 & 18.6915 & 0.3461 & Taurus & 140. & 2, 2 & - & -\\
\hline
CoKu Tau/1 & M0 & 3850 & 0.691 & 35. & 7. & $<$12 & - & $> 2.52$ & 2.0191 & 0.4038 & Taurus & 140. & 2, 2 & - & -\\
\hline
CoKu Tau/3 & M1 & 3705 & 0.596 & $<$8. & - & $<$16 & - & - & $<$0.4615 & - & Taurus & 140. & 2, 2 & w & 4 \\
\hline
CoKu Tau/4 & M2 & 3560 & 0.579 & 9.0 & 2.9 & $<$15 & - & $> -1.20$ & 0.5192 & 0.1673 & Taurus & 140. & 2, 2 & c & 4 \\
\hline
CW Tau & K2 & 4900 & 1.971 & 66. & 6. & 96 & 8. & $-0.88 \pm 0.29$ & 3.8075 & 0.3461 & Taurus & 140. & 2, 2 & - & -\\
\hline
CX Tau & M3 & 3415 & 0.478 & 25. & 6. & $<$40 & - & $> -1.11$ & 1.4422 & 0.3461 & Taurus & 140. & 2, 2 & - & -\\
\hline
CZ Tau & M2 & 3560 & 0.579 & $<$9. & - & $<$30 & - & - & $<$0.5192 & - & Taurus & 140. & 2, 2 & c & 4 \\
\hline
DD Tau & M1 & 3705 & 0.596 & $<$42. & - & 17 & 4. & $< 2.13$ & 2.9786 & 0.7008 & Taurus & 140. & 2, 2 & c & 4 \\
\hline
DE Tau & M2 & 3560 & 0.579 & 90. & 7. & 36 & 5. & $2.16 \pm 0.37$ & 5.1921 & 0.4038 & Taurus & 140. & 2, 2 & - & -\\
\hline
DF Tau & M1 & 3705 & 0.596 & 8.8 & 1.9 & $<$25 & - & $> -2.46$ & 0.5077 & 0.1096 & Taurus & 140. & 2, 2 & c & 4 \\
\hline
$^{\ddag}$ DH Tau & M1 & 3705 & 0.596 & 57. & 9. & $<$57 & - & $> 0.00$ & 3.2883 & 0.5192 & Taurus & 140. & 2, 2 & w & 2 \\
\hline
DK Tau & K7 & 4060 & 0.785 & 80. & 10. & 35 & 7. & $1.95 \pm 0.56$ & 4.6152 & 0.5769 & Taurus & 140. & 2, 2 & w & 4 \\
\hline
DL Tau & K7 & 4060 & 0.785 & 440. & 40. & 230 & 14. & $1.53 \pm 0.26$ & 25.3835 & 2.3076 & Taurus & 140. & 2, 2 & - & -\\
\hline
DM Tau & M1 & 3705 & 0.596 & 237. & 12. & 109 & 13. & $1.83 \pm 0.30$ & 13.6725 & 0.6923 & Taurus & 140. & 2, 2 & - & -\\
\hline
DN Tau & M0 & 3850 & 0.691 & 201. & 7. & 84 & 13. & $2.05 \pm 0.37$ & 11.5957 & 0.4038 & Taurus & 140. & 2, 2 & - & -\\
\hline
DO Tau & M0 & 3850 & 0.691 & 258. & 42. & 136 & 11. & $1.51 \pm 0.43$ & 14.8840 & 2.4230 & Taurus & 140. & 2, 2 & - & -\\
\hline
DP Tau & M1 & 3705 & 0.596 & $<$10. & - & $<$27 & - & - & $<$0.5769 & - & Taurus & 140. & 2, 2 & c & 4 \\
\hline
DQ Tau & M0 & 3850 & 0.691 & 208. & 8. & 91 & 9. & $1.95 \pm 0.25$ & 11.9995 & 0.4615 & Taurus & 140. & 2, 2 & sb & 3 \\
\hline
DR Tau & K5 & 4350 & 1.000 & 533. & 7. & 159 & 11. & $2.85 \pm 0.17$ & 30.7487 & 0.4038 & Taurus & 140. & 2, 2 & - & -\\
\hline
FM Tau & M0 & 3850 & 0.691 & 32. & 8. & $<$36 & - & $> -0.28$ & 1.8461 & 0.4615 & Taurus & 140. & 2, 2 & - & -\\
\hline
FN Tau & M5 & 3125 & 0.150 & - & - & $<$17.5 & - & - & $<$3.0662 & - & Taurus & 140. & -, 6 & - & -\\
\hline
FO Tau & M2 & 3560 & 0.579 & 13. & 3. & $<$14 & - & $> -0.17$ & 0.7500 & 0.1731 & Taurus & 140. & 2, 2 & c & 4 \\
\hline
FP Tau & M4 & 3270 & 0.286 & - & - & $<$9.3 & - & - & $<$1.6295 & - & Taurus & 140. & -, 6 & - & -\\
\hline
$^{\ddag}$ FV Tau & K5 & 4350 & 1.000 & 48. & 5. & 15 & 4. & $2.74 \pm 0.67$ & 2.7691 & 0.2884 & Taurus & 140. & 2, 2 & w & 4 \\
\hline
$^{\ddag}$ FV Tau/c & M4 & 3270 & 0.286 & $<$25. & - & $<$16 & - & - & $<$1.4422 & - & Taurus & 140. & 2, 2 & w & 4 \\
\hline
FX Tau & M1 & 3705 & 0.596 & 17. & 3. & $<$30 & - & $> -1.34$ & 0.9807 & 0.1731 & Taurus & 140. & 2, 2 & w & 4 \\
\hline
$^{\ddag}$ FY Tau & K7 & 4060 & 0.785 & $<$27. & - & 16 & 5. & $< 1.23$ & 2.8034 & 0.8761 & Taurus & 140. & 2, 2 & - & -\\
\hline
$^{\ddag}$ GG Tau A & K7 & 4060 & 0.785 & 1255. & 57. & 593 & 53. & $1.76 \pm 0.24$ & 72.4007 & 3.2883 & Taurus & 140. & 2, 2 & c & 4 \\
\hline
$^{\ddag}$ GH Tau & M2 & 3560 & 0.579 & 15. & 3. & $<$30 & - & $> -1.63$ & 0.8653 & 0.1731 & Taurus & 140. & 2, 2 & c & 4 \\
\hline
$^{\ddag}$ GK Tau & K7 & 4060 & 0.785 & 33. & 7. & $<$21 & - & $> 1.06$ & 1.9038 & 0.4038 & Taurus & 140. & 2, 2 & - &  -\\
\hline
GM Aur & K3 & 4730 & 1.538 & - & - & 253 & 12. & - & 44.3283 & 2.1025 & Taurus & 140. & -, 2 & - & -\\
\hline
$^\ast$GM Tau & M6.5 & 2935 & 0.074 & 0.86 & 0.87 & $<$4.8 & - & - & 0.0496 & 0.0502 & Taurus & 140. & 5, 6 & - & -\\
\hline
GN Tau & M2.5 & 3488 & 0.571 & 12. & 3. & $<$50 & - & $> -3.36$ & 0.6923 & 0.1731 & Taurus & 140. & 2, 2 & c & 4 \\
\hline
GO Tau & M0 & 3850 & 0.691 & 173. & 7. & 83 & 12. & $1.73 \pm 0.35$ & 9.9803 & 0.4038 & Taurus & 140. & 2, 2 & - & -\\
\hline
\end{tabular}}
\end{sidewaystable}
\begin{sidewaystable}
\centering
{\scriptsize
\begin{tabular}{|l|l|l|l|l|l|l|l|l|l|l|l|l|l|l|l|l|l|l|l|l|l|l|l|l|}
\hline
ID\tablenotemark{(a)} & SpT & $T_{eff}$ & $M_{\ast}$ & $F_{[850]}$\tablenotemark{(b)} & 1$\sigma$ error\tablenotemark{(b)} & $F_{[1300]}$\tablenotemark{(b)} & 1$\sigma$ error\tablenotemark{(b)} & $\alpha_{850-1.3}$\tablenotemark{(c)} & $M_{d,\nu}$\tablenotemark{(d)} & 1$\sigma$ error & region & dist & Ref\tablenotemark{(e)} & mult\tablenotemark{(f)} & Ref\tablenotemark{(g)} \\
 &  & (K) & ($\msun$) & (mJy) & (mJy) & (mJy) & (mJy) & & ($M_{Jup}$) & ($M_{Jup}$) &  & (pc) & & &  \\
\hline
Haro 6-37 & K6 & 4205 & 0.850 & 245. & 7. & $<$88 & - & $> 2.41$ & 14.1340 & 0.4038 & Taurus & 140. & 2, 2 & m & 4 \\
\hline
HN Tau & K5 & 4350 & 1.000 & 29. & 3. & $<$15 & - & $> 1.55$ & 1.6730 & 0.1731 & Taurus & 140. & 2, 2 & w & 4 \\
\hline
HO Tau & M1 & 3705 & 0.596 & 44. & 6. & $<$30 & - & $> 0.90$ & 2.5384 & 0.3461 & Taurus & 140. & 2, 2 & w & 3 \\
\hline
IQ Tau & M1 & 3705 & 0.596 & 178. & 3. & 87 & 11. & $1.68 \pm 0.30$ & 10.2688 & 0.1731 & Taurus & 140. & 2, 2 & - & -\\
\hline
IS Tau & K7 & 4060 & 0.785 & 30. & 3. & $<$20 & - & $> 0.95$ & 1.7307 & 0.1731 & Taurus & 140. & 2, 2 & c & 4 \\
\hline
IT Tau & K2 & 4900 & 1.971 & 22. & 3. & $<$33 & - & $> -0.95$ & 1.2692 & 0.1731 & Taurus & 140. & 2, 2 & w & 4 \\
\hline
J041411+2811 & M6.25 & 2963 & 0.086 & - & - & 0.91 & 0.65 & - & 0.1594 & 0.1139 & Taurus & 140. & -, 4 & - & -\\
\hline
J043814+2611 & M7.25 & 2838 & 0.049 & - & - & 2.29 & 0.75 & - & 0.4012 & 0.1314 & Taurus & 140. & -, 4 & - & -\\
\hline
J043903+2544 & M7.25 & 2838 & 0.049 & - & - & 2.86 & 0.76 & - & 0.5011 & 0.1332 & Taurus & 140. & -, 4 & - & -\\
\hline
J044148+2534 & M7.75 & 2753 & 0.035 & - & - & 2.64 & 0.64 & - & 0.4626 & 0.1121 & Taurus & 140. & -, 4 & - & -\\
\hline
$^\ast$J044427+2512 & M7.25 & 2838 & 0.049 & 9.85 & 0.76 & 7.55 & 0.89 & $0.63 \pm 0.33$ & 0.5682 & 0.0438 & Taurus & 140. & 5, 4 & - & -\\
\hline
JH 112 & K6 & 4205 & 0.850 & 30. & 10. & $<$18 & - & $> 1.20$ & 1.7307 & 0.5769 & Taurus & 140. & 2, 2 & m & 4 \\
\hline
JH 223 & M2 & 3560 & 0.579 & $<$7. & - & $<$19 & - & - & $<$0.4038 & - & Taurus & 140. & 2, 2 & w & 4 \\
\hline
KPNO Tau 6 & M8.5 & 2555 & 0.020 & - & - & -0.66 & 0.79 & - & -0.1156 & 0.1384 & Taurus & 140. & -, 4 & - & -\\
\hline
KPNO Tau 7 & M8.25 & 2633 & 0.025 & - & - & 0.70 & 0.88 & - & 0.1226 & 0.1542 & Taurus & 140. & -, 4 & - & -\\
\hline
KPNO Tau 12 & M9 & 2400 & 0.014 & - & - & -0.92 & 0.70 & - & -0.1612 & 0.1226 & Taurus & 140. & -, 4 & - & -\\
\hline
LkCa-15 & K5 & 4350 & 1.000 & 428. & 11. & 167 & 6. & $2.22 \pm 0.10$ & 24.6912 & 0.6346 & Taurus & 140. & 2, 2 & - & -\\
\hline MHO 5 & M6 & 2990 & 0.096 & - & - & $<$9.0 & - & - & $<$1.5769 & - & Taurus & 140. & -, 6 & - & -\\
\hline RW Aur & K3 & 4730 & 1.538 & 79. & 4. & 42 & 5. & $1.49 \pm 0.30$ & 4.5575 & 0.2308 & Taurus & 140. & 2, 2 & w & 4 \\
\hline
RY Tau & K1 & 5080 & 2.481 & 560. & 30. & 229 & 17. & $2.10 \pm 0.22$ & 32.3063 & 1.7307 & Taurus & 140. & 2, 2 & - & -\\
\hline
St 34 & M3 & 3415 & 0.478 & $<$11. & - & $<$15 & - & - & $<$0.6346 & - & Taurus & 140. & 2, 2 & sb,w & 3, 4 \\
\hline
SU Aur & G2 & 5860 & 3.267 & 74. & 3. & $<$30 & - & $> 2.12$ & 4.2690 & 0.1731 & Taurus & 140. & 2, 2 & - & -\\
\hline
T Tau & K0 & 5250 & 2.847 & 628. & 17. & 280 & 9. & $1.90 \pm 0.10$ & 36.2292 & 0.9807 & Taurus & 140. & 2, 2 & m & 3 \\
\hline
UX Tau & K2 & 4900 & 1.971 & 173. & 3. & 63 & 10. & $2.38 \pm 0.38$ & 9.9803 & 0.1731 & Taurus & 140. & 2, 2 & m & 4 \\
\hline
UY Aur & K7 & 4060 & 0.785 & 102. & 6. & 29 & 6. & $2.96 \pm 0.51$ & 5.8844 & 0.3461 & Taurus & 140. & 2, 2 & w & 4 \\
\hline
UZ Tau & M1 & 3705 & 0.596 & 560. & 7. & 172 & 15. & $2.78 \pm 0.21$ & 32.3063 & 0.4038 & Taurus & 140. & 2, 2 & m & 3, 4 \\
\hline
V410 Anon 13 & M5.75 & 3024 & 0.109 & - & - & $<$9.0 & - & - & $<$1.5769 & - & Taurus & 140. & -, 6 & - & -\\
\hline
$^{\ddag}$ V710 Tau & M1 & 3705 & 0.596 & 152. & 6. & 60 & 7. & $2.19 \pm 0.29$ & 8.7689 & 0.3461 & Taurus & 140. & 2, 2 & w & 3, 4\\
\hline
V836 Tau & K7 & 4060 & 0.785 & 74. & 3. & 37 & 6. & $1.63 \pm 0.39$ & 4.2690 & 0.1731 & Taurus & 140. & 2, 2 & - & -\\
\hline
V892 Tau & A0 & 9520 & 3.791 & 638. & 54. & 234 & 19. & $2.36 \pm 0.28$ & 36.8061 & 3.1153 & Taurus & 140. & 2, 2 & w & 3 \\
\hline
$^{\ddag}$ V955 Tau & M0 & 3850 & 0.691 & 14. & 2. & $<$19 & - & $> -0.72$ & 0.8077 & 0.1154 & Taurus & 140. & 2, 2 & c & 4 \\
\hline
VY Tau & M0 & 3850 & 0.691 & $<$10. & - & $<$17 & - & - & $<$0.5769 & - & Taurus & 140. & 2, 2 & c & 4 \\
\hline
\hline
$^\ast$2M1207-3932 & M8 & 2710 & 0.035 & 0.07 & 1.75 & - & - & - & 0.0006 & 0.0141 & TWA & 52.4 & 5, - & c & 5 \\
\hline
Hen3-600A & M4 & 3270 & 0.218 & 65. & 5. & - & - & - & 0.4783 & 0.0368 & TWA & 50. & 7, - & c,sb & 6 \\
\hline
$^\ast$SSPM1102-3431 & M8 & 2710 & 0.035 & -0.94 & 1.57 & - & - & - & -0.0084 & 0.0141 & TWA & 55.2 & 5, - & - & -\\
\hline
$^\ast$TWA 30A & M5 & 3125 & 0.114 & 1.70 & 1.63 & - & - & - & 0.0088 & 0.0085 & TWA & 42. & 5, - & w & 7 \\
\hline
$^\ast$TWA 30B & M4 & 3270 & 0.218 & 0.013 & 1.86 & - & - & - & 6.7e-5 & 0.0097 & TWA & 42. & 5, - & w & 7 \\
\hline
TW Hya & K6 & 4205 & 0.949 & 1450. & 310. & 874.\tablenotemark{h} & 54. & $1.19 \pm 0.52$ & 13.3840 & 2.8614 & TWA & 56. & 8, 8 & - & -\\
\hline
\end{tabular}}
{\tiny
\begin{flushleft}
\footnotetext[1] { \,\,`$\ast$' marks sources observed in this paper, at 850$\mu$m on SCUBA-2. $^{\ddag}$ marks sources with more binarity information in footnote [13] of the main text.}
\footnotetext[2]{Measured fluxes and 1$\sigma$ errors, or 3$\sigma$ upper limits (for sources from the literature where the mesured value is not cited). } 
\footnotetext[3]{ \,\,$M_{d,[850]}$ if either a measured value of the 850\,$\mu$m flux is available, or the upper limit on $M_{d,[850]}$ is smaller than the upper limit on $M_{d,[1300]}$; otherwise $M_{d,[1300]}$.  Computed using equation (9).  These are the values used in our Bayesian analysis. } 
\footnotetext[4]{Measured values of $\alpha$ with 1$\sigma$ error bars, or 3$\sigma$ upper or lower limits. See equation (8) and discussion, in \S5. }
\footnotetext[5]{References for fluxes at 850\,$\mu$m (1$^{st}$ ref) and 1.3\,mm (2$^{nd}$ ref). 1: AW07 (and references therein); 2: AW05 (and references therein); 3: Klein et al.\,(2003); 4: Scholz et al.\,(2006); 5: this study (these sources are also marked with an asterisk); 6: Schaefer et al.\,(2009); 7: Zuckerman (2001); 8: \citet{1989ApJ...340L..69W}.   }
\footnotetext[6]{\,\,c = close binary (sep $<$ 100\,AU), w = wide binary (sep $\geq$ 100\,AU), sb = spectroscopic binary, m = multiple.  }
\footnotetext[7]{ References for multiplicity information. 1: R10a (and references therein); 2: \citet{1993A&A...278...81R}; 3: AW05 (and references therein); 4: \citet[and references therein]{2011ApJ...731....8K}; 5: \citet{2005A&A...438L..25C}; 6: Andrews et al.\,(2010, and references therein); 7: Looper et al.\,(2010a). See also footnote [13] in the main text, for sources IDs marked with $^{\ddag}$.  } 
\footnotetext[8]{ For TW Hya, this is the flux (and 1$\sigma$ error) at 1.1\,mm (Weintraub et al.\,1989); we have cited it here for completeness, since close to 1.3\,mm. }
\end{flushleft}
}
\end{sidewaystable}

\clearpage
\begin{appendix}
\section{Disk Spectral Energy Distribution}
We outline the theory of disk SEDs, initially following the exposition by B90.  The disk flux density at a frequency $\nu$, measured by an observer a distance $D$ away, is
$$ F_\nu = \frac{{\rm cos}\,i}{D^2}\int_{r_0}^{R_d}B_\nu (T) (1 - e^{-\tau_\nu})\,2\pi r dr \eqno({\rm A}1) $$
where $i$ is the inclination of the disk relative to the observer (with $i = 90^\circ$ for edge-on), $B_\nu(T)$ is the Planckian specific intensity emitted by a locally blackbody disk with temperature $T(r)$ at a radial distance $r$ from the central star, $\tau_\nu(r)$ is the line-of-sight optical depth of the emitting material, and $r_0$ and $R_d$ are the disk inner and outer edge radii respectively.  This formula is only valid if the source function (=$B_\nu(T)$) is roughly constant with optical depth through the disk, and thus inapplicable for very large viewing angles ($i \rightarrow 90^\circ$); we assume that the disk is sufficiently far from edge-on to satisfy this constraint.  The optical depth is then $\tau_\nu(r) = \kappa_\nu \Sigma(r)/{{\rm cos}\,i}$, where $\Sigma(r)$ is the surface density of the disk at $r$ (so that the total disk mass is $M_d = \int_{r_0}^{R_d} \Sigma(r) \,2\pi r dr$), and $\kappa_\nu$ is the total opacity (gas$+$dust) of the emitting material (i.e., the actual opacity of the emitting dust grains scaled by the gas-to-dust ratio).  We assume the surface density and temperature are power-laws in radius, $\Sigma(r) = \Sigma_0\left({r}/{r_0}\right)^{-p}$ and $T(r) = T_0\left({r}/{r_0}\right)^{-q}$, where $\Sigma_0$ and $T_0$ are the values at the disk inner edge $r_0$.  

Since the surface density declines with increasing radius, the inner regions of the disk are more optically thick than the outer parts at any frequency, by the definition of $\tau_\nu$.  For a given $\nu$, the change from optically thick ($\tau_\nu \gg 1$) to optically thin ($\tau_\nu \ll 1$) conditions occurs at a radius $r_1$, found by setting $\tau_\nu (r_1) \sim 1$:
$$ r_1 \,=\, \tau_0^{1/p} r_0 \,\approx\, \left[\frac{(2-p){\bar{\tau}}_\nu}{2}\right]^{1/p}R_d \eqno({\rm A}2) $$
Here $\tau_0$ is the optical depth at the inner edge $r_0$, and ${\bar{\tau}}_\nu$ is the average optical depth in the disk at $\nu$: $ {\bar{\tau}}_\nu \equiv \kappa_\nu \,M_d/(\pi R_d^2\,{\rm cos}\,i)$.  The latter uses the definition of $M_d$ in terms of the integral of $\Sigma(r)$, and assumes that $R_d \gg r_0$ (i.e., we have a disk, not a narrow annulus).  

Lastly, for a disk with some global minimum temperature $T_{min}$, the flux densities emitted from all radii lie in the Rayleigh-Jeans (RJ) limit for all frequencies $\nu \ll kT_{min}/h$. {\it If} the observed frequencies are in this regime, then the Planck spectrum reduces to $B_\nu (T)_{\rm{RJ}} \approx (2\nu^2/c^2)kT$.     

Expressing equation (A1) as the sum of two integrals representing the flux densities from optically thick radii ($r_0 \leq r \leq r_1$) and optically thin ones ($r_1 \leq r \leq R_d$), noting that the attenuation factor $(1-e^{-\tau_\nu})$ becomes $\sim$1 for $\tau_\nu \gg 1$ and $\sim$$\tau_\nu$ for $\tau_\nu \ll 1$, and {\it stipulating} that we are in the RJ regime, finally yields
$$ F_\nu \approx \nu^2 \left(\frac{4\pi k}{c^2}\right) f_0 \left(\frac{{\rm cos}\,i}{D^2}\right) \left[T_0 r_0^q \frac{(2-p){\bar\tau}_\nu}{2}\,R_d^p\right]\left[\frac{R_d^{2-p-q}-r_1^{2-p-q}}{2-p-q}\right] (1+\Delta) \eqno({\rm A}3) $$
where the dimensionless factor $f_0$ corrects for our having neglected the curvature of $(1-e^{-\tau_\nu})$ over the transition from optically thick to thin regimes; $f_0 \sim 0.8$ yields good agreement between equation (A3) and numerical integration of equation (A1) in the RJ limit (B90).  The product of all the terms outside the last parentheses is the optically thin contribution to the flux density, while the term $\Delta$ is the ratio of the optically thick to thin contributions:
$$ \Delta \equiv \left[\frac{2}{(2-p){\bar\tau}_\nu\,R_d^p}\right] \left(\frac{2-p-q}{2-q}\right) \left(\frac{r_1^{2-q} - r_0^{2-q}}{R_d^{2-p-q} - r_1^{2-p-q}}\right) \eqno({\rm A}4) $$
Note that B90 further simplify the product of the two square brackets in equation (A3); their expression implicitly assumes that the $r_1$ derived from equation (A2) satisfies $r_0 \leq r_1 \leq R_d$.  If however equation (A2) formally yields $r_1 < r_0$, then the entire disk is optically thin, and we must set $r_1 \equiv r_0$ for a physical solution to equation (A3); similarly, if formally $r_1 > R_d$, then all radii are optically thick, and we must set $r_1 \equiv R_d$.  In either case, B90's contraction does not hold, but our explicit equation (A3) remains valid.  

If the opacity is moreover a power-law in frequency, $\kappa_\nu = \kappa_f \,\left({\nu}/{\nu_f}\right)^\beta$, where $\kappa_f$ is the opacity at some fiducial frequency $\nu_f$, and the spectral index is defined as $\alpha \equiv {d({\rm ln}F_\nu)}/{d({\rm ln}\,\nu)}$, then equation (A3) can be manipulated to find $ \alpha \approx 2 + \left({\beta}/{1+\Delta}\right)$ in the RJ limit. 

We now depart from B90's discussion.  If the RJ approximation is {\it not} valid, then equations (A3) and (A4) fail.  General expressions for $F_{\nu}$ and $\Delta$ can still be derived by dividing the disk into optically thick and thin parts, as above, but now using the general Planck formula instead of its RJ limit.  The results are cited in equations (5) and (6) in \S4.

\section{Fiducial Disk Parameters}
We assume $i \approx 60^\circ$, corresponding to the mean value of cos($i$) for a random distribution of orientations (note that for optically thin disks, $F_\nu$ becomes independent of $i$).  We also adopt $r_0 \approx 5\rstar$, the usual value for a magnetospherically truncated inner edge (the precise location of the hot inner edge has negligible impact on the long wavelength emission).  The location of the outer edge $R_d$ is harder to constrain; we adopt a fiducial value of $R_d = 100$\,AU, but also examine a large range of $R_d = 10-300$\,AU, consistent with most sources in resolved observations of stars (e.g., AW07; Isella et al.\,2009; Andrews et al.\,2010), and with the few available constraints for BDs (Scholz et al.\,2006; Luhman et al.\,2009).  

For the Minimum Mass Solar Nebula, Hayashi (1981) derived a surface density exponent of $p = 1.5$.  More recent spatially resolved observations of disks all imply shallower radial profiles, mostly in the range 0.5--$<$1.5 (e.g., Wilner et al.\,2000; Isella et al.\,2009; Andrews \& Wiiliams 2007b; Andrews et al.\,2009, 2010).  Specifically, Andrews et al.\,(2010) find that the surface densities are consistent with a uniform exponent of $\sim$1 independent of stellar mass, for solar-type to intermediate-mass stars ($\sim$0.3--4\,$\msun$).  We thus adopt a fiducial $p = 1$, and discuss the effects of varying this over the range 0.5--1.5 at appropriate junctures.  

Finally, both the normalization of the temperature ($T_0$ here) and its radial exponent $q$ can be estimated from SED fits to optically thick emission in the mid- to far-infrared (see B90).  AW05 do this for sources in their sample with infrared data, to find median values of $q = 0.58$ and $T$(1AU) = 148\,K (they choose to normalize at 1\,AU instead of at the inner-edge $r_0$).  This $q$ is intermediate between that for flat disks, $q = 3/4$, and fully flared ones, $q = 3/7$ (Chiang \& Goldreich 1997, hereafter CG97), suggesting some degree of dust-settling.  While the majority of stars AW05 examine in this regard are solar-types, a significant fraction of the small number of intermediate mass stars they analyse also evince intermediate $q$.  Similarly, VLMS/BD disks exhibit mid-infrared signatures of disk flattening as well (Apai et al.\,2005; Scholz et al.\,2006).  We thus adopt AW05's median $q = 0.58$ as a reasonable fiducial value.  

Furthermore, we scale AW05's temperature normalization, applicable mainly to solar-type stars, to the wide range of stellar/BD masses in our sample as follows.  We assume that the disks are passive, i.e., heated predominantly by stellar irradiation instead of by accretion (a good assumption for average stellar/BD accretion rates in the Class II phase).  The general expression for the disk effective temperature at a radius $r$ is then (CG97): $T(r) = (\alpha/2)^{1/4} (\rstar/r)^{1/2}\, \tstar$, where $\alpha$ is the grazing angle at which stellar photons impinge on the disk at $r$.  {\it Assuming} now that $\alpha$ is roughly constant at the disk inner edge for all stellar masses (justified below), and recalling that we adopt $r_0/\rstar \approx 5$ in all cases, then yields $T_0 \propto \tstar$, with the constant of proportionality independent of stellar parameters.  Moreover, for solar-type stars with fiducial values $\tstar \sim 4000$\,K and $\rstar \sim 2\rsun$ (\S3), AW05's median $T$(1\,AU) and $q$ (derived mainly from solar-types) imply $T_0 ({\rm solar \,\, types}) = 148 \,(5\rstar/1\,{\rm AU})^{-0.58}$\,K $\approx 880$\,K.  The disk temperature normalization at $r_0$ for any stellar temperature $\tstar$ is then 
$$T_0 \approx 880 \,(\tstar/4000\,{\rm K})\,{\rm K} \eqno({\rm B1}) $$   
In fact, the $T_0$ derived this way turn out to be very close to those expected for a flat disk at $r_0$ ($T_0 \approx 0.2^{1/4} (\rstar/r_0)^{3/4} \,\tstar$ ; CG97).  In the latter case, $\alpha \approx 0.4\rstar/r_0$, which is constant for all stars with our choice of $r_0$.  Since passive disks should indeed be nearly flat very close to the star (where the flaring term in $\alpha$ is negligible; CG97), we see that our assumption of a constant $\alpha$ at our chosen $r_0$, and hence adoption of equation (B1), is self-consistent.  

Note that for the range of $\tstar$ covering most of our sample, $\sim$2500--5500\,K, the $T_0$ we derive are lower than the dust sublimation temperature $\sim$1500\,K; our adoption of $r_0 = 5\rstar$ (set by magnetospheric truncation) as the inner edge of the dust disk is thus formally self-consistent (in the sense that dust can exist at these temperatures).  In reality, the disk may be significantly hotter here due to accretion and/or frontal heating of the disk inner wall; the real inner edge  of the dust disk may then be set by sublimation instead, and lie at somewhat larger radii (e.g., Eisner et al.\,2005).  However, the hot emission from regions close to $r_0$ contributes negligibly to $F_\nu$ in the sub-mm/mm, so the precise temperature and location of $r_0$ is not crucial in itself to our calculations.  The true importance of $T_0$ for us is as a {\it scaling factor} for the temperatures at {\it larger} (cooler) radii, which do contribute to the sub-mm/mm emission.  Heating at {\it these} radii is expected to be dominated by stellar irradiation instead of accretion (CG97, D'Alessio et al.\,1999), and not subject to frontal irradiation effects either, so the temperatures here can be self-consistently derived by scaling under the assumption of a passive disk {\it throughout} (tapering to a small grazing angle at $r_0$), as we have done.  

Lastly, the disk temperature cannot continue decreasing as a power-law indefinitely; eventually, other heating sources, such as cosmic rays, radionuclides and the interstellar radiation field (ISRF) must dominate over the stellar irradiation.  Cosmic rays and radionuclides do not appear important in this context (e.g., for a disk around a solar-type star, heating from these two sources becomes important only when the temperature due to stellar irradiation has formally dropped to $<$3\,K, i.e., below the cosmic microwave background temperature; see D'Alessio et al.\,1998). The ISRF, however, is critical.  Specifically, Mathis et al.\,(1983) calculate the temperature of grains in both the interstellar medium (ISM) and in Giant Molecular Clouds (GMCs) due to ISRF heating.  Given the very high visual extinction to the midplane of accretion disks, the GMC case is more germane to our analysis.  For their calculated ISRF near the galactocentric distance of the Sun, they find a grain temperature of $\sim$8--15\,K, for silicate and graphite grains respectively.  We therefore adopt a minimum disk temperature of 10\,K due to the ISRF.  Note that the gas does not influence this grain temperature at all; when the density is sufficiently high (as it is in accretion disks), the gas simply reaches thermal equilibrium with the dust at the same temperature.  

Thus, our final adopted temperature profile is one that declines as a power-law in radius, due to stellar irradiation, until the 10\,K threshold is reached, and then remains constant at this temperature, due to ISRF heating, at all larger radii.   


\section{Bayesian Analysis}
Consider a specified model $\mathcal{M}_i$ invoked to explain a set of data $\mathcal{D}$.  Then the fundamental equation of Bayesian statistics states that:
$$ P({\mathcal M}_i|{\mathcal D}) = \frac{P({\mathcal M}_i)\,P({\mathcal D}|{\mathcal M}_i)}{P({\mathcal D})} \eqno({\rm C1}) $$ 
where $P({\mathcal M}_i|{\mathcal D})$ is the posterior probability that the model is correct given the data ({\it posterior} for short); $P({\mathcal M}_i)$ is the prior probability that we assign to the model's veracity (before comparing to the data), given all our pre-existing information/biases about the world ({\it prior} for short); $P({\mathcal D}|{\mathcal M}_i)$ is the probability of obtaining the data given this model (known as the {\it likelihood}); and $P({\mathcal D}) \equiv \sum_i{P({\mathcal M}_i)\,P({\mathcal D}|{\mathcal M}_i)}$ -- where the summation is over all possible models -- is a normalization factor (known as the {\it evidence}) that ensures $\sum_i{P({\mathcal M}_i|{\mathcal D})} = 1$.    

Equation (C1) elegantly expresses the intuitive notion that the probability of a given model being correct, in light of the data, is proportional to both our a priori preference for the model and the likelihood of actually obtaining the observed data if the model were true.  Thus a model which fits the data well but is judged outlandish for whatever other reason must still be considered improbable, and so must a model which seems reasonable to start with but fits the data very poorly.  The normalization factor simply expresses the condition that the data must be explicable by some subset of the available models.       

This technique can be used for either model comparison (evaluating the posterior probabilities of different models) or parameter-estimation (evaluating the posteriors for different parameter values within the context of a single model).  In this work, we are concerned with the latter; we will consider one model (in particular a lognormal, discussed further below), specified by $N$ parameters.  Our task is to estimate the probability of these parameters taking on any particular set of values \{$\theta_{n=1..N}$\} (in our case, a specific mean and standard deviation), given a set of actual measurements, upper limits and uncertainties.  Thus consider data with $N_d$ measured values \{${\hat{m}}_{d=1..N_d}$\} and associated errors \{$\sd$\}, and $N_u$ upper limits \{${\hat{m}}_{lim,u=1..N_u}$\} with errors \{$\sl$\}.  Then, from equation (C1), the posterior probability of the model parameters taking on the values \{$\theta_n$\} may be explicitly written as:
$$ P(\{\theta_n\} \,|\, \{\hatm\},\{\hatml\}) = \frac{P(\{\theta_n\})\,\, P(\{\hatm\},\{\hatml\} \,|\,\{\theta_n\})}{\int{P(\{\theta_n\}) P(\{\hatm\},\{\hatml\} \,|\, \{\theta_n\})}} \eqno({\rm C2})$$
where the evidence in the denominator is expressed as an integral, instead of a summation, for a continuous instead of discrete range for each model parameter.  

The evidence integral is immaterial for inferring the relative probabilities of model parameters.  We will also assume here that the prior is a uniform distribution, expressing our lack of any strong preference for one set of model parameters over another.  Then the posterior is simply proportional to the likelihood:
$$ P(\{\theta_n\} \,|\, \{\hatm\},\{\hatml\}) \propto P(\{\hatm\},\{\hatml\} \,| \,\{\theta_n\}) \eqno({\rm C3})$$

The likelihood is is given by the product of the likelihood of each measured value or upper limit, i.e.:
$$ P(\{\hatm\},\{\hatml\} \,| \,\{\theta_n\}) = \left[\prod_{d=1}^{N_d}P(\hatm \,|\, \{\theta_n\})\right] \times \left[\prod_{u=1}^{N_u}P(\hatml \,|\, \{\theta_n\})\right]  \eqno({\rm C4})$$
The one subtlety in evaluating the above is that the true value of each data point (as opposed to the measured or upper limit value, which is the true value scattered by the measurement noise) must be taken into account (i.e., marginalised over), since we want the {\it underlying} distribution of the population.  This is accomplished by making use of the known uncertainties (noise), as follows.

\subsubsubsection{Measured Values}

We will assume that the true value $m_d$ is non-negative, and that the noise is additive and Gaussian.  Then, for objects with measurements, the individual likelihoods are given by:
$$ P(\hatm \,|\, \{\theta_n\}) = \int_0^\infty P(m_d \,|\, \{\theta_n\}) \,\frac{1}{(2\pi)^{1/2}\sd} {\rm exp}\left[-\frac{1}{2}\left(\frac{\hatm - m_d}{\sd}\right)^2\right] \,{\rm d}m_d  \eqno({\rm C5})$$
In general, the integral must be evaluated numerically once the model is specified, which is nevertheless usually straightforward since the integrand is only significant in the approximate range ${\rm MAX}(0,\hatm - 3\sd) \lesssim m_d \lesssim (\hatm + 3\sd)$. 
Note that the {\it measured} value $\hatm$ may well be negative; it is only the true value $m_d$ that is required to be non-negative.  

\subsubsubsection{Upper Limits}

For objects without reported measured values, the individual likelihood is complicated by the fact that the upper limit is consistent with any case in which the unreported measurement $\hatmu$ (i.e., the true value $m_u$ scattered by the measurement noise $\sl$) is less than or comparable to the reported limit $\hatml$.  Hence it is necessary to marginalise not only over the unknown true value $m_u$, but also over the unknown measurement $\hatmu$.  Assuming again that the noise is additive and Gaussian, the likelihood for each object with a reported upper limit becomes:
$$ P(\hatml \,|\, \{\theta_n\}) = \int_0^\infty P(m_u \,|\, \{\theta_n\}) \,\left\{\int_{-\infty}^{\hatml}\frac{1}{(2\pi)^{1/2}\sl} {\rm exp}\left[-\frac{1}{2}\left(\frac{\hatmu - m_u}{\sl}\right)^2\right] {\rm d}\hatmu\right\}\,{\rm d}m_u  \eqno({\rm C6})$$
The inner integral is given by:
$$ \int_{-\infty}^{\hatml}\frac{1}{(2\pi)^{1/2}\sl} {\rm exp}\left[-\frac{1}{2}\left(\frac{\hatmu - m_u}{\sl}\right)^2\right] {\rm d}\hatmu = \frac{1}{2}\,{\rm erfc}\left(\frac{\hatml - m_u}{\sl}\right) \eqno({\rm C7}) $$
where erfc is the complementary error function, so that the likelihood simplifies to: 
$$ P(\hatml \,|\, \{\theta_n\}) = \int_0^\infty P(m_u \,|\, \{\theta_n\}) \, \frac{1}{2}\,{\rm erfc}\left(\frac{\hatml - m_u}{\sl}\right) {\rm d}m_u \eqno({\rm C8}) $$
This integral must be evaluated numerically, but this is again straightforward, with the integrand appreciable only over the range $0 \le m_u \lesssim (\hatml + 3\sl)$.  

\subsubsubsection{Model}

We assume that the underlying distribution of $M_{d,\nu}/\mstar$ is a lognormal (justified at the end of \S8), specified by 2 parameters: $\theta_1 \equiv \mu$ (the mean), and $\theta_2 \equiv s$ (the standard deviation).  For any given $\mu$ and $s$, the probability of any particular positive value $m$ ($\equiv M_{d,\nu}/\mstar$) in this distribution is:
$$ P(m \,|\, \{\mu, s\}) \equiv \frac{1}{(2\pi)^{1/2}\,m\,s}{\rm exp}\left[-\frac{1}{2}\left(\frac{{\rm ln}(m) - \mu}{s}\right)^2\right] \eqno({\rm C9}) $$
Substituting this into equation (C5) with $m \equiv m_d$, and into equation (C8) with $m \equiv m_u$, then gives us the individual likelihoods for any measured value $\hatm$ or upper limit $\hatml$.  The product of these, as specified in equation (C4), yields the total likelihood of the data given any particular parameter set \{$\mu$, $s$\}, which leads to the posterior probability of this set via equation (C3).  Repeating this procedure over the whole range of model parameters then gives the posterior probability distribution of the model parameters (in reality, we carry this out over a fine but discrete and finite mesh covering the range of plausible $\mu$ and $s$). This method of calculating posterior probabilities is graphically illustrated in Fig.\,17.    

\end{appendix}



\clearpage

\begin{figure}
\includegraphics[scale=0.6]{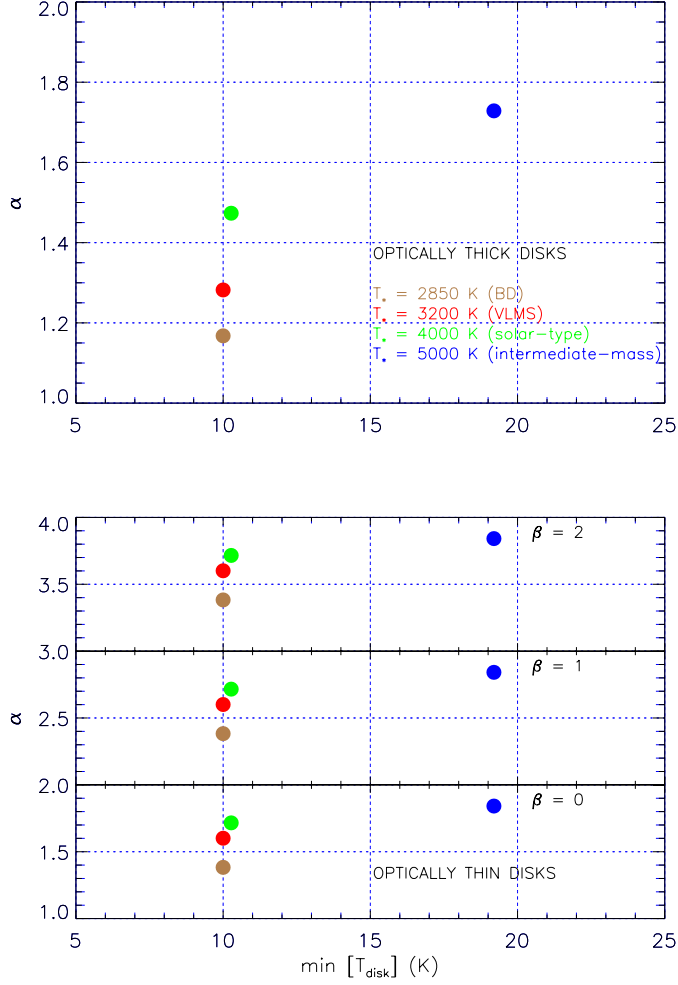}
\caption{$\alpha$ (derived from the general equation (8)) versus minimum disk temperature ($T_{disk}$ at the outermost radius $R_d = 100$\,AU), for a fiducial disk around a representative BD ({\it brown}), VLMS ({\it red}), solar-type star ({\it green}) and intermediate-mass star ({\it blue}).  The minimum disk temperature flattens at 10\,K for VLMS and BDs, due to heating by the ISRF.  {\bf Top panel:} Optically thick disk.  {\bf Bottom panel:} Optically thin disks with $\beta$ = 0, 1 and 2.  For both optically thick and thin disks, the RJ approximation ($\alpha$=2 for thick and $\alpha = 2+\beta$ for thin) becomes increasingly poor for cooler (less massive) objects.  See \S6.     }
\end{figure}

\begin{figure}
\includegraphics[scale=0.6, angle=90]{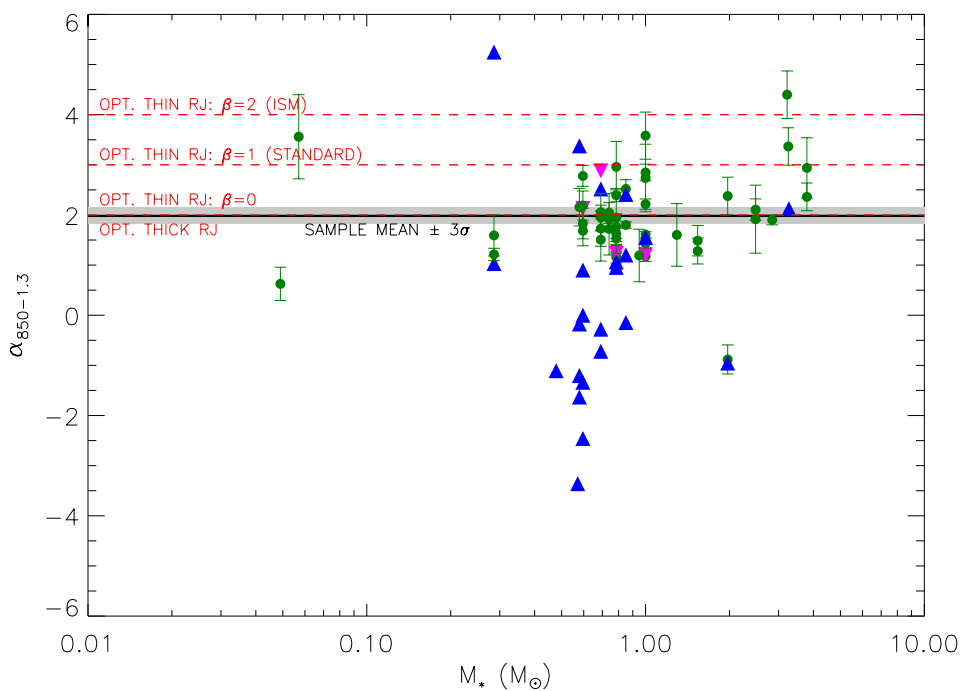}
\caption{$\alpha$ for sources in our sample observed at both 850\,$\mu$m and 1.3\,mm and detected in at least one of these bands, versus stellar mass.  {\it Green filled circles} indicate $\alpha$ for sources detected at both wavelengths, with {\it green vertical lines} denoting the $\pm 1\sigma$ uncertainty in these values; {\it purple downward triangles} are 3$\sigma$ upper limits on $\alpha$, for sources detected only at 1.3\,mm; {\it blue upward triangles} are 3$\sigma$ lower limits on $\alpha$, for detections only at 850\,$\mu$m.  The four {\it red horizontal dashed lines} show the expected $\alpha$ in the RJ limit, for optically thin disks with $\beta$ = 0, 1 and 2 as well as optically thick disks (note that $\alpha=2$ for both optically thick disks and optically thin ones with $\beta=0$, in the RJ limit; note also from Fig.\,1 that the RJ approximation becomes less valid with decreasing stellar mass).  The mean slope for all our sources detected in both bands is $\langle\alpha\rangle \approx 2$ ({\it solid black horizontal line}); the {\it grey zone} delimits the $\pm 3\sigma$ ($\pm 0.18$) errors on the latter.  See \S7.1. }
\end{figure}

\begin{figure}
\includegraphics[scale=0.7, angle=90]{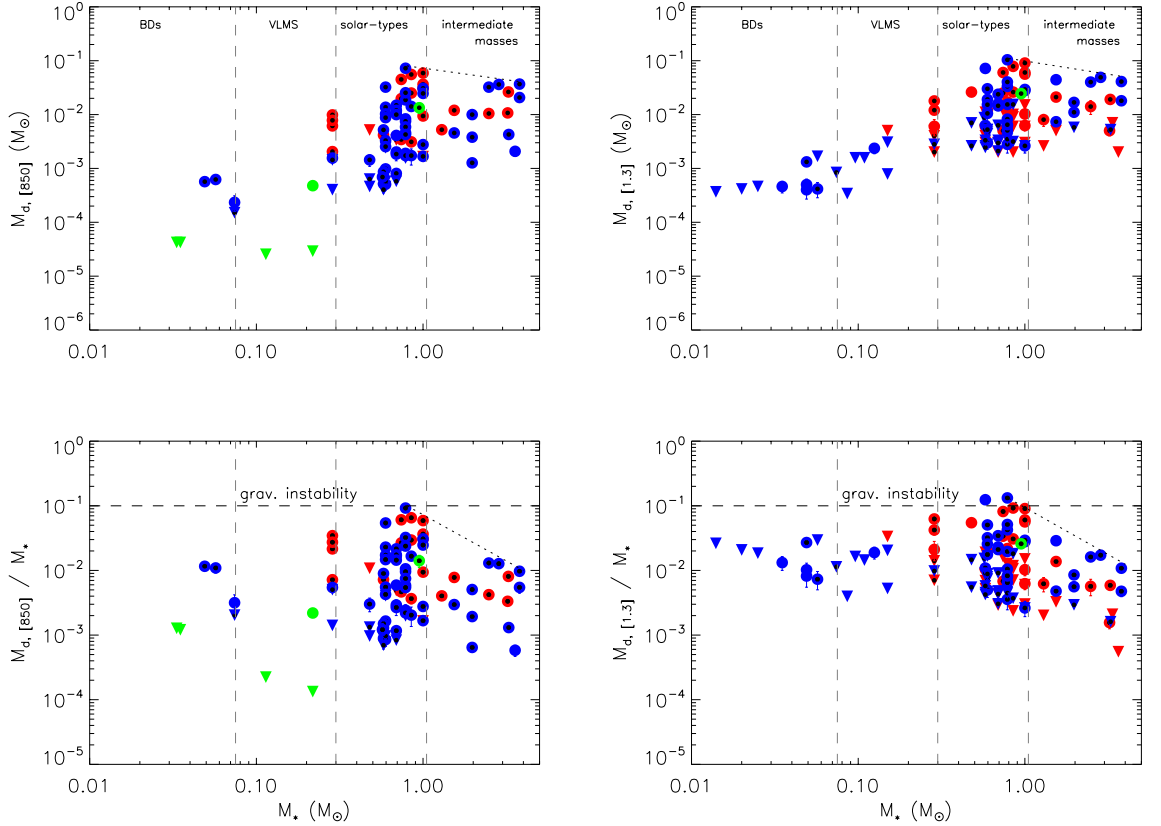}
\caption{Apparent disk mass $M_{d,\nu}$ (from equation (9), using the flux at either 850\,$\mu$m or 1.3\,mm) versus stellar mass.  {\bf Top panels:} $M_{d,[850]}$ versus $\mstar$ ({\it left}); $M_{d,[1300]}$ versus $\mstar$ ({\it right}).  {\bf Bottom panels:} $M_{d,[850]}/\mstar$ versus $\mstar$ ({\it left}); $M_{d,[1300]}/\mstar$ versus $\mstar$ ({\it right}).  {\it Circles} are measured values, {\it triangles} are upper limits; either symbol with a {\it central black dot} represents a source observed at both wavelengths.  Taurus objects in {\it blue}, $\rho$ Oph in {\it red}, TWA in {\it green}.  {\it Vertical dashed lines} distinguish between intermediate-mass stars, solar-types, VLMS and BDs.  A {\it sloping dotted line} marks the falling upper envelope from solar-types to intermediate-mass stars ($M_{d,\nu} \propto \mstar^{-1/2}$ in the upper panels, $M_{d,\nu}/\mstar \propto \mstar^{-3/2}$ in the lower ones).  The bottom panels also show the boundary above which disks are expected to become gravitationally unstable, $M_d/\mstar \sim 0.1$ ({\it horizontal dashed line}). See \S7.1.         }
\end{figure}

\begin{figure}
\includegraphics[scale=0.5, angle=90]{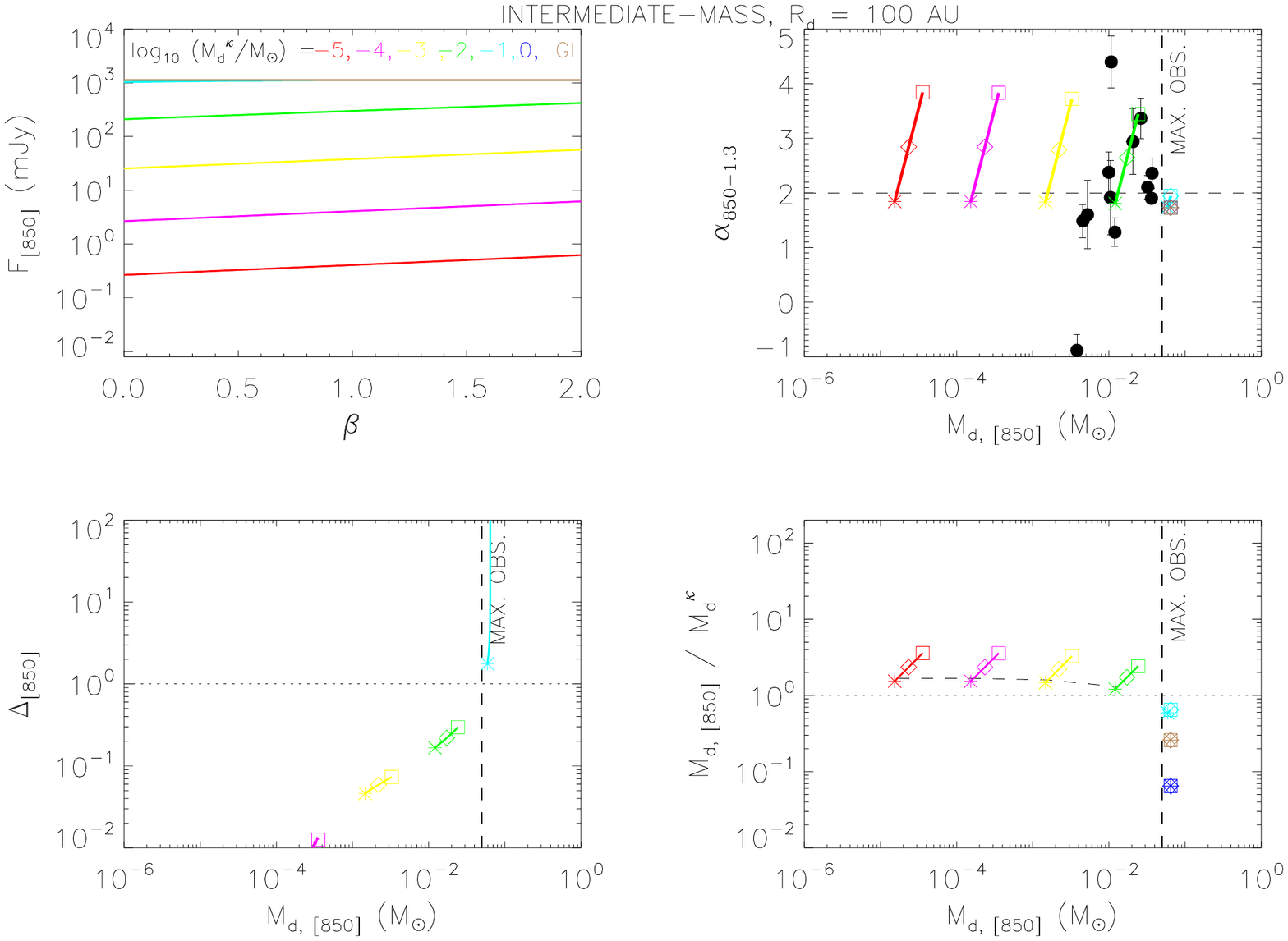}
\includegraphics[scale=0.5, angle=90]{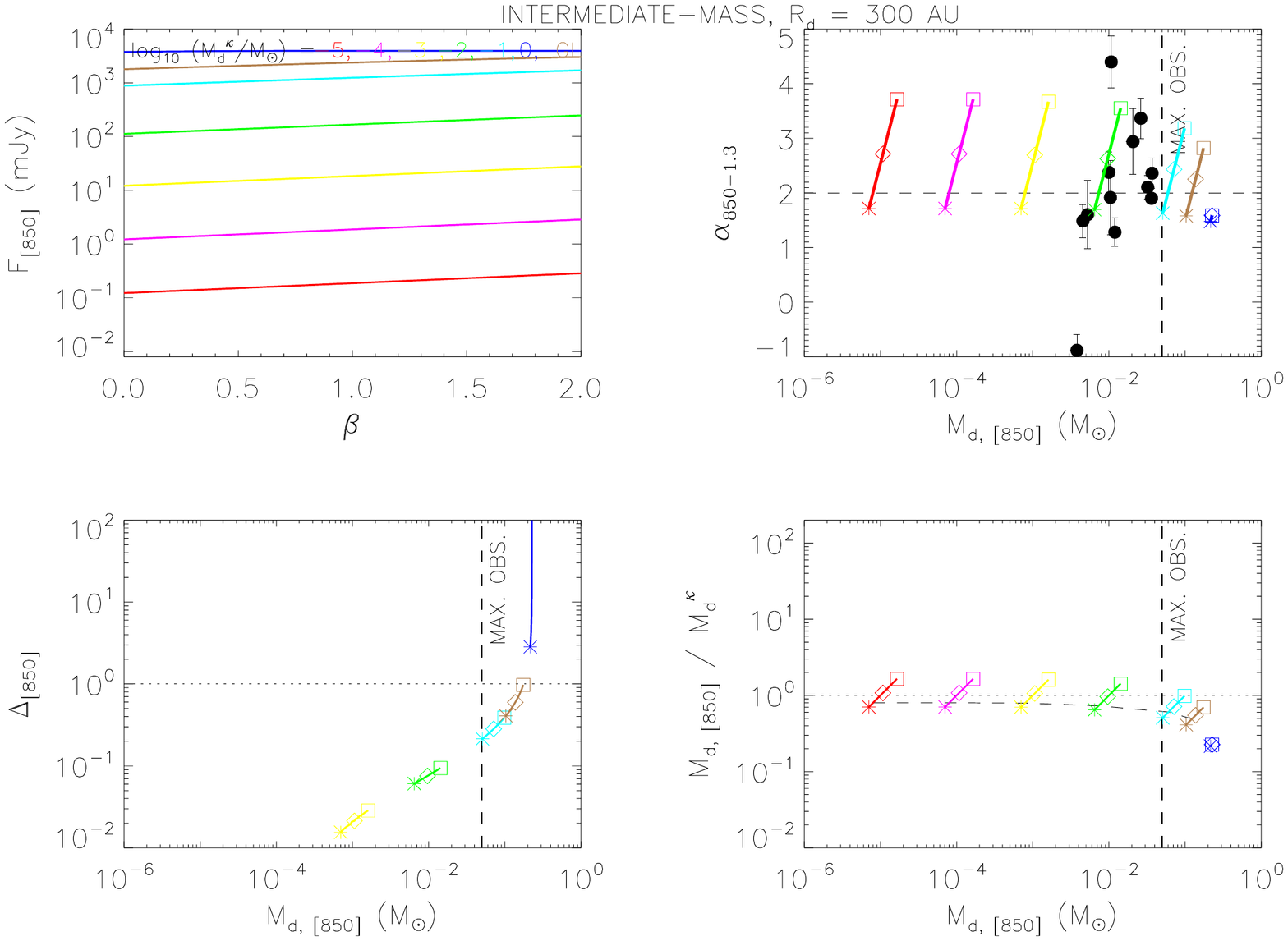}
\end{figure}
\begin{figure}
\caption{Our generalized model predictions (using equations (5), (6) and (8)) for our fiducial disk parameters (equation (7)), for the case of an intermediate-mass star at $\sim$1\,Myr (2.5\,$\msun$, 4\,$\rsun$, 5000\,K). {\bf Top set of four plots} is for $R_d = 100$\,AU, {\bf bottom set} is for $R_d = 300\,$AU.  Each set of plots shows the predictions for various input $\beta$ (ranging over 0--2; {\it asterisk:} $\beta = 0$, {\it diamond:} 1, {\it square:} 2) and $M_d^\kappa$ ({\it red:} $M_d^{\kappa}/\msun = 10^{-5}$, {\it magenta:} $10^{-4}$, {\it yellow:} $10^{-3}$, {\it green:} $10^{-2}$, {\it aqua:} $10^{-1}$, {\it blue:} $10^{0}$, {\it brown:} gravitational instability limit $M_{d,GI}^\kappa/\mstar = 0.1 \Rightarrow M_{d,GI}^\kappa/\msun = 0.25$; note that the real instability limit may occur anywhere in the range 0.1--10\,$M_{d,GI}^{\kappa}$).  The four panels in each set of plots show the following.  {\bf Top left panel:} Predicted 850\,$\mu$m flux (in mJy, scaled to a distance of 140\,pc) versus $\beta$.  {\bf Top right panel:} Predicted $\alpha$ versus predicted $M_{d,[850]}$ (in units of $\msun$, independent of distance).  The {\it dashed horizontal line} shows the mean value $\alpha \approx 2$ for our sample.  Sources with measured $\alpha$ and $M_{d,[850]}$ are overplotted as {\it black filled circles} with error bars.  The {\it thick vertical line} marks the maximum observed $M_{d,[850]}$ in our data for this stellar mass bin.  {\bf Bottom left panel:} Predicted ratio of optically thick to thin emission at 850\,$\mu$m versus $M_{d,[850]}$.  The {\it horizontal dotted line} marks equal contributions from each.  {\bf Bottom right panel:} Ratio of $M_{d,[850]}$ to the opacity-normalized disk mass $M_d^\kappa$ of the model, as a function of $M_{d,[850]}$.  The ratio is unity along the {\it horizontal dotted line}.  The ratio corresponding to the mean value $\alpha \approx 2$ in our sample is shown by the {\it dashed curve}.  See \S\S7.2, 7.3.1.                    }
\end{figure}

\begin{figure}
\includegraphics[scale=0.5, angle=90]{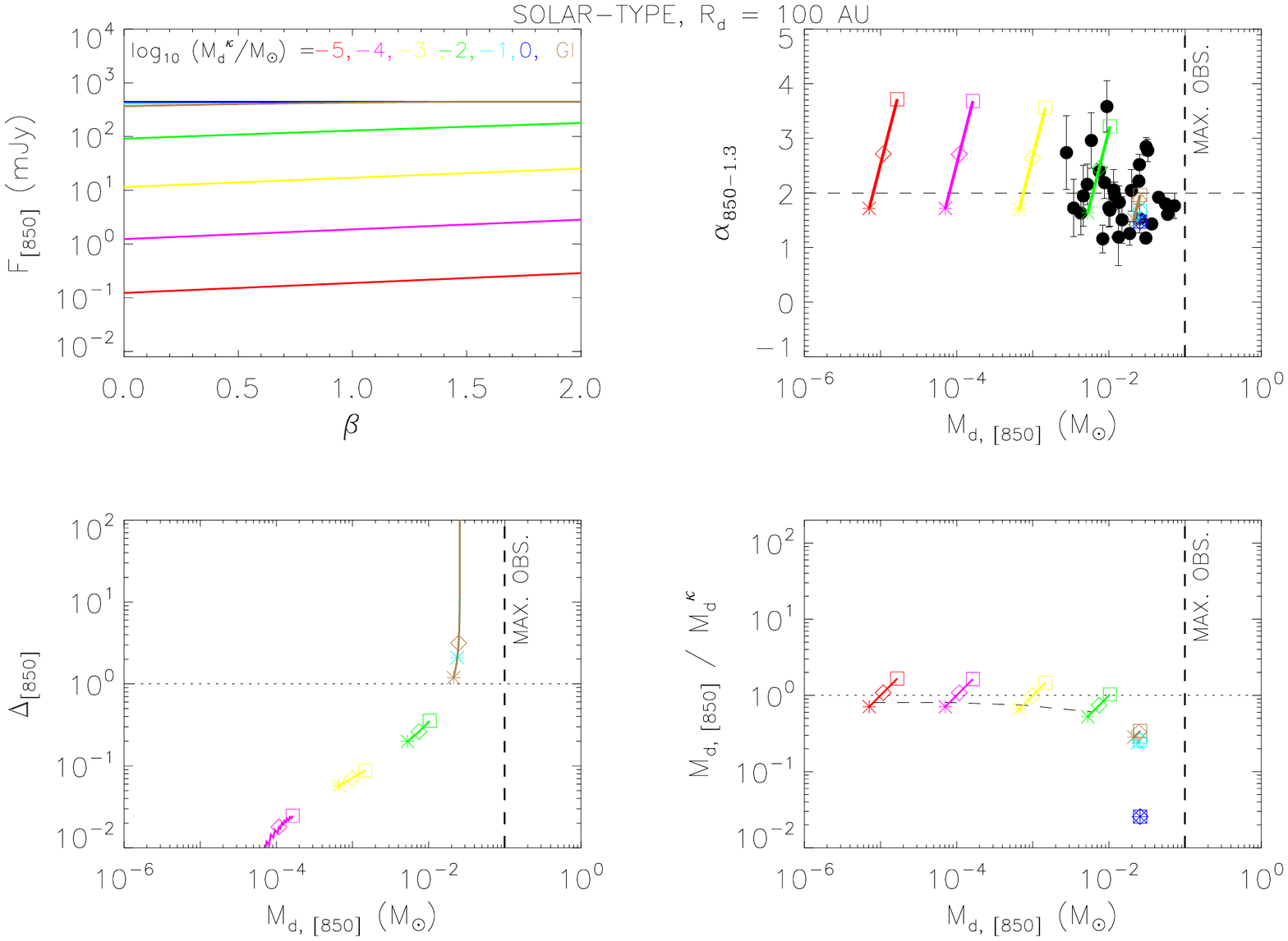}
\includegraphics[scale=0.5, angle=90]{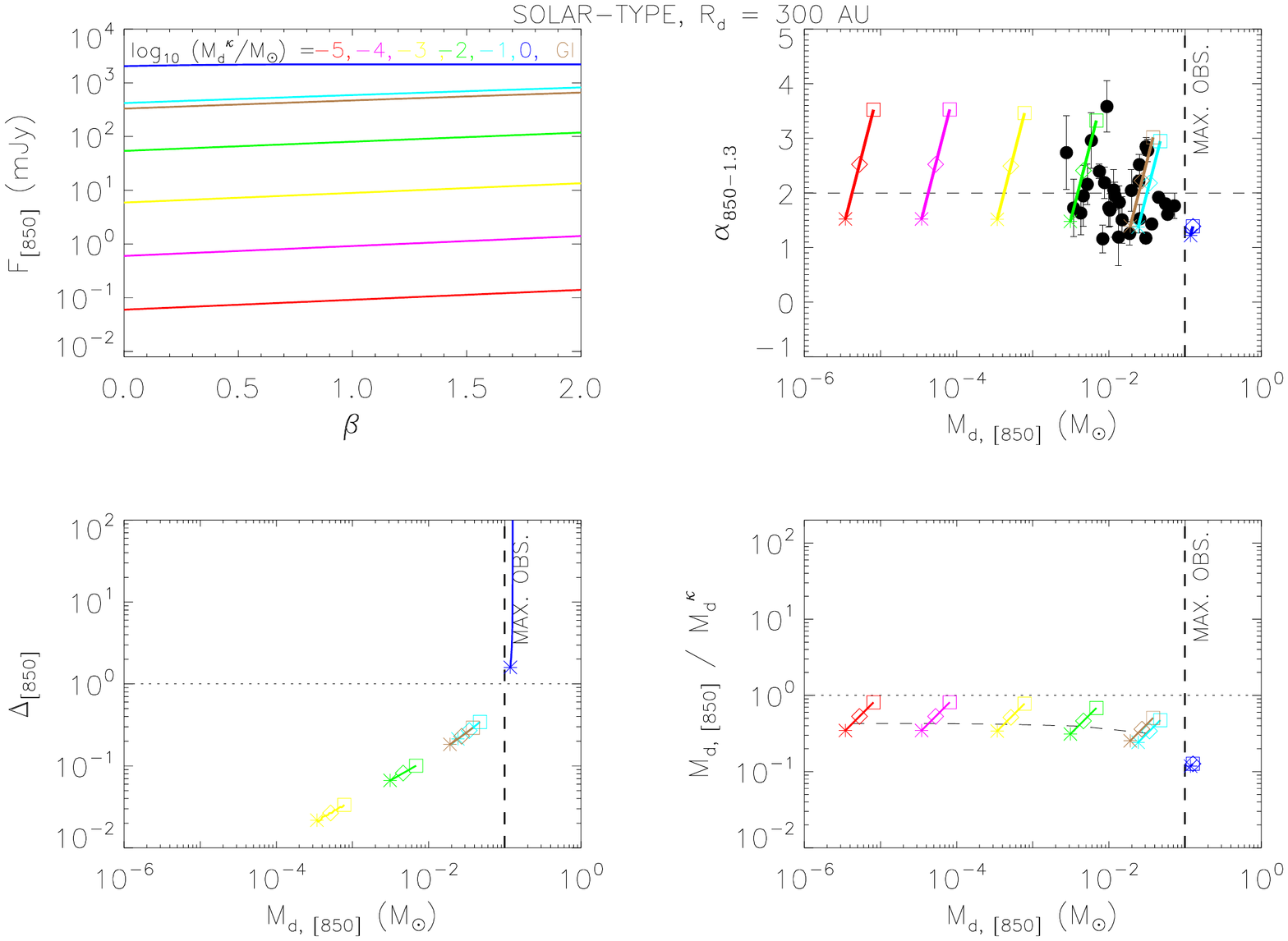}

\caption{Same as Fig.\,4, for solar-type stars (0.75\,$\msun$, 2\,$\rsun$, 4000\,K).  The gravitational instability limit ({\it brown}) is now at $M_{d,GI}^\kappa/\mstar = 0.1 \Rightarrow M_{d,GI}^\kappa/\msun = 0.075$.  }
\end{figure}

\begin{figure}
\includegraphics[scale=0.5, angle=90]{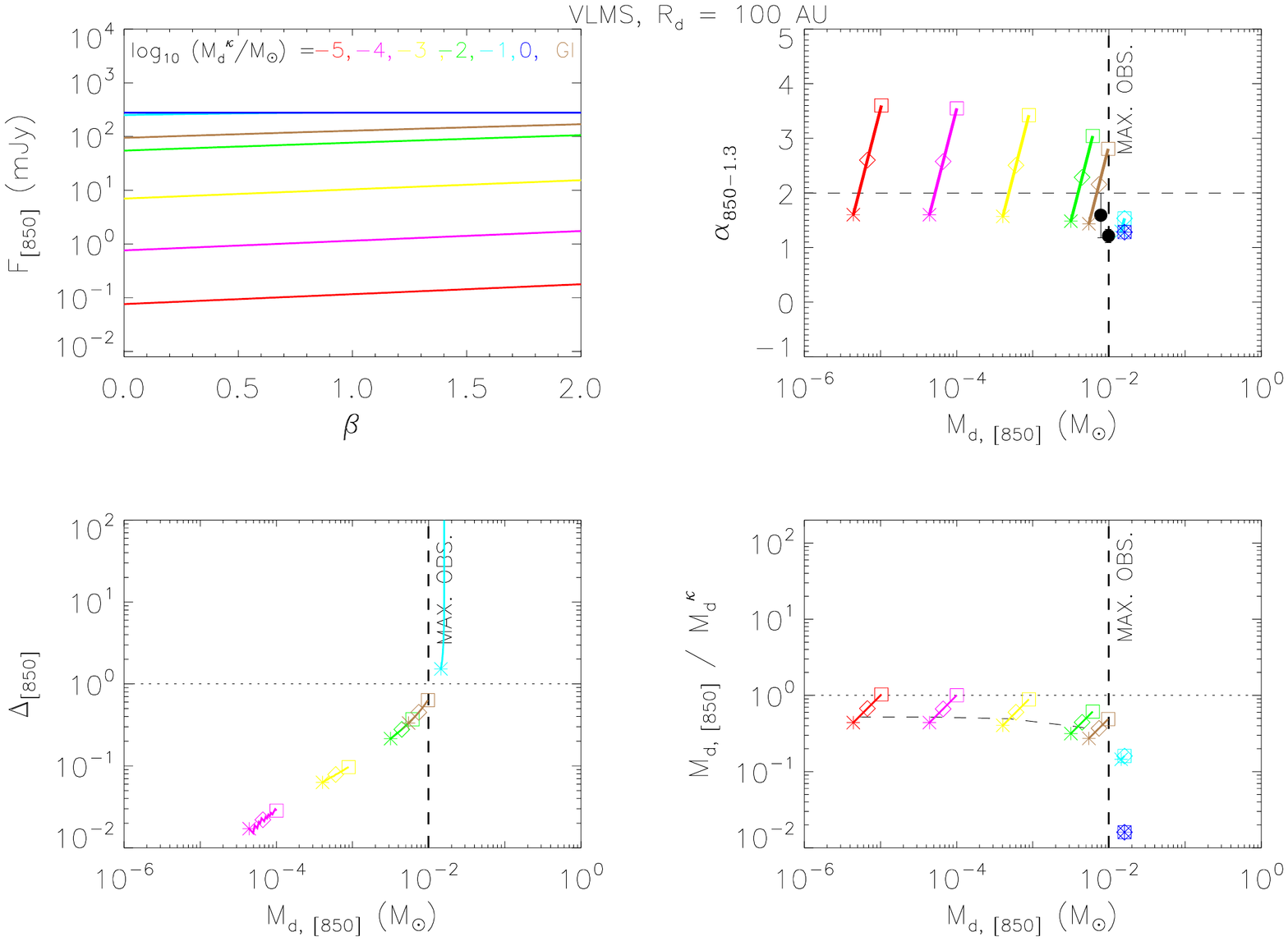}
\includegraphics[scale=0.5, angle=90]{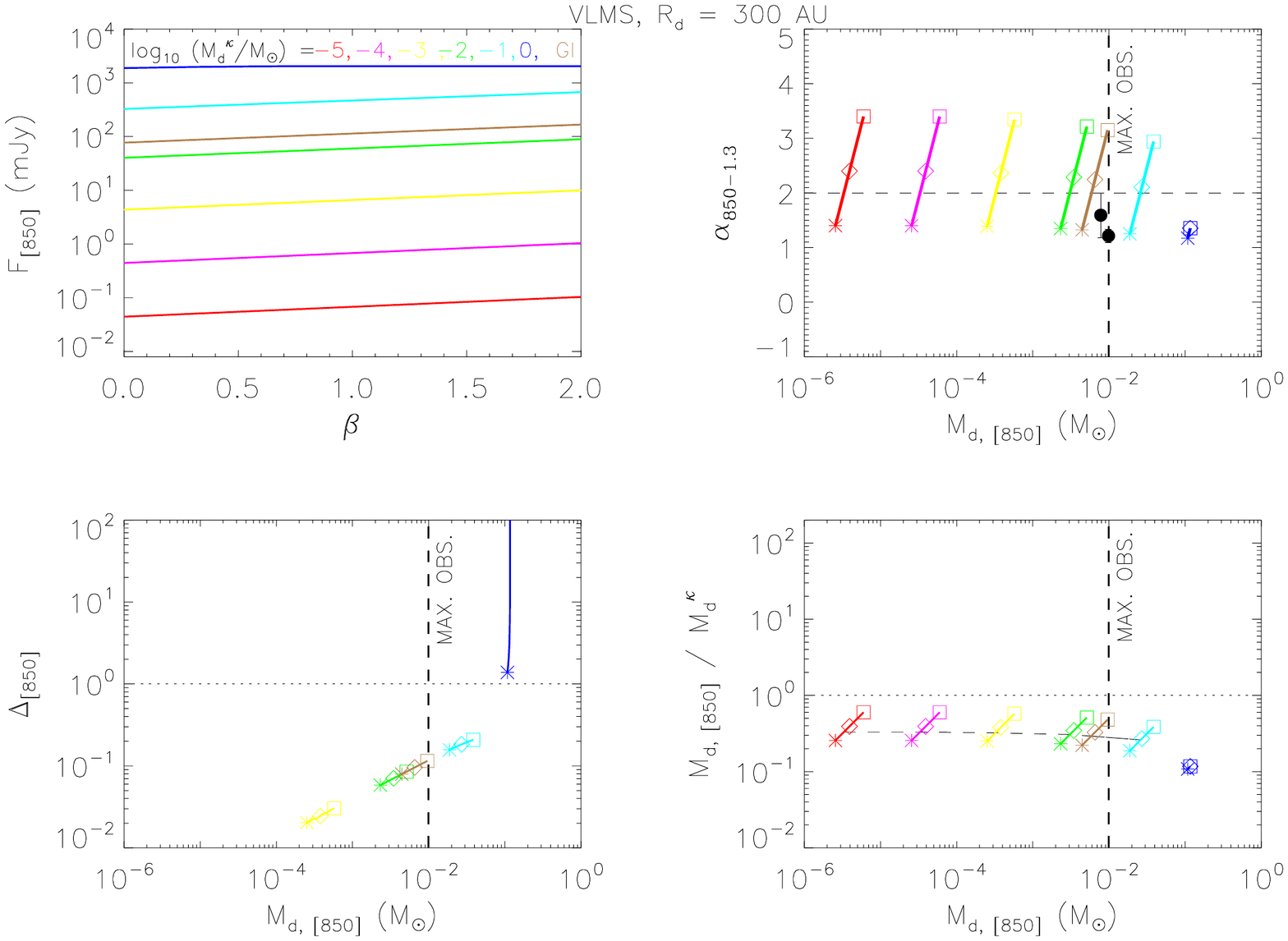}
\caption{Same as Fig.\,4, for VLMS (0.2\,$\msun$, 1.5\,$\rsun$, 3200\,K).  The gravitational instability limit ({\it brown}) is now at $M_{d,GI}^\kappa/\mstar = 0.1 \Rightarrow M_{d,GI}^\kappa/\msun = 0.02$.}
\end{figure}

\begin{figure}
\includegraphics[scale=0.5, angle=90]{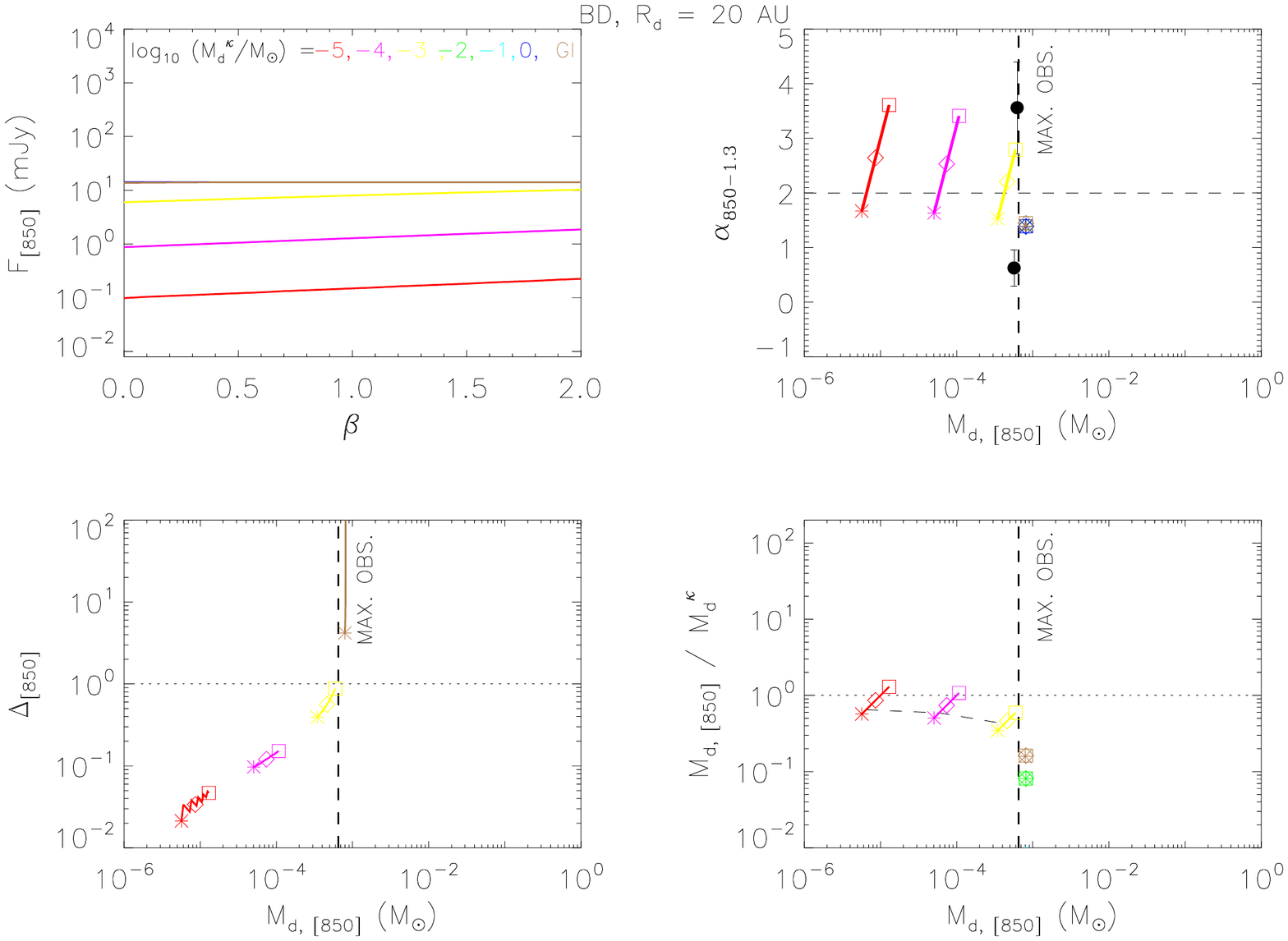}
\includegraphics[scale=0.5, angle=90]{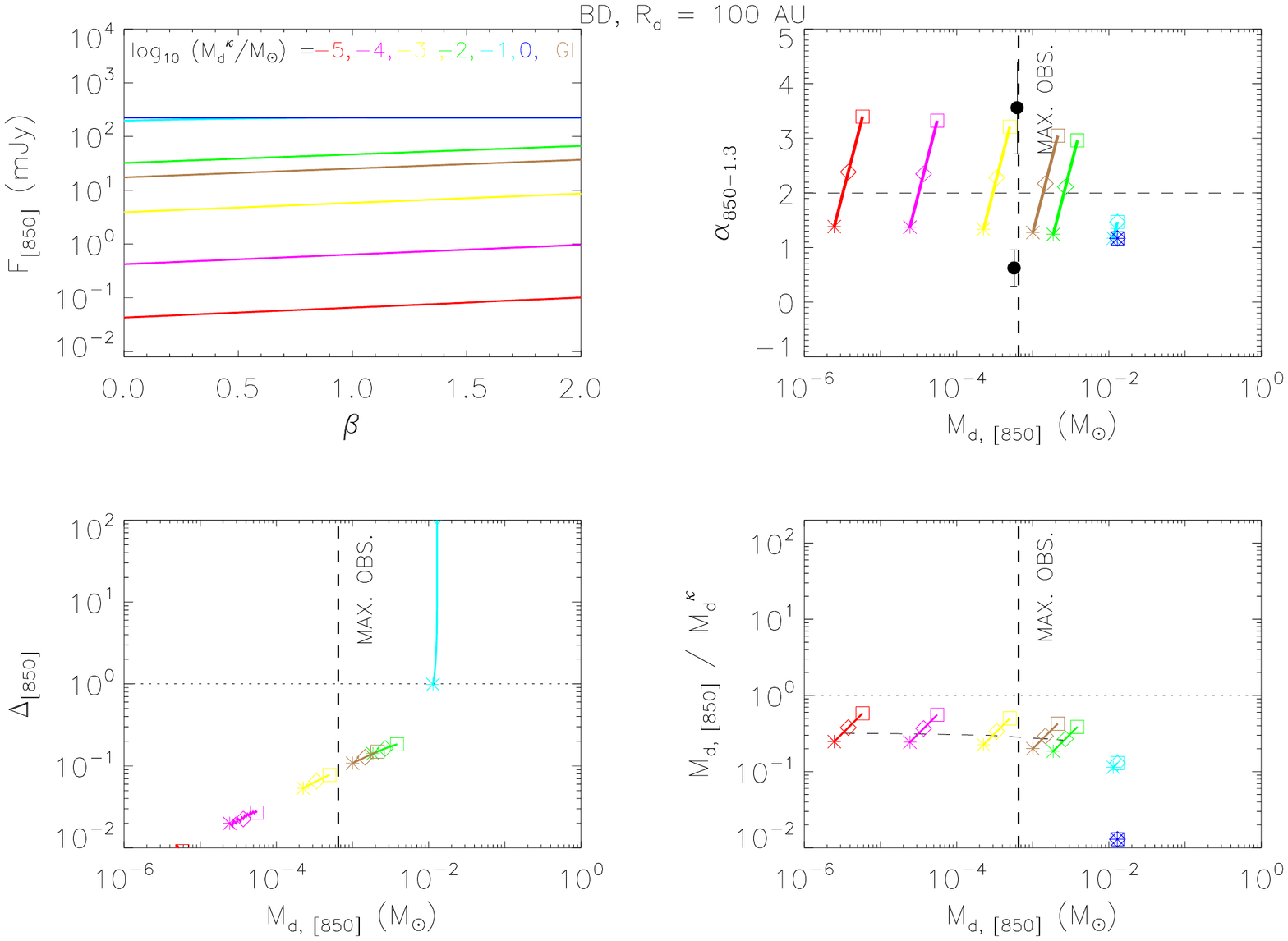}
\caption{Same as Fig.\,4, for BDs (0.05\,$\msun$, 0.55\,$\rsun$, 2850\,K).  The {\bf top set of plots} is now for $R_d = 20$\,AU, and the {\bf bottom set of plots} for $R_d = 100$\,AU. The gravitational instability limit ({\it brown}) is now at $M_{d,GI}^\kappa/\mstar = 0.1 \Rightarrow M_{d,GI}^\kappa/\msun = 0.005$.}
\end{figure}

\begin{figure}
\includegraphics[scale=0.7]{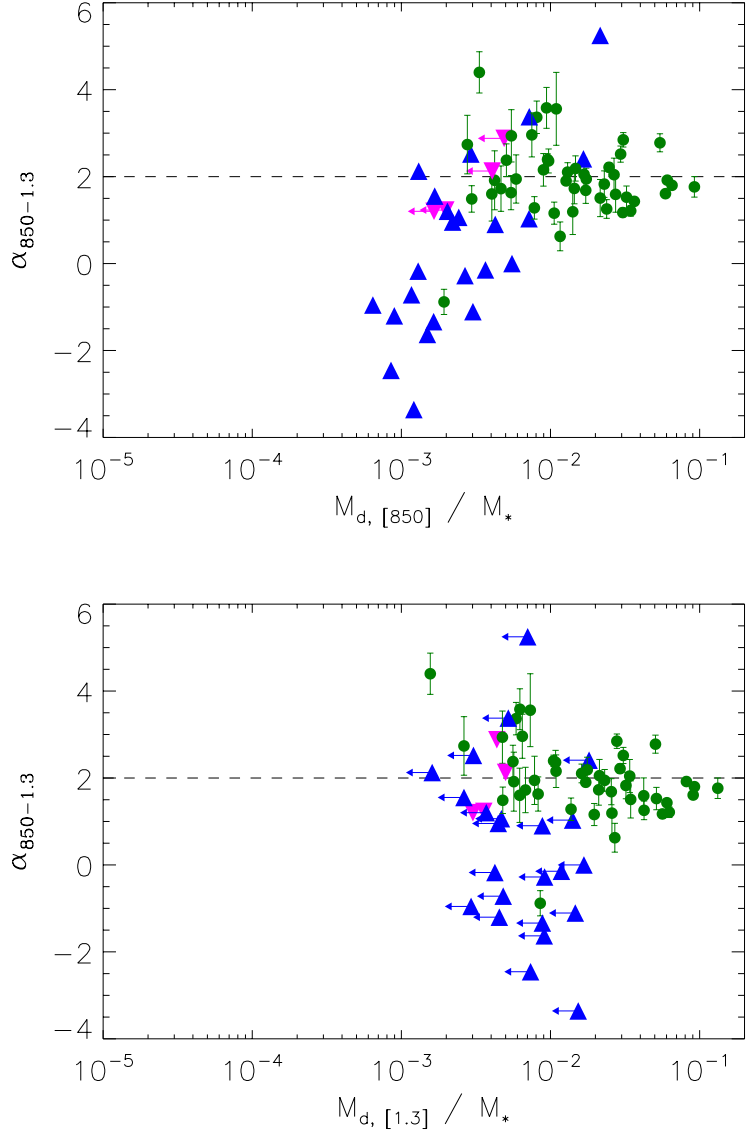}
\caption{Analogous to Fig.\,2, except now for $\alpha$ versus $M_{d,\nu}/\mstar$.  {\bf Top panel:} $\alpha$ versus $M_{d,[850]}/\mstar$.  {\bf Bottom panel:} $\alpha$ versus $M_{d,[1300]}/\mstar$.  All symbols same as in Fig.\,2.  Arrows mark upper limits in $M_{d,[850]}$ (i.e., in 850\,$\mu$m flux, and hence also upper limits in $\alpha$) or upper limits in $M_{d,[1300]}$ (and hence lower limits in $\alpha$).  Sources with measured $\alpha$ ({\it circles}) are spread evenly around the sample mean $\alpha \approx 2$, with no clear trend of $\alpha$ increasing with $M_{d,\nu}/\mstar$ as might be expected if grain growth causes spuriously low $M_{d,\nu}$ estimates.  For sources with lower limits on $\alpha$, the real distribution of $\alpha$ with $M_{d,\nu}/\mstar$ is unknown.  See \S7.3.1.     }
\end{figure}

\begin{figure}
\includegraphics[scale=0.7]{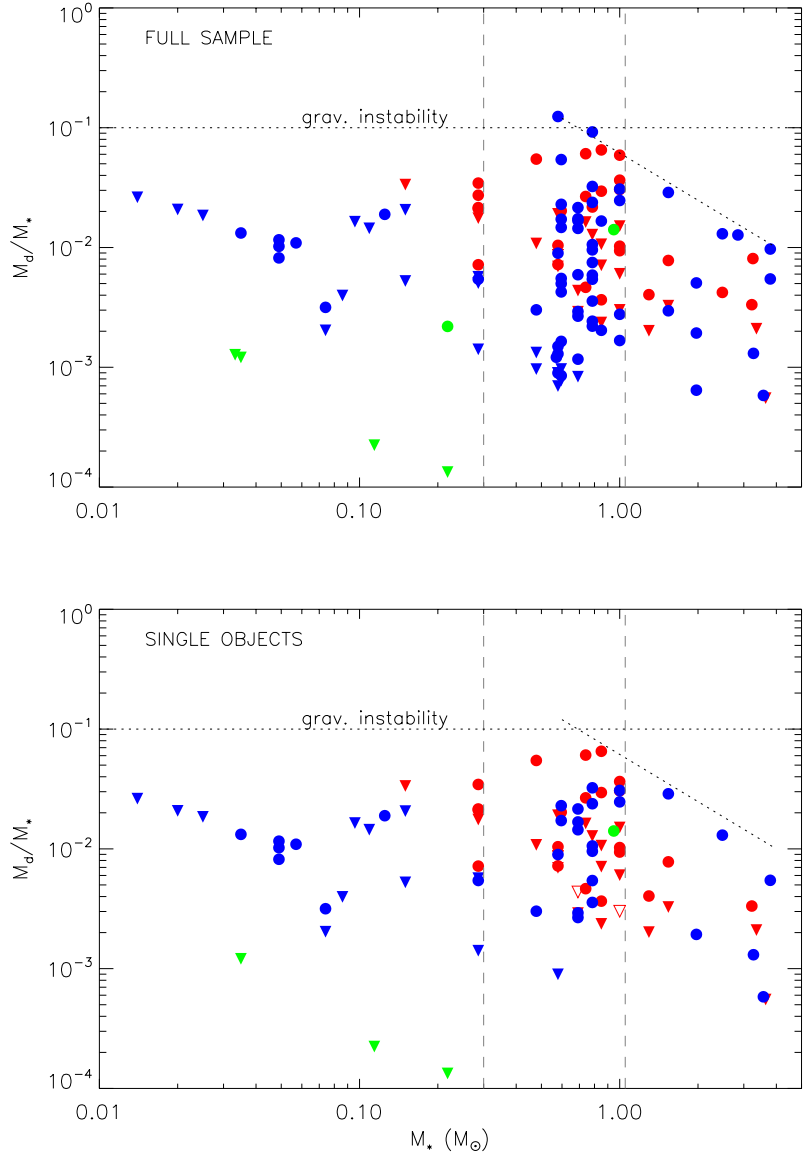}
\caption{$M_{d,\nu}/\mstar$ for our sample (same as bottom panels of Fig.\,3, but now with the 850\,$\mu$m and 1.3\,mm data merged into a single set, as given in Table I).  Same symbols as in Fig.\,3.  {\bf Top panel:} Entire sample.  {\bf Bottom panel:} Sample with known binaries and multiples systems excluded (except the two TWA VLMS TWA 30A and B, which are sufficiently widely separated for individual SCUBA-2 measurements).  Note that this sub-sample, though nominally denoted as `single objects', may contain undiscovered binaries/multiples.  See \S8. }
\end{figure}

\begin{figure}
\includegraphics[scale=0.7, angle=90]{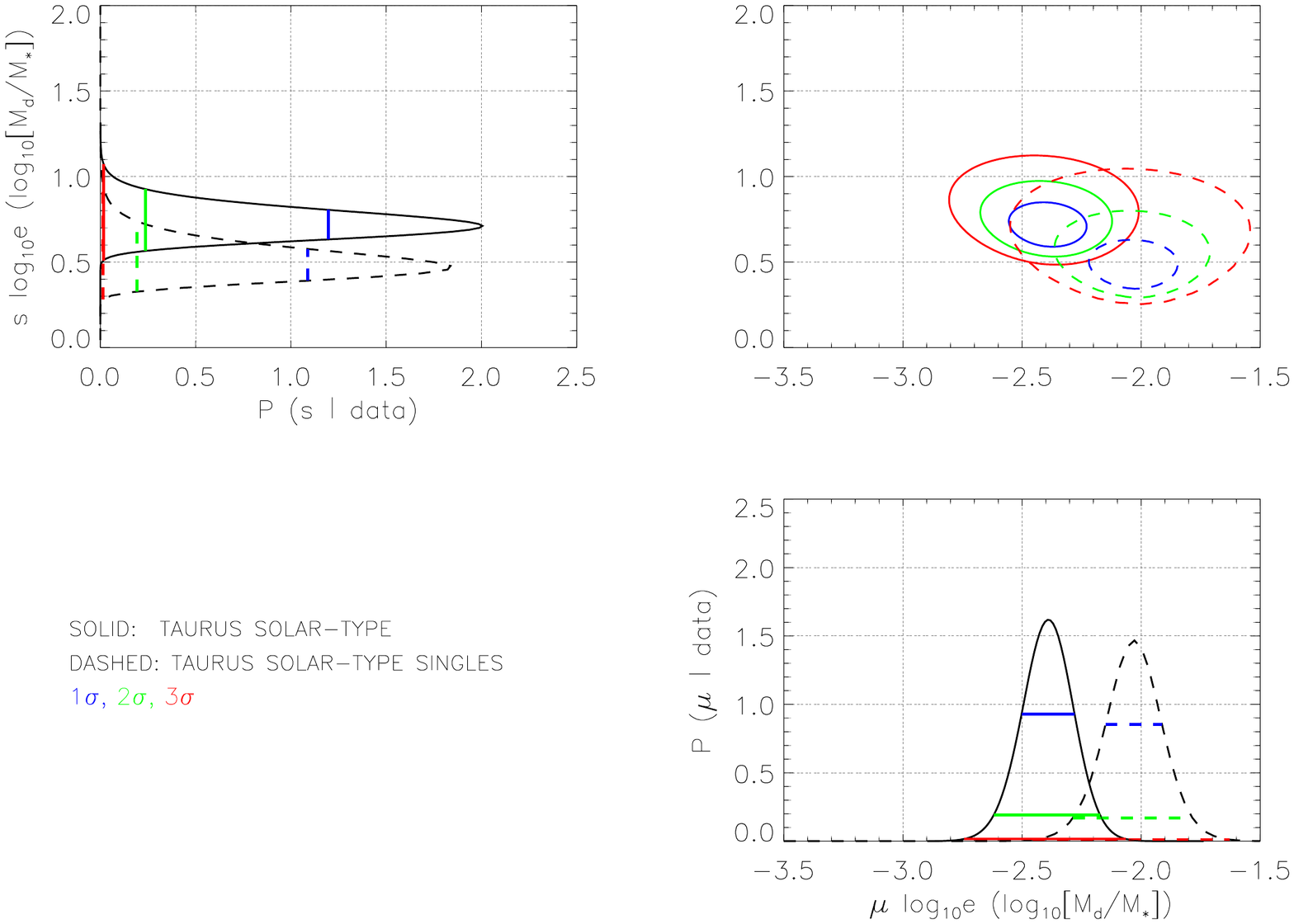}
\end{figure}
\begin{figure}
\caption{Distribution of posterior probability densities for lognormals, for the full Taurus sample of solar-type stars ({\it solid curves}) and the Taurus sample of `single' solar-type stars (i.e., sample with known binaries/multiples excluded; {\it dashed curves}).  {\bf Top right panel}:  2-D distribution of posterior probability densities of the lognormal mean ($\mu$) and standard deviation ($s$) in our data (i.e., P($\mu$,$s$ $|$ data)), with the range of lognormal means along the X-axis and the range of lognormal standard deviations along the Y-axis.  For greater intuition, the means ($\mu$) and standard deviations ($s$) are expressed here in base 10 units (log$_{10}$$[M_{d,\nu}/\mstar]$): $\mu$\,log$_{10}$e and $s$\,log$_{10}$e.  {\it Blue}, {\it green} and {\it red} contours enclose 68.27\%, 95.45\% and 99.73\% (i.e., 1, 2 and 3$\sigma$) of the posterior probability respectively.  {\bf Top left panel}:  1-D distribution of posterior probability densities for $s$, marginalised over all $\mu$.  The colors again represent 1, 2 and 3$\sigma$ levels: [68.27\%, 95.45\% and 99.73\%] of the posterior probability is within the intervals indicated by the [{\it blue},{\it green} and {\it red}] horizontal lines.  {\bf Bottom right panel}: 1-D distribution of posterior probability densities for $\mu$, marginalised over all $s$; colors have the same meaning, except now pertain to $\mu$.  The highest posterior density model (i.e., the most probable model) for the full sample of Taurus solar-types has a mean log$_{10}$$[M_{d,\nu}/\mstar]$ of $\mu$\,log$_{10}$e = $-2.4$ and standard deviation $s$\,log$_{10}$e = 0.7, while the corresponding values for the Taurus solar-type singles are $\mu$\,log$_{10}$e $\approx$ $-2$ and $s$\,log$_{10}$e = 0.5. The distribution of means for the two samples are separated by nearly $2\sigma$, suggesting that binarity may influence the disk mass distribution.  This is explored explicitly in the next two figures. See \S8. }
\end{figure}

\begin{figure}
\includegraphics[scale=0.7, angle=90]{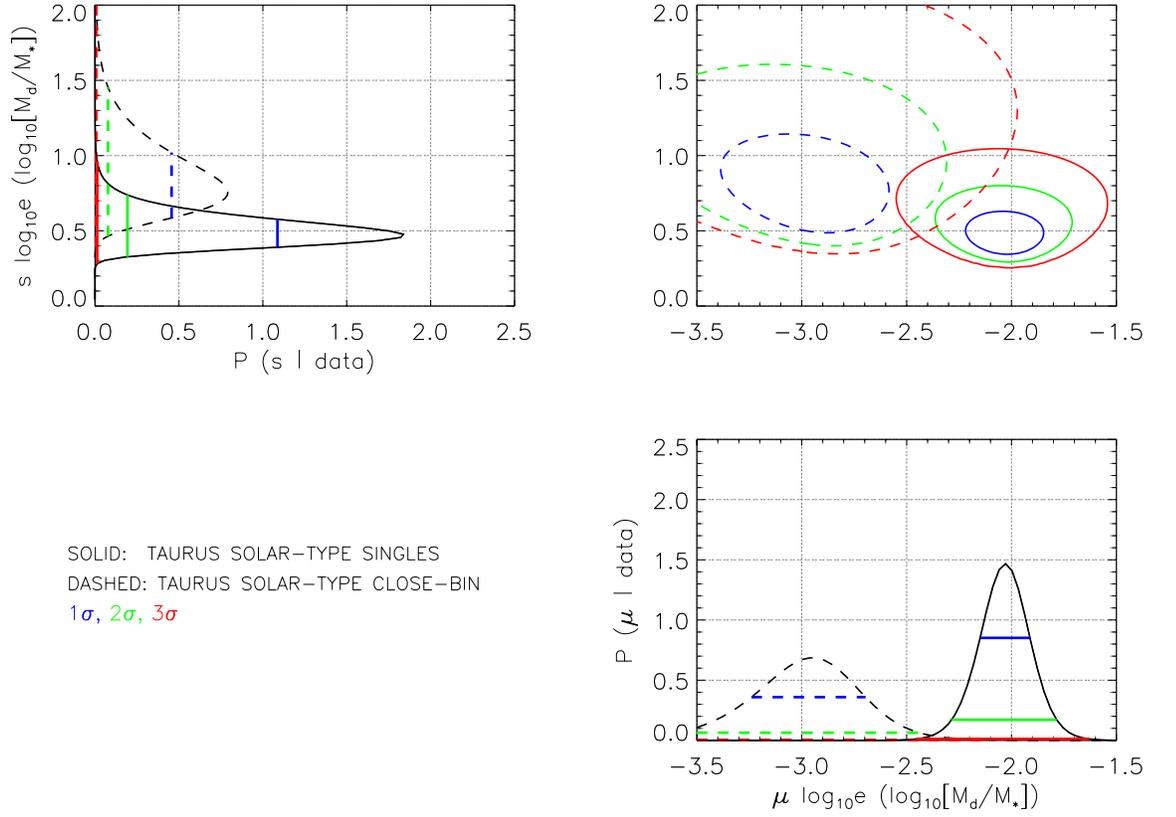}
\caption{Same as Fig.\,10, except the Taurus solar-type sample of single stars ({\it solid curves}) is now compared to the Taurus solar-type sample of close binaries ({\it dashed curves}).  The distribution of means for the two samples are separated by $\sim$$3\sigma$, implying a significant difference in disk mass between close binaries and singles. The most probable mean disk mass in the close binaries (log$_{10}$$[M_{d,\nu}/\mstar] \approx -3$) is 10$\times$ lower than in the singles. See \S8.  }
\end{figure}

\begin{figure}
\includegraphics[scale=0.7, angle=90]{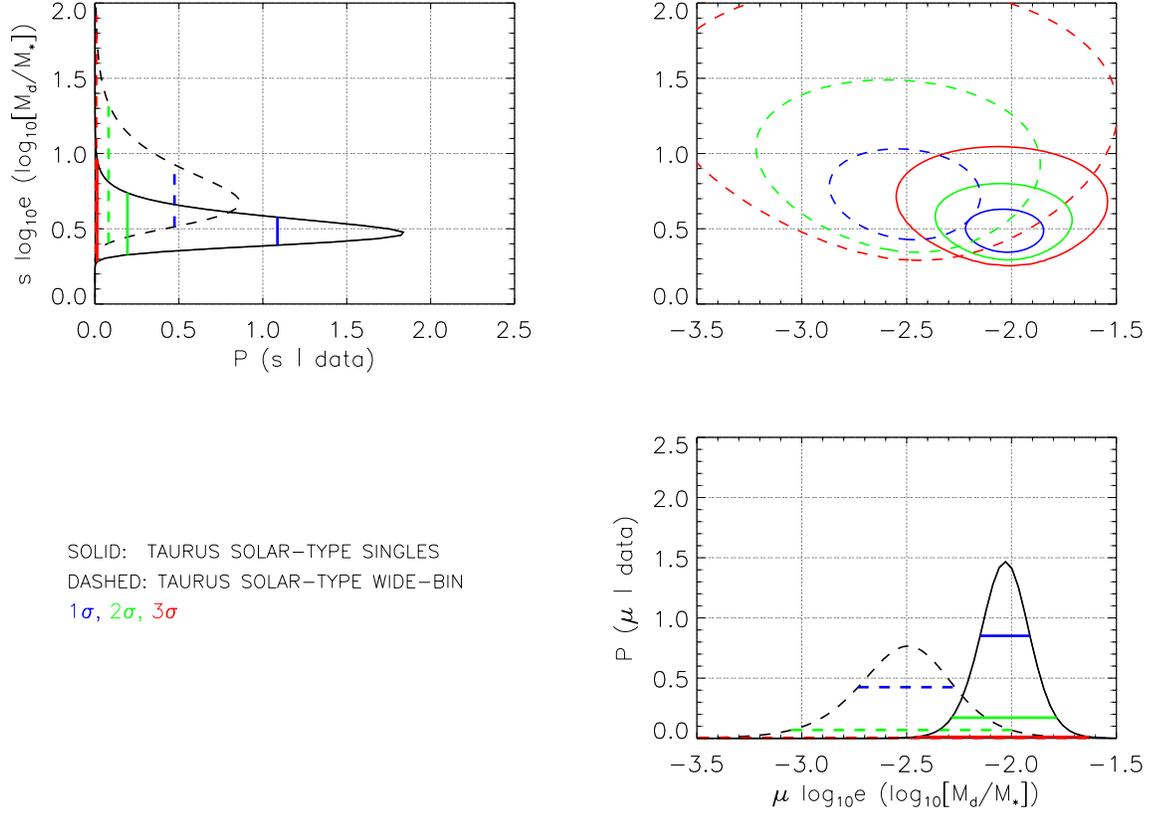}
\caption{Same as Fig.\,10, except the Taurus solar-type sample of single stars ({\it solid curves}) is now compared to the Taurus solar-type sample of wide binaries ({\it dashed curves}).   The distribution of means for the two samples are separated by $<$$2\sigma$, implying that the disk mass in wide binaries is much more similar to that in the singles, compared to the close binaries in the previous figure.  The most probable mean disk mass in the wide binaries (log$_{10}$$[M_{d,\nu}/\mstar] \approx -2.5$) is 3$\times$ lower than in the singles. See \S8. }
\end{figure}

\begin{figure}
\includegraphics[scale=0.7, angle=90]{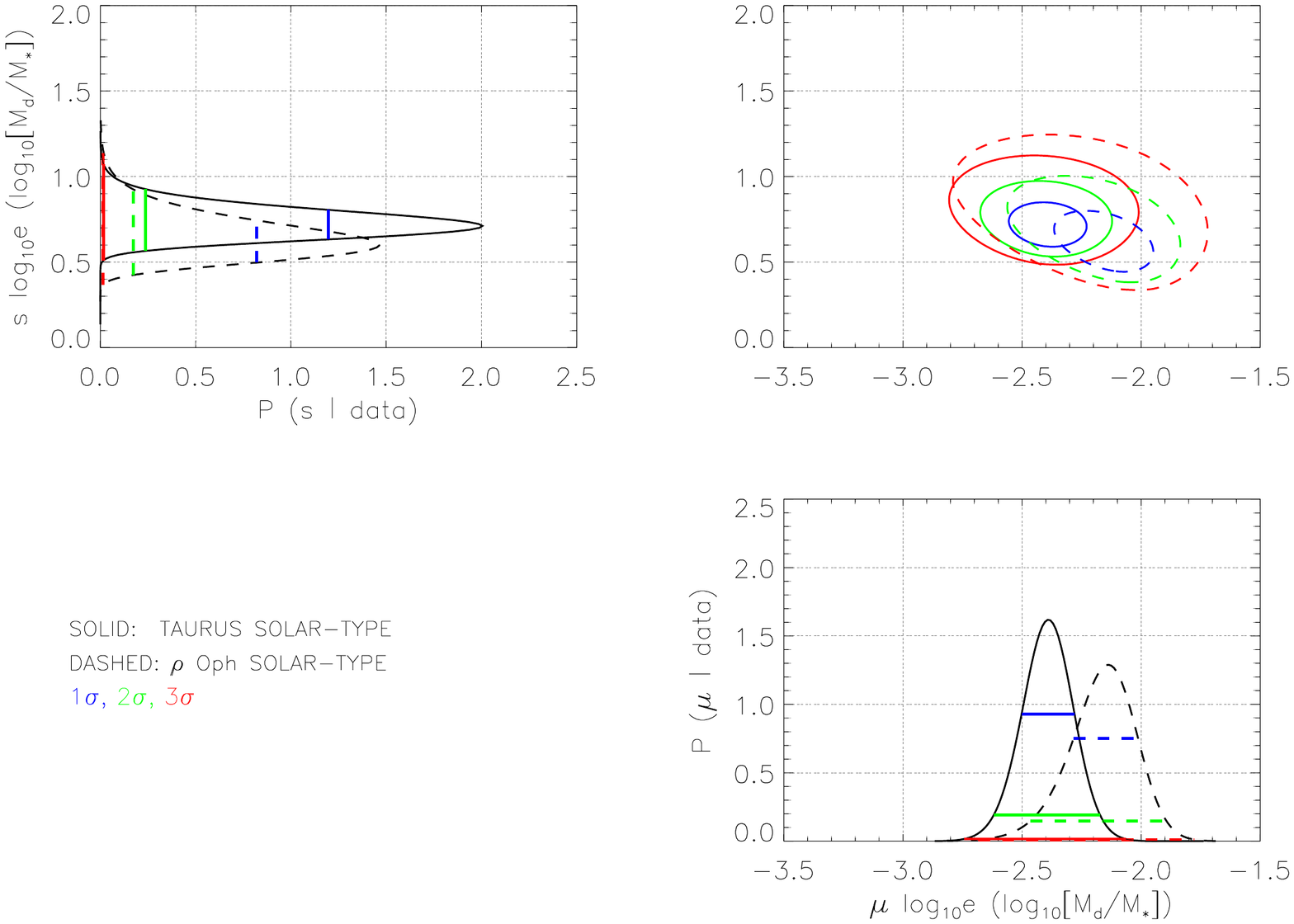}
\caption{Same as Fig.\,10, except the full Taurus sample of solar-type stars ({\it solid curves}) is now compared to the full $\rho$ Oph sample of solar-type stars ({\it dashed curves}).  The distributions of means and standard deviations in the two populations lie within 1$\sigma$ of each other. See \S8. }
\end{figure}

\begin{figure}
\includegraphics[scale=0.7, angle=90]{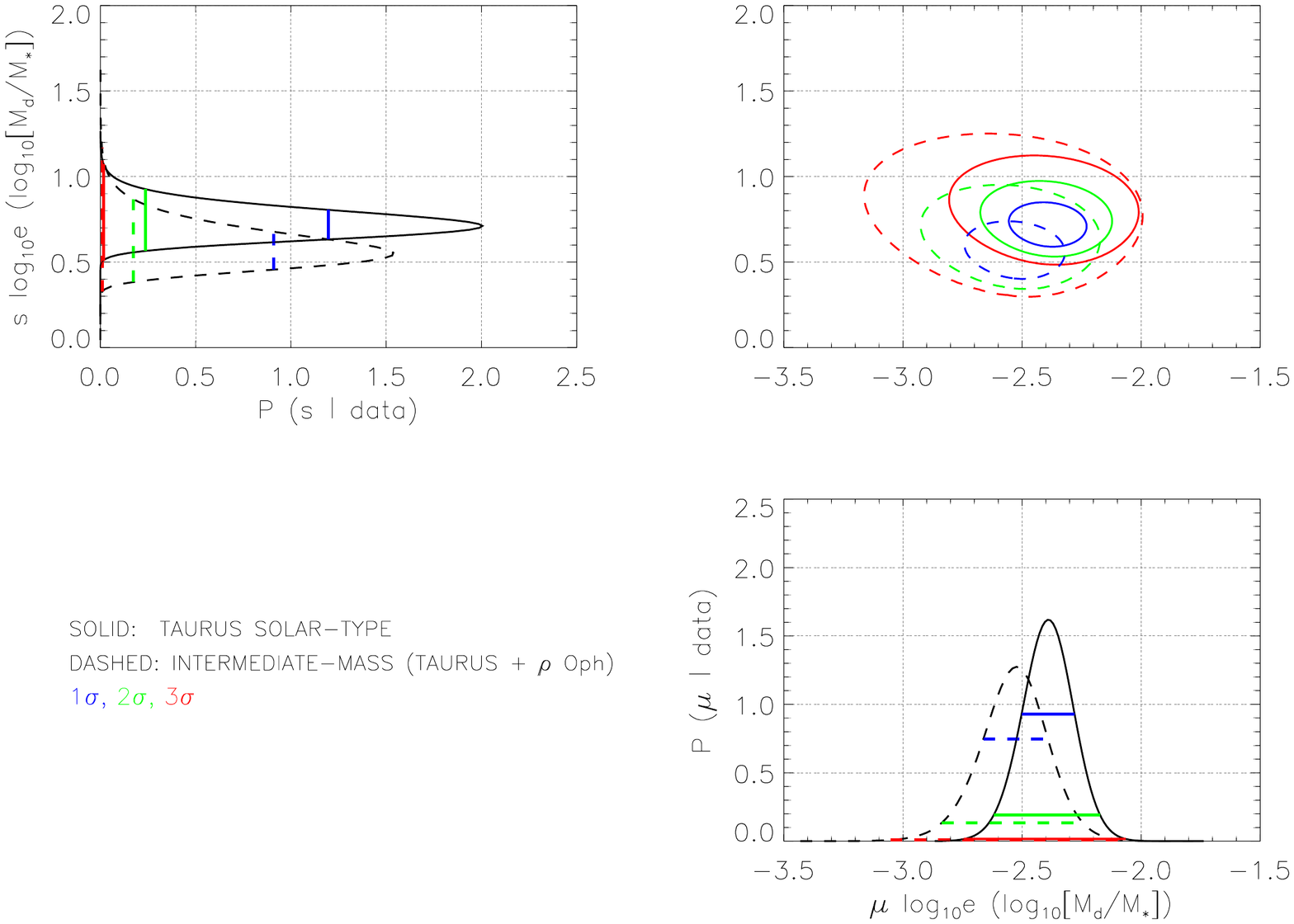}
\caption{Same as Fig.\,10, except the full Taurus sample of solar-type stars ({\it solid curves}) is now compared to the full sample of Taurus + $\rho$ Oph intermediate-mass stars ({\it dashed curves}).  The distributions of means and standard deviations in the two populations lie within 1$\sigma$ of each other. See \S8. }
\end{figure}

\begin{figure}
\includegraphics[scale=0.7, angle=90]{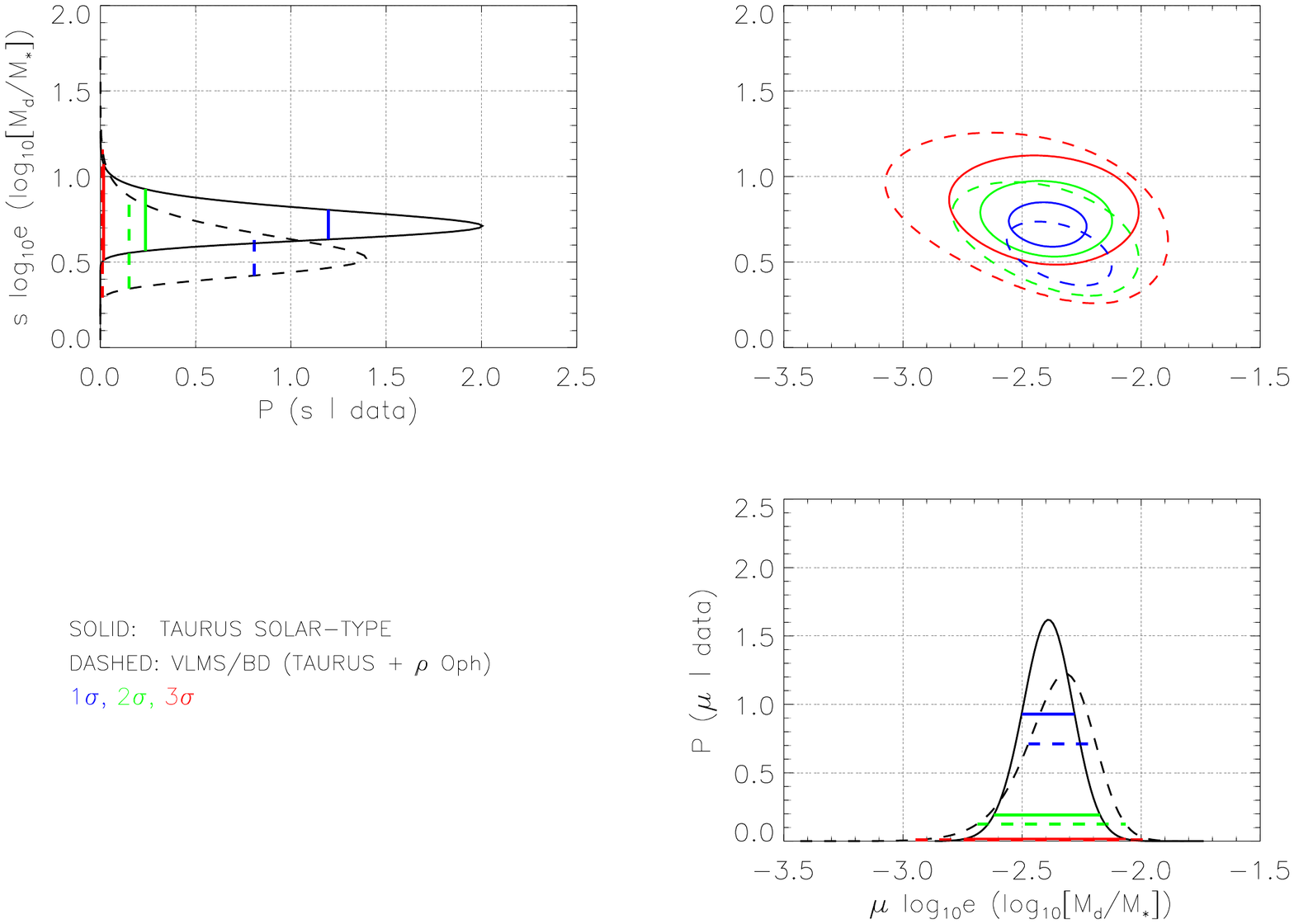}
\caption{Same as Fig.\,10, except the full Taurus sample of solar-type stars ({\it solid curves}) is now compared to the full sample of Taurus + $\rho$ Oph VLMS/BDs ({\it dashed curves}). The distributions of means and standard deviations in the two populations lie within 1$\sigma$ of each other. See \S8. }
\end{figure}

\begin{figure}
\includegraphics[scale=0.7, angle=90]{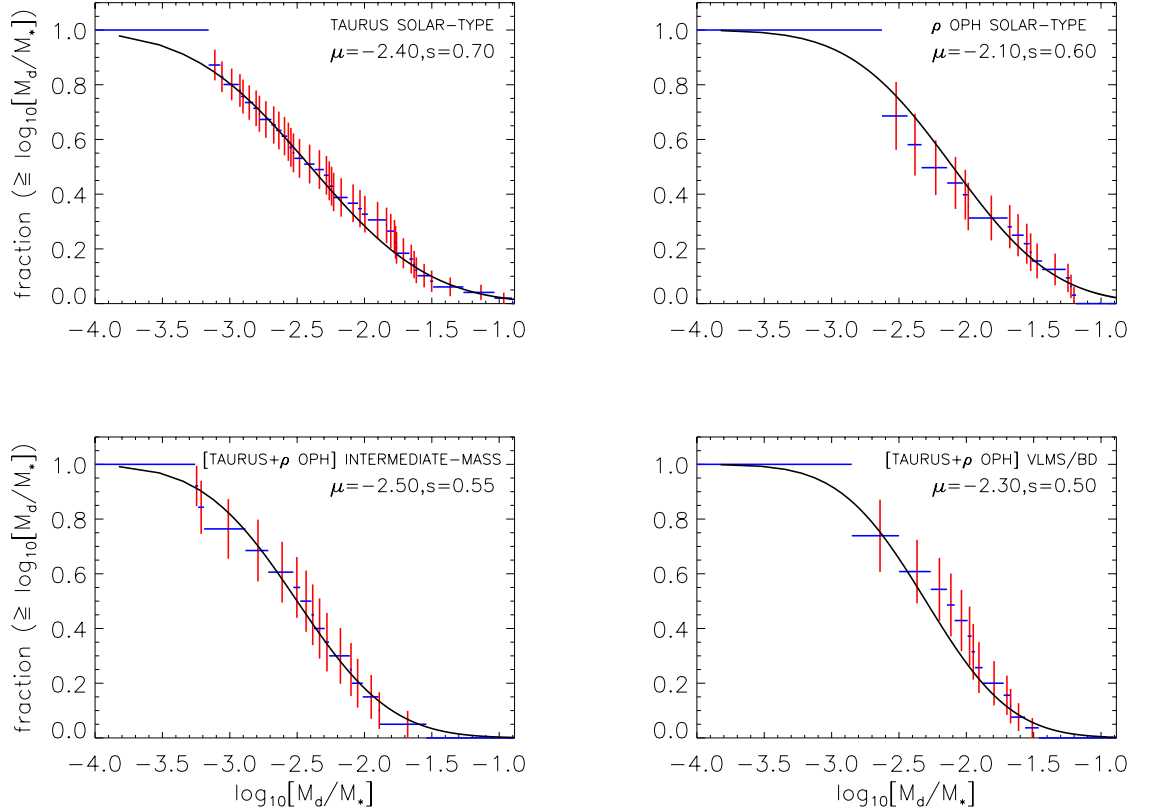}
\caption{Comparison of the cumulative distribution of objects implied by our Bayesian analysis assuming a lognormal distribution ({\it black curves}) to the cumulative distribution implied by a K-M survival analysis (denoted by the {\it crosses}; the {\it blue horizontal} and {\it red vertical} bar of each cross represent the errors implied by the K-M analysis assuming that upper limits are randomly distributed; they do not include the actual noise in the data).  {\it Top left panel}:  Comparison for full sample of Taurus solar-type stars.  {\it Top right panel}: Comparison for full sample of $\rho$ Oph solar-type stars.  {\it Bottom left panel}:  Comparison for full sample of Taurus + $\rho$ Oph intermediate-mass stars.  {\it Bottom right panel}:  Comparison for full sample of Taurus + $\rho$ Oph VLMS/BDs.  The two distributions agree very well for the Taurus solar-type and Taurus + $\rho$ Oph intermediate-mass samples where the measurement noise is least and upper limits are relatively few; they deviate more for the $\rho$ Oph solar-type stars and most for the Taurus + $\rho$ Oph VLMS/BDs, as the noise increases and upper limits become abundant. See \S 8. }
\end{figure}

\begin{figure}
\includegraphics[scale=0.7, angle=90]{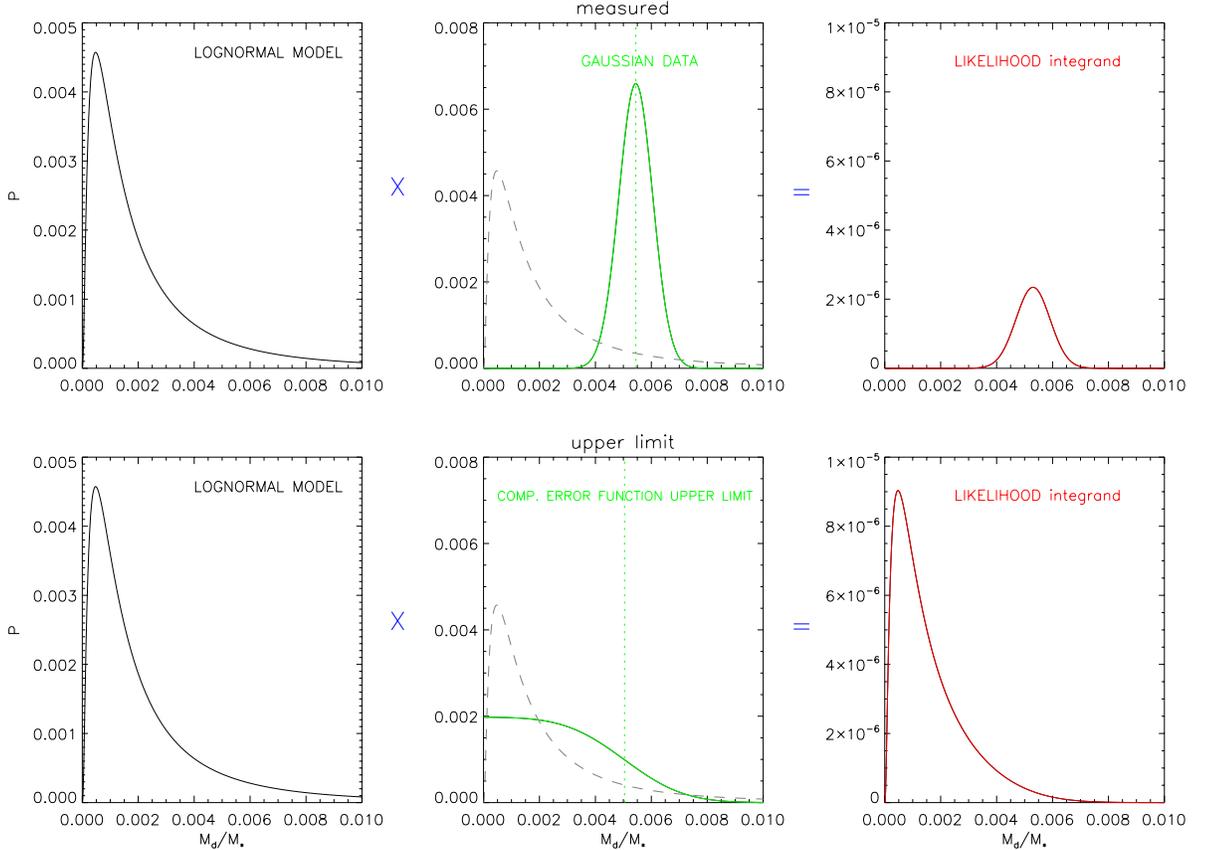}
\caption{(see Appendix C): Illustration of our posterior probability calculation.  {\bf Top row}:  Case where actual measured value is known.  {\it Left panel}:  Specific lognormal model to be tested (equation (C9), with some specified $\mu$ and $s$).  {\it Middle panel}:  Gaussian distribution of possible true values (in {\it green}), centred on the measured value (shown by the {\it vertical dotted line}).  The lognormal from the left panel is overplotted as a {\it dashed line}.  {\it Right panel}:  Product of the lognormal and Gaussian, giving the integrand of the likelihood integral (equation (C5), proportional to the posterior).  {\bf Bottom row}:  Case where only a 3$\sigma$ upper limit is known. {\it Left panel}:  Same lognormal model as in top row.  {\it Middle panel}:  Distribution of possible true values, given by the complementary error function.  The value of the 3$\sigma$ upper limit is shown by the {\it vertical dotted line}.  Note that the distribution rapidly flattens to a constant value below the upper limit (since the true value may be any value below this limit), and quickly goes to zero above the upper limit (since the probability of the true value being greater than this limit rapidly diminishes).  {\it Right panel}:  Product of the left and middle panels, giving the integrand of the likelihood integral in this case (equation (C8)).           }
\end{figure}

\end{document}